\documentclass[fleqn,usenatbib]{mnras}

\usepackage{newtxtext,newtxmath}

\usepackage[T1]{fontenc}

\DeclareRobustCommand{\VAN}[3]{#2}
\let\VANthebibliography\thebibliography
\def\thebibliography{\DeclareRobustCommand{\VAN}[3]{##3}\VANthebibliography}

\usepackage{graphicx}	
\usepackage{amsmath}	
\usepackage{mathrsfs} 
\usepackage{stfloats}
\usepackage{orcidlink}

\newcommand{\angstrom}{\mbox{\normalfont\AA}}

\newcommand{\lya}{Ly$\alpha$}

\newcommand{\SiIII}{Si\,\textsc{III}}

\title[WEAVE-QSO forecast for the \lya\ power spectrum]{Forecast for the detectability of patchy hydrogen reionization in WEAVE-QSO measurements of the Lyman-$\alpha$ forest power spectrum at redshift $\mathbf{z\geq 4}$}

\author[Ke Ma et al.]{Ke Ma$^{1}\,\orcidlink{0000-0002-0564-891X}$\thanks{E-mail: ke.ma@nottingham.ac.uk},
James S. Bolton$^{1}\,\orcidlink{0000-0003-2764-8248}$,
Vid Ir\v{s}i\v{c}$^{2}\,\orcidlink{0000-0002-5445-461X}$,
Prakash Gaikwad$^{3}\,\orcidlink{0000-0002-2423-7905}$,
Matthew M. Pieri$^{4}$\,$\orcidlink{0000-0003-0247-8991}$,
\newauthor
Trystyn A. M. Berg$^{5,6}\,\orcidlink{0000-0002-2606-5078}$,
Rajeshwari Dutta$^{7}\,\orcidlink{0000-0002-6095-7627}$,
Matteo Fossati$^{5,8}\,\orcidlink{0000-0002-9043-8764}$,
Michele Fumagalli$^{5,9}\,\orcidlink{0000-0001-6676-3842}$,
\newauthor
Emanuel Gafton$^{10}\,\orcidlink{0000-0003-0781-6638}$,
Ignasi P{\'e}rez R{\`a}fols$^{11}\,\orcidlink{0000-0001-6979-0125}$,
\& Francesco Pistis$^{12,8}\,\orcidlink{0000-0003-1189-2617}$
\\
\\
$^{1}$School of Physics and Astronomy, The University of Nottingham, University Park, Nottingham, NG7 2RD, UK\\
$^{2}$Centre for Astrophysics Research, Department of Physics, Astronomy and Mathematics, University of Hertfordshire, College Lane, Hatfield, AL10 9AB, UK\\
$^{3}$Department of Astronomy, Astrophysics and Space Engineering, Indian Institute of Technology Indore, Simrol, MP 453552, India\\
$^{4}$Aix Marseille Universit{\'e} , CNRS, CNES, Laboratoire d’Astrophysique de Marseille, Marseille 13388, France\\
$^{5}$Dipartimento di Fisica ``G. Occhialini'', Universit\`{a} degli Studi di Milano-Bicocca, Piazza della Scienza 3, 20126 Milano, Italy\\
$^{6}$Camosun College, 3100 Foul Bay Rd, Victoria, BC V8P 5J2, Canada\\
$^{7}$IUCAA, Postbag 4, Ganeshkhind, Pune 411007, Maharashtra, India\\
$^{8}$INAF - Osservatorio Astronomico di Brera, Via Brera 28, 20122 Milano, via E. Bianchi 46, 23807 Merate, Italy\\
$^{9}$INAF - Osservatorio Astronomico di Trieste, via G.B. Tiepolo 11, I-34143 Trieste, Italy\\
$^{10}$Isaac Newton Group of Telescopes, Apartado 321, 38700 Santa Cruz de la Palma, Tenerife, Spain\\
$^{11}$Departament de Física, EEBE, Universitat Politècnica de Catalunya, c/Eduard Maristany 10, 08930 Barcelona, Spain\\
$^{12}$National Centre for Nuclear Research, ul. Pasteura 7, 02-093 Warsaw, Poland\\
}

\date{Accepted XXX. Received YYY; in original form ZZZ}

\pubyear{2026}

\begin{document}
\maketitle

\begin{abstract}
    We present the first detailed forecasts for the detectability of patchy hydrogen reionization in the one-dimensional Ly$\alpha$ forest power spectrum to be measured by the WEAVE-QSO survey.  Using the Sherwood-relics reionization simulations and a WEAVE-QSO survey configuration, we generate mock spectra in four redshift bins, $z=4.0,4.2,4.4,$ and $4.6$, in which relic ionization and temperature fluctuations from patchy hydrogen reionization enhance the \lya\ forest power spectrum on large scales (i.e., at wavenumber $k\sim 10^{-3}\rm\,s\,km^{-1}$).   Our \lya\ forest pipeline forecasts the power spectrum covariance by considering sample size, spectral resolution, noise subtraction, continuum placement, metal contamination, and damping wings from high-column density absorbers.  Applying our covariance forecast within a Bayesian parameter inference framework, we find that the signature of patchy hydrogen reionization should be detectable at a significance of $\simeq4.5\sigma$.  The forthcoming WEAVE-QSO 1D power spectrum measurements should therefore be able to directly detect and characterize the large-scale relic imprint of patchy hydrogen reionization in the \lya\ forest power spectrum at $z\geq 4$.
\end{abstract}



\section{Introduction}
\label{sec:intro}

The intergalactic medium (IGM), as the largest reservoir of baryons in the Universe, plays a crucial role in tracing large-scale cosmic structure. The IGM is primarily studied through Lyman-$\alpha$ (Ly$\alpha$) forest absorption \citep{Meiksin2009, Mcquinn2016} detected in the spectra of distant quasi-stellar objects (QSOs).  Statistical analysis of the \lya\ forest offers insights into the underlying matter density fluctuations \citep{Croft2002, Viel2004, McDonald2005} as well as the thermal and ionization state of the IGM \citep{Rauch1997, Schaye2000, Bolton2005, Gaikwad2021} across a range of scales and redshifts.

The statistical properties of the \lya\ forest are often quantified by the one-dimensional flux power spectrum, $P_{\rm Ly\alpha}$, which measures the amplitude of \lya\ absorption fluctuations as a function of wavenumber, $k$. Over the past two decades, $P_{\rm Ly\alpha}$ has been established as a flexible observable with distinct scientific applications spanning various scales and redshifts \citep{Palanque-Delabrouille2013, Irsic2017b, Karacayli2020, Karacayli2022, Karacayli2024, Ravoux2023}. On small scales (wavenumbers $\log_{10}(k/{\rm s\ km^{-1}}) \gtrsim -1.5$), the shape of the power spectrum is sensitive to the Jeans smoothing of the IGM and the free-streaming length of dark matter particles. This sensitivity allows constraints to be placed on warm dark matter (WDM) models \citep{Irsic2017a, Hooper2022, Villasenor2023, Irsic2024, Rogers2025, Garcia2025}, primordial magnetic fields \citep{Pavicevic2025} and isocurvature models \citep{GarciaGallego2026}.

Another key aspect of $P_{\rm Ly\alpha}$ at high redshifts ($z \gtrsim 4$) lies in its connection to the epoch of hydrogen reionization. During reionization, the first stars and galaxies photoionized and photoheated their surrounding environments, driving the formation and expansion of ionized bubbles. As these bubbles grew and overlapped, the photoheated gas slowly cooled and imprinted spatial inhomogeneities in the IGM temperature distribution \citep{Meiksin2009, Mcquinn2016}. Due to the long timescales associated with adiabatic cooling \citep{Theuns2002, HuiHaiman2003}, these thermal inhomogeneities persist well after the completion of reionization at $z\sim 6$, remaining detectable down to $z \sim 4$ \citep{Keating2018,Wu2019,Molaro2022}. At fixed gas density and for photoionization equilibrium, hotter gas has a lower hydrogen recombination coefficient and hence a lower neutral fraction, resulting in weaker \lya\ absorption. These large-scale “thermal relics” of reionization introduce additional fluctuations in the Ly$\alpha$ optical depth, thereby enhancing the large-scale (i.e., $\log_{10}(k/{\rm s\ km^{-1}}) \lesssim -1.5$) $P_{\mathrm{Ly\alpha}}$ at $z \gtrsim 4$ \citep{Cen2009, DAloisio2015, Molaro2023}. In particular, \cite{Molaro2023} investigated this enhancement using the high resolution $P_{\rm Ly\alpha}$ measurements from \cite{Karacayli2022}. Although \citet{Molaro2023} only found tentative ($2.7\sigma$) evidence for enhanced large scale power in $P_{\rm Ly\alpha}$, the limited precision of the \citet{Karacayli2022} measurements from the Keck Observatory Database of Ionized Absorption toward Quasars (KODIAQ) \citet{OMeara2015}, Spectral Quasar Absorption Database (SQUAD) \citet{Murphy2019} and XQ-100 survey \citep{Lopez2016} (with a redshift path length of only $\Delta z=17.2$ at $z>4$) meant that the expected signature of inhomogeneous reionization could not be recovered at high significance. \citet{Molaro2023} concluded that improved statistical precision and good control of the systematic uncertainties at large scales were required for further progress.

In this context, the $P_{\rm Ly\alpha}$ measurements from massive spectroscopic surveys with resolution $1500<R<7000$, such as the Baryon Oscillation Spectroscopic Survey of the Sloan Digital Sky Survey \citep[SDSS/BOSS, ][]{Palanque-Delabrouille2013}, extended-BOSS \citep[SDSS/eBOSS, ][]{Chabanier2019}, and the Dark Energy Spectroscopic Instrument \citep[DESI, ][]{Ravoux2023, Karacayli2024, Karacayli2025, ChavesMontero2026} represent an exciting opportunity. Although restricted in their ability to access small scales due to spectral resolution ($\Delta v \sim 40$--$200\,\rm km\,s^{-1}$ for $R=7000$--$1500$), the large sample sizes result in smaller statistical uncertainties compared to the high resolution $P_{\rm Ly\alpha}$ measurements. Ongoing surveys such as DESI \citep{DESI2016}, the William Herschel Telescope Enhanced Area Velocity Explorer \citep[WEAVE, ][]{Jin2024}, the BarYon Cycle project \citep[ByCycle, ][]{Szakacs2023} on the 4-meter
Multi-Object Spectroscopic Telescope \citep[4MOST, ][]{DeJong2019}, along with planned facilities including the Prime Focus Spectrograph \citep[PFS, ][]{Greene2022}, the MUltiplexed Survey Telescope \citep[MUST, ][]{zhao2025}, and the wide-field spectroscopic telescope \citep[WST, ][]{Mainieri2024}, are expected to further refine these measurements and provide new insights into the epoch of reionization through higher spectral resolution, larger sample sizes, and increased QSO number densities.  Realizing this potential, however, requires robust synthetic data pipelines that include patchy reionization modelling to forecast systematic uncertainties, optimize survey strategies, and validate analysis frameworks.

In this paper, we therefore introduce a new pipeline for generating detailed  \lya\ forest mocks tailored to large-scale spectroscopic surveys.  Our approach is based on the Sherwood-Relics simulation suite \citep{Bolton2017, Puchwein2023} -- a set of cosmological hydrodynamical simulations specifically designed to study the IGM and the \lya\ forest during and after reionization. In contrast to other recent spectroscopic survey pipelines, we use Sherwood-Relics to model the impact of inhomogeneous reionization on the thermal and ionization history of the IGM.  Furthermore, the mocks we construct here are designed to replicate the instrumental properties of the WEAVE-QSO survey \citep{Pieri2016}, as part of WEAVE \citep{Jin2024}, which is optimized for quasar absorption science.  We use these mocks to obtain the first detailed forecast of the precision to which WEAVE-QSO will measure $P_{\rm Ly\alpha}$ at redshift $z\geq 4$, and to assess the detectability of large-scale power arising from patchy hydrogen reionization.  We note that some recent work \citep{Montero2021,Zheng2026} has also explored the Ly$\alpha$ forest power spectrum as a probe of reionization and cosmic-dawn astrophysics.  However, these studies focus on temperature fluctuations seeded by the hydrodynamical response of gas during reionization that may remain detectable at $z<4$ \citep{Hirata2018,Cain2024}; here we instead focus on the large-scale thermal relic imprint predicted by patchy reionization models at $z>4$ \citep[e.g.,][]{Keating2018,Wu2019,Molaro2022}.

This paper is structured as follows: Section 2 details the method for generating QSO spectra mocks. Section 3 outlines the computation of $P_{\rm Ly\alpha}$ and describes how instrumental effects and absorption line contamination are modelled to construct the associated overall uncertainty.  In Section 4, we present forecasts for detecting the relic signature of inhomogeneous reionization using our $P_{\rm Ly\alpha}$ pipeline at redshifts $z \geq 4$. Finally, a summary is provided in Section 5.


\section{Construction of mock \lya\ forest spectra}

\subsection{Hydrodynamical simulation}

 In this work we employ the Sherwood-Relics simulation suite \citep{Puchwein2023} for generating mock \lya\ forest spectra.  Sherwood-Relics was performed with a customised version of the cosmological hydrodynamical code P-Gadget-3 \citep{Springel2005}.  The underlying cosmology in the simulation used here is flat $\Lambda$CDM, parameterized by $\Omega_{\Lambda} = 0.692$, $\Omega_{\rm m} = 0.308$, $\Omega_{\rm b}=0.0482$, $\sigma_{8} = 0.829$, $n_{\rm s} = 0.961$, $h = 0.678$, with a primordial helium abundance by mass $Y = 0.24$.  The dark matter particle mass is $M_{\rm dm}=5.37\times 10^{5}h^{-1}\,\rm M_{\odot}$ and the gas particle mass is $M_{\rm gas}=9.97\times 10^{4}h^{-1}\,M_{\odot}$, which are chosen to adequately resolve the small-scale structure of the \lya\ forest \citep{BoltonBecker2009}.  We use simulated \lya\ optical-depth skewers extracted from a $40h^{-1}\rm\,cMpc$ box with $2 \times 2048^{3}$ particles, following the procedure described in \cite{ma2025}.


\subsection{The WEAVEify algorithm}
\label{sec:QSM_WEAVEify}

We construct our mock spectra assuming survey properties that resemble those of WEAVE-QSO \citep{Pieri2016}, one of the eight core surveys of WEAVE \citep{Jin2024}. The WEAVE-QSO survey will obtain spectra of approximately $450\,000$ QSOs at redshifts $z_{\rm QSO} > 2.2$, covering a nominal survey area of approximately $10\,000\rm\,deg^2$ to $m_{\rm r}<21.5$, with $z>4$ QSOs additionally targeted over $21.5<m_{\rm r}<22$. Also a further nested tier in the HETDEX spring field will provide $100$ QSOs per deg$^2$.

We have developed a software package, WEAVEify, to fully forward model spectra representative of typical observations from the WEAVE-QSO survey. It has been rigorously tested with operation rehearsal (OpR) mock spectra of the WEAVE survey, and accurately models noise at the detector level. Further details on WEAVEify will be presented in a forthcoming publication (Gaikwad et al., in prep).  Similar WEAVE-like mock quasar spectra have also been used in recent WEAVE continuum-modelling work \citep{Pistis2025}.  Here we focus on describing the core functionality necessary for obtaining \lya\ forest mocks.

Starting from input optical-depth skewers drawn from the Sherwood-Relics hydrodynamical simulation, WEAVEify maps the rest frame optical depths to a common set of wavelength bins based on the systemic redshift of the QSO.  Once aligned, the total optical depth is computed by summing the
individual optical depths, $\tau$, from any of the transitions required in the mock spectrum (e.g. \lya, \SiIII).  The transmission is computed as $F={\rm e}^{-\tau}$.  WEAVEify then applies wavelength-dependent instrumental broadening by convolving the transmission with a Gaussian function whose standard deviation varies
with wavelength. WEAVEify then resamples the broadened transmission from the simulation wavelength grid onto the WEAVE wavelength solution using linear interpolation.  For continuum placement, WEAVEify
employs a Principal Component Analysis (PCA)-based method, with PCA components
derived from a large set of SDSS QSO continua as outlined in \citet{paris2011} \citep[also see][]{davies2018}. These components are used to project the continuum from the red to the blue side of the rest frame
\lya\ wavelength.

In the final step, WEAVEify adds realistic noise to the spectrum. As WEAVE is a
fibre-fed spectrograph, the spectra are photon-limited, with the noise governed
by Poisson statistics. Photon counts in each fibre arise from four primary
sources: (i) photons from the source within the fibre’s field of view (FoV),
(ii) sky photons, (iii) dark current, and (iv)
detector readout noise.  WEAVEify models the photon counts for the source and
sky by incorporating key parameters such as source and sky brightness, exposure
time, and the spatial extent of the source (point-like or extended). It further accounts for atmospheric observing conditions, including the seeing profile (with a Moffat function with FWHM $0.75''$ and $\beta=2.5$) and an airmass of $1.107$. The OpR3 noise-model defaults here use a fixed exposure time of $3060\,\rm s$ and a sky brightness of $m_B=20.92$.   Noise characteristics dependent on the telescope's
engineering design, such as fibre diameter, fibre offset, CCD size, and
telescope throughput, are also included to accurately estimate photon counts.
Detector dark current and readout noise are included as separate noise terms. The overall noise is then calculated as a quadrature sum of these terms.


\begin{figure*}
    \centering
    \includegraphics[width=2\columnwidth]{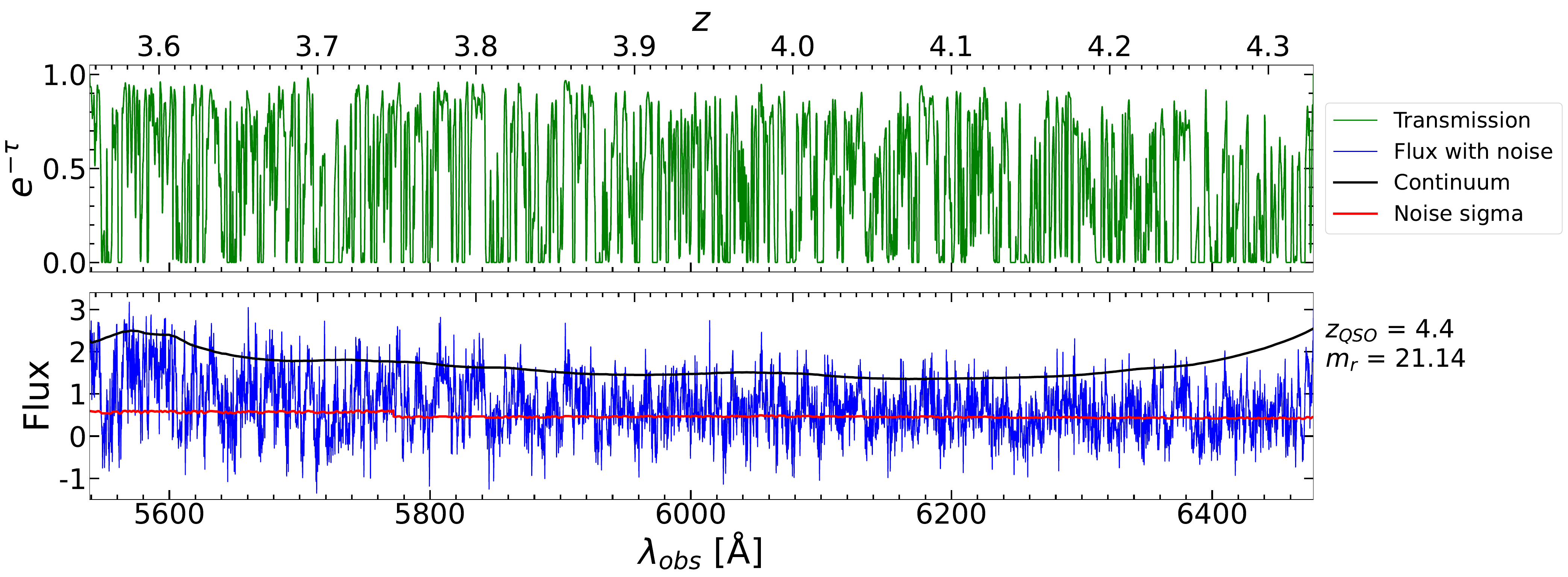}
    \vspace{-3mm}
    \caption{An example mock QSO spectrum constructed using Sherwood-Relics and WEAVEify. The top panel shows the Ly$\alpha$ forest transmission constructed from one of  the Sherwood-Relics hydrodynamical simulations \citep{Puchwein2023}. The bottom panel shows the WEAVEified spectrum when taking the \lya\ transmission in the top panel as input, for a QSO with systemic redshift $z_{\rm QSO} = 4.4$ and r-band magnitude $21.14$. Note that the \lya\ transmission is only modelled between the blue edge of the QSO proximity region of extent $4\,000\rm\,km\,s^{-1}$ and the rest frame Ly$\beta$ wavelength (see text for details). The signal-to-noise per pixel, shown as $\sigma = (\rm S/N)^{-1}$, is displayed by the orange curve.}
    \label{fig:seeFlux}
\end{figure*}

\subsection{WEAVEifying the Ly$\alpha$ forest transmission}
\label{sec:QSM_TLya}

The mock Ly$\alpha$ optical depths used as the input to WEAVEify are obtained by combining different snapshots of the Sherwood-Relics simulation.  The snapshots have a redshift cadence of $\Delta z = 0.1$ and contain \lya\ optical depths along $5\,000$ skewers drawn parallel to the simulation box boundaries. A full synthetic QSO line of sight is generated by connecting multiple such flux skewers end-to-end.

The top panel of Figure \ref{fig:seeFlux} shows an example of the resulting Ly$\alpha$ forest transmission for a mock QSO with systemic redshift $z_{\rm QSO} = 4.4$.  The optical depths are rescaled by a common multiplicative factor within each redshift interval, such that the mean transmission averaged over all skewers and pixels in that interval matches the effective optical depth measured by \cite{Becker2013}. This reference model may underestimate $\tau_{\rm eff}$ at $z>4$ \citep[see the discussion in][]{Boera2019}; in our parameter inference, however, $\tau_{\rm eff}$ is varied independently in each redshift bin and marginalized over, rather than fixed to this prescription (Section~\ref{sec:inference_framework}). Each skewer is periodically shifted to ensure continuity at the points where the skewers are joined.  For each mock QSO spectrum, the Ly$\alpha$ forest absorption is modelled only at wavelengths redward of the Ly$\beta$ emission line, below which contamination from higher order Lyman series lines becomes significant.  Regions blueward of the Ly$\alpha$ emission line that are affected by the QSO proximity effect \citep{Carswell1982,Murdoch1986} are also excluded. We account for this proximity effect over a velocity interval of $4\,000\,\rm km\,s^{-1}$ \citep[e.g.,][]{Kim2007}.   The \lya\ transmission is then converted into a mock observed spectrum using the WEAVEify software. An example of the resulting mock QSO spectrum (blue) is shown in the lower panel of Figure \ref{fig:seeFlux}.



\subsection{Generating the mock WEAVE-QSO survey sample}
\label{sec:QSM_mock}

\begin{figure*}
    \centering
    \includegraphics[width=2\columnwidth]{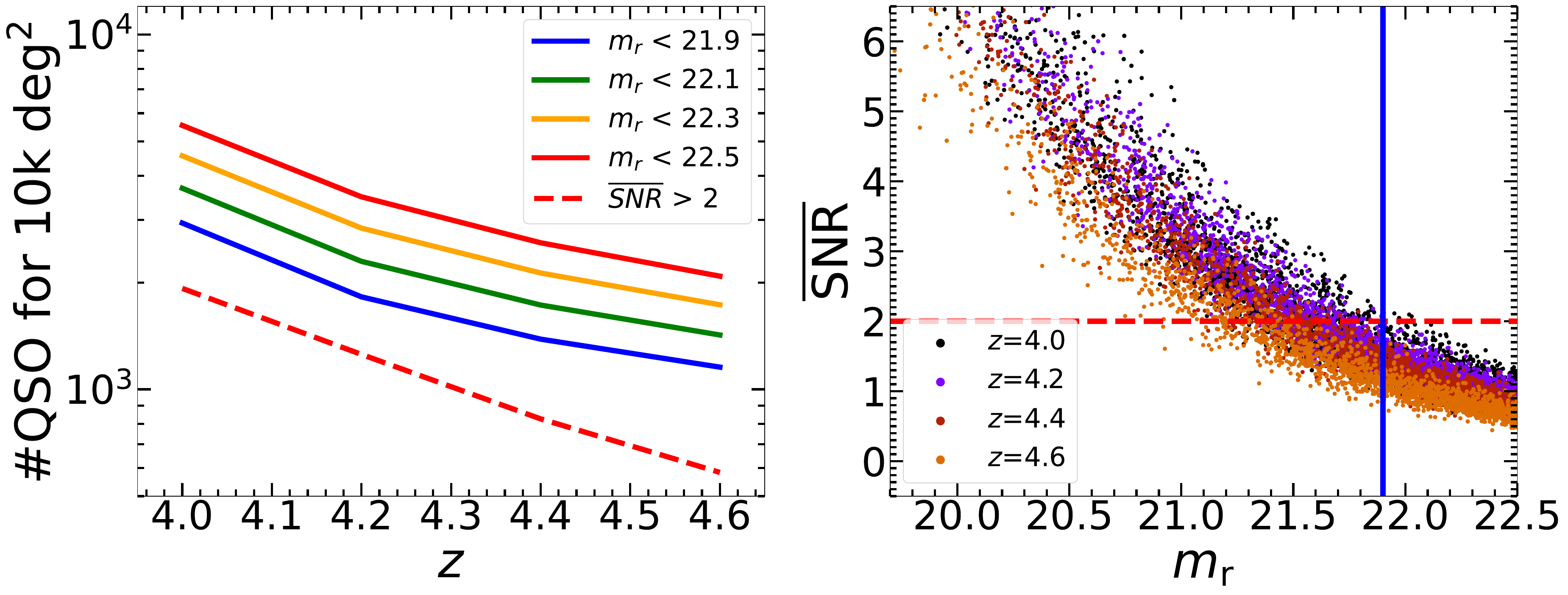}
    \vspace{-3mm}
    \caption{Left panel: The cumulative number of QSOs available in a $10\,000$ deg$^2$ survey for four \lya\ forest power spectrum bins centred at $z=4.0,\,4.2,\,4.4$ and $z=4.6$. The numbers are derived using the QSO luminosity function measurements presented by \citep{Palanque-Delabrouille2016}.  The results are shown for different r-band magnitude cuts, $m_{\rm r}$ (solid curves), or a mean signal-to-noise ratio cut, $\overline{\rm SNR}$ (dashed curve).  Right panel: Scatter plot of $\overline{\rm SNR}$ against the r-band magnitude for the QSOs in each redshift bin. The blue solid line corresponds to a $m_{\rm r}$ cut at 21.9 and the red dashed line corresponds to a $\overline{\rm SNR}$ cut at 2. We adopt an r-band magnitude cut $m_{\rm r} < 21.9$ throughout this paper.}
    \label{fig:NQSOs}
\end{figure*}

Using the procedure described in the previous section, for each of four $P_{\rm Ly\alpha}$ redshift bins centered at $z=4.0,\,4.2,\,4.4$ and $4.6$ we generate $10\,000$ mock QSO spectra that have systemic redshifts selected to have full spectral coverage within the bin. These bins target the redshift range where relic fluctuations from patchy reionization are expected to remain observable while retaining sufficient WEAVE-QSO sightlines for a useful forecast.   The distributions of the QSO r-band magnitude, $m_{\rm r}$, and systemic redshift, $z_{\rm QSO}$, are sampled using a Monte Carlo method based on the QSO number density derived from the \cite{Palanque-Delabrouille2016} luminosity function using their PLE+LEDE model predictions at $0<z<6$. Our reference forecast will therefore assume a complete magnitude-limited sample subject to the forest-coverage requirements above, without additional losses from target selection, fibre allocation or unsuccessful observations. Uncertainties in the luminosity function and departures from this completeness assumption may change the available QSO counts and their magnitude and redshift distributions, and hence the forecast precision. Note also that the r-band magnitude and systemic redshift are assigned to the simulated sightlines independently of the \lya\ absorption in our mocks. We then scale this luminosity-function-based number density distribution to a reference forecast sample covering $10\,000$ $\rm deg^{2}$, applying the relevant $m_{\rm r}$ and $z_{\rm QSO}$ selections to obtain the expected QSO counts used below.   For each set of $10\,000$ mock QSO spectra, we also determine the fraction of QSOs in each redshift bin satisfying specific magnitude and signal-to-noise criteria (described below).  These fractions are then scaled to obtain the total expected number of suitable QSOs across the full $10\,000\,\rm deg^2$ survey, as illustrated in the left panel of Figure \ref{fig:NQSOs}.

The QSO sample cuts for $P_{\rm Ly\alpha}$ are determined as follows.  The r-band magnitude $m_{\rm r}$ influences the noise level in the Ly$\alpha$ forest region, which directly affects the measurement of $P_{\rm Ly\alpha}$.  We quantify the noise in our mocks using $\overline{\rm SNR}$, defined as the signal-to-noise ratio per $\rm \angstrom$ averaged over all pixels within a given redshift bin.  Previous studies of $P_{\rm Ly\alpha}$ at $2\lesssim z \lesssim 4$ \citep[e.g.,][]{Chabanier2019, Ravoux2023} have often applied $\overline{\rm SNR}$ cuts of 2 or 3 to mitigate noise.  However, at the comparatively high redshifts considered in this work where the number of sufficiently bright QSOs is smaller, such cuts would drastically reduce the number of usable QSOs and increase the statistical uncertainty on $P_{\rm Ly\alpha}$.  Hence, to maintain sample size while controlling noise in our mocks, we therefore instead implement a cut in r-band magnitude. Specifically, we adopt $m_{\rm r} < 21.9$ for all redshift bins, but without an additional SNR cut; the SNR cuts shown in Figure~\ref{fig:NQSOs} are included for comparison only. Note again that because the assigned magnitudes are independent of the simulated absorption, this selection does not preferentially retain higher-transmission sightlines in our mocks, as might be expected if using an observed r-band-selected sample at $z>4$. In practice, rather than the simple r-band cut we use here, for WEAVE-QSO the planned selection strategy will use prior redshift estimates and the forest SNR will be optimised using a spectral or narrow-band information redward of \lya\ emission at each redshift.

As shown in the left panel of Figure \ref{fig:NQSOs}, this magnitude cut retains over $1\,000$ QSOs even in our highest redshift bin at $z = 4.6$. After applying the cut $m_{\rm r} < 21.9$, we find the number of suitable QSOs in a $10\,000\,\rm deg^2$ survey is $N_{\rm QSO}=2\,942$, $1\,822$, $1\,385$, and $1\,155$ for $P_{\rm Ly\alpha}$ redshift bins $z = 4.0,\,4.2,\,4.4$ and $4.6$ respectively. We use these numbers throughout the rest of the paper, but in Section~\ref{sec:inference_p1d_comparison} we will briefly discuss how a smaller numbers of suitable QSOs (e.g., through incompleteness or reduced survey area) may impact on our forecast).



\section{Power spectrum estimation}


\subsection{FFT estimator}
\label{sec:P1D_FFT}

We now discuss the calculation of the 1D power spectrum and its statistical and systematic uncertainties using our QSO survey mocks.  The two approaches for computing the P$_{\rm Ly\alpha}$ commonly used in the literature are the Fast Fourier Transform \citep[FFT, e.g.][]{Palanque-Delabrouille2013, Chabanier2019, Ravoux2023} and the Quadratic Maximum Likelihood Estimator \citep[QMLE, e.g.][]{Karacayli2022, Karacayli2024}. Although the latter benefits from being less sensitive to masking of the \lya\ forest spectrum, the former method is more straightforward. In addition, \cite{Ravoux2023} and \cite{Karacayli2024} measured the P$_{\rm Ly\alpha}$ on the same DESI data sets using FFT and QMLE respectively, and found that the results are consistent within 1 per cent up to the half Nyquist frequency. We therefore use the FFT approach to measure $P_{\rm Ly\alpha}$.

When performing power spectra analysis, the commonly used estimator is the flux contrast $\delta_{\rm Ly\alpha}(\lambda)$, where
\begin{equation}
    \delta_{\rm Ly\alpha}(\lambda) = \frac{F_{\rm Ly\alpha}(\lambda)}{\overline{F}(\lambda)} - 1.
    \label{eqn:deltaF_Lya}
\end{equation}
Here $F_{\rm Ly\alpha}(\lambda)$ is the \lya\ transmission and $\overline{F}(\lambda)$ is the mean transmission over some wavelength interval (which also depends on redshift).  In this work we assume a mean transmission based on the \lya\ forest effective optical depth measurement presented by \cite{Becker2013}, where $\tau_{\rm eff}=-\ln\bar{F}$. The power spectrum $P_{\rm Ly\alpha}(k)$ is then given by
\begin{equation}
    P_{\rm Ly\alpha}(k) = | \mathcal{F}( \delta_{\rm Ly \alpha}(\lambda)) |^2,
    \label{eqn:FFT}
\end{equation}
where $\mathcal{F}$ represents the Fourier transform.  In practice, $P_{\rm Ly\alpha}$ is obtained by averaging the individual power spectra obtained from each of the QSO spectra within the redshift bin under consideration.

However, $\delta_{\rm Ly\alpha}(\lambda)$ is not directly observable.  Instead, the observed flux contrast, $\delta_{\rm obs}(\lambda)$, is given by
\begin{equation}
    \delta_{\rm obs}(\lambda) = \frac{F_{\rm obs}(\lambda)}{C_{\rm QSO}(\lambda)\ \overline{F}(\lambda)} - 1,
    \label{eqn:deltaF_obs}
\end{equation}
where $C_{\rm QSO}(\lambda)$ is the intrinsic QSO emission obtained from the continuum fitting procedure. For now we shall assume the QSO continuum is perfectly known, but we will return to modelling continuum fitting uncertainties later.

The underlying Ly$\alpha$ flux contrast, $\delta_{\rm Ly\alpha}(\lambda)$, differs from the observed flux contrast, $\delta_{\rm obs}(\lambda)$, due to instrumental resolution effects as well contamination by metal lines present in the \lya\ forest.  The combination of these effects along with noise yields the observed flux. The relation between $\delta_{\rm Ly\alpha}(\lambda)$ and $\delta_{\rm obs}(\lambda)$ can be expressed as,
\begin{equation}
    \delta_{\rm obs}(\lambda) = [\delta_{\rm Ly\alpha}(\lambda) + \delta_{\rm metals}(\lambda)] \ast W(\lambda,R,\Delta \lambda) + \delta_{\rm noise}(\lambda),
    \label{eqn:delta_components}
\end{equation}
where $W$ is the spectrograph line spread function as a function of wavelength $\lambda$, spectral resolution $R$, and the pixel size $\Delta \lambda$. The terms $\delta_{\rm metals}$ and $\delta_{\rm noise}$ represent the contribution to the observed flux contrast by intervening metal lines and noise, respectively. The contribution from high column density systems is treated separately in Section~\ref{sec:P1D_HCD}.

After Fourier transforming Equation (\ref{eqn:delta_components}) using Equation (\ref{eqn:FFT}), we obtain the relation between the observed power spectrum, $P_{\rm obs}$, and the \lya\ forest power spectrum, $P_{\rm Ly\alpha}$:
\begin{equation}
    P_{\rm obs}(k) = [P_{\rm Ly\alpha}(k) + P_{\rm metals}(k)]W^2(k,R,\Delta \lambda) + P_{\rm noise}(k).
    \label{eqn:P1D_components}
\end{equation}
One again, $P_{\rm metals}$ and $P_{\rm noise}$ represent the contribution to the observed power spectrum by metals and noise.  We now turn to consider each of these components and their contribution to the measurement uncertainties in turn.


\subsection{Spectrograph resolution correction}
\label{sec:P1D_res}

\begin{figure}
    \centering
    \includegraphics[width=1\columnwidth]{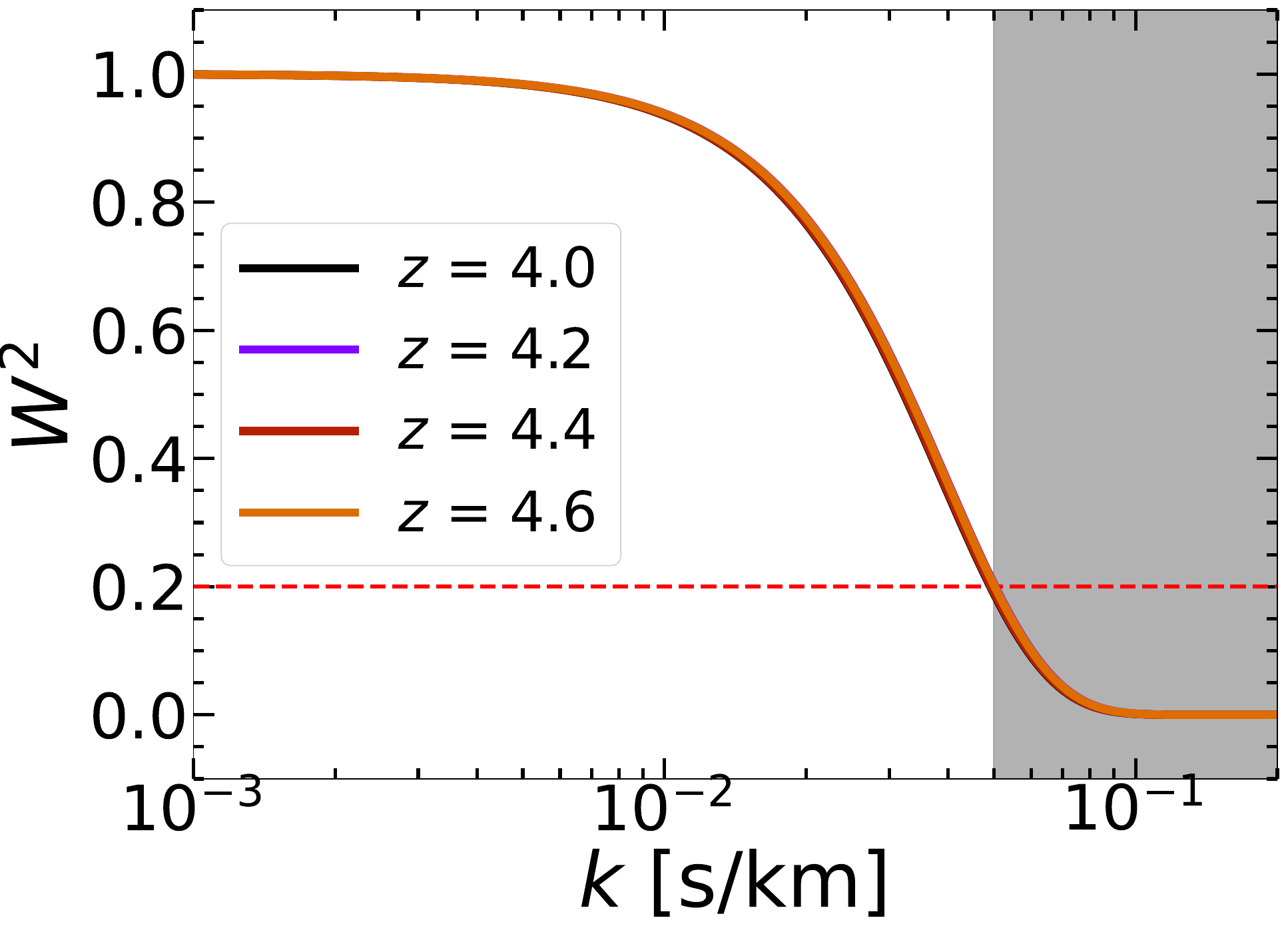}
    \vspace{-6mm}
    \caption{The instrumental resolution correction factor, $W^2(k,R,\Delta \lambda)$, for the four $P_{\rm Ly\alpha}$ redshift bins considered in this work.   Note the curves are almost indistinguishable.   The red dashed horizontal line marks the threshold $W^2=0.2$, and the grey shaded region shows the maximum wavenumber for the $P_{\rm Ly\alpha}$ measurements that we consider in this work, $k_{\rm max} = 0.05\ \rm s\ km^{-1}$, where the resolution correction factor falls below this value.  This choice follows the threshold used by the DESI collaboration \citep{Ravoux2023}.}
    \label{fig:ResCorr}
\end{figure}

The WEAVE spectrograph low-resolution mode covers $366$--$959\,\rm nm$ at a representative resolving power of $R\sim5\,000$, while the high-resolution mode provides restricted wavelength coverage at $R\sim20\,000$ \citep{Pieri2016, Jin2024}. The majority of QSO spectra in WEAVE-QSO will be obtained using the low-resolution mode.  We assume $R\sim 5\,000$ in this work.

The WEAVEify package characterizes the effect of the spectrograph resolution using the line spread function matrix $W(\lambda,R,\Delta \lambda)$.  As a forecast-level approximation, a Gaussian kernel is assumed with a standard deviation that varies as a function of wavelength, as detailed in Section~\ref{sec:QSM_WEAVEify}.  The square of the Fourier transform of this resolution matrix corresponds to the resolution correction factor $W^2(k,R,\Delta \lambda)$ that is applied to the power spectrum in Equation (\ref{eqn:P1D_components}). The corresponding resolution windows for the four redshift bins we consider are shown in Figure~\ref{fig:ResCorr}; these remain very similar in all four bins.   We use these correction factors to determine the $k$-range impacted by instrument resolution and to estimate the contribution of the instrumental resolution correction to the final systematic uncertainty estimate.   The maximum wavenumber, $k_{\rm max}$, for the $P_{\rm Ly\alpha}$ measurements that we consider in this work is selected following the same criterion as the DESI collaboration \citep{Ravoux2023}, where $k_{\rm max}$ is defined as the point beyond which the resolution correction falls below $0.2$. This yields $k_{\rm max} = 0.05\,\rm s\ km^{-1}$.

To estimate the uncertainty associated with applying this resolution correction, we draw $1\,000$ bootstrap realizations of $N_{\rm QSO}$ sight-lines from our mock spectra. For each realization, we compare the resolution corrected estimate with the underlying model and measure the scatter in their ratio across the bootstrap samples. The resulting term quantifies the residual uncertainty in the correction procedure. However, we note that this uncertainty quantifies the robustness of the correction for a fixed, known resolution window function. In a real WEAVE-QSO analysis, the resolving power will vary with wavelength and observing conditions, and uncertainty in $R$ will propagate into uncertainty in $W^2(k,R,\Delta\lambda)$. This effect can become important near the high-$k$ limit of the measurement and should ultimately be treated using a wavelength-dependent resolution matrix or window-function uncertainty model.  We discuss this further in Section~\ref{sec:P1D_err}.


\subsection{Noise subtraction}
\label{sec:P1D_nos}

\begin{figure}
    \centering
    \includegraphics[width=1\columnwidth]{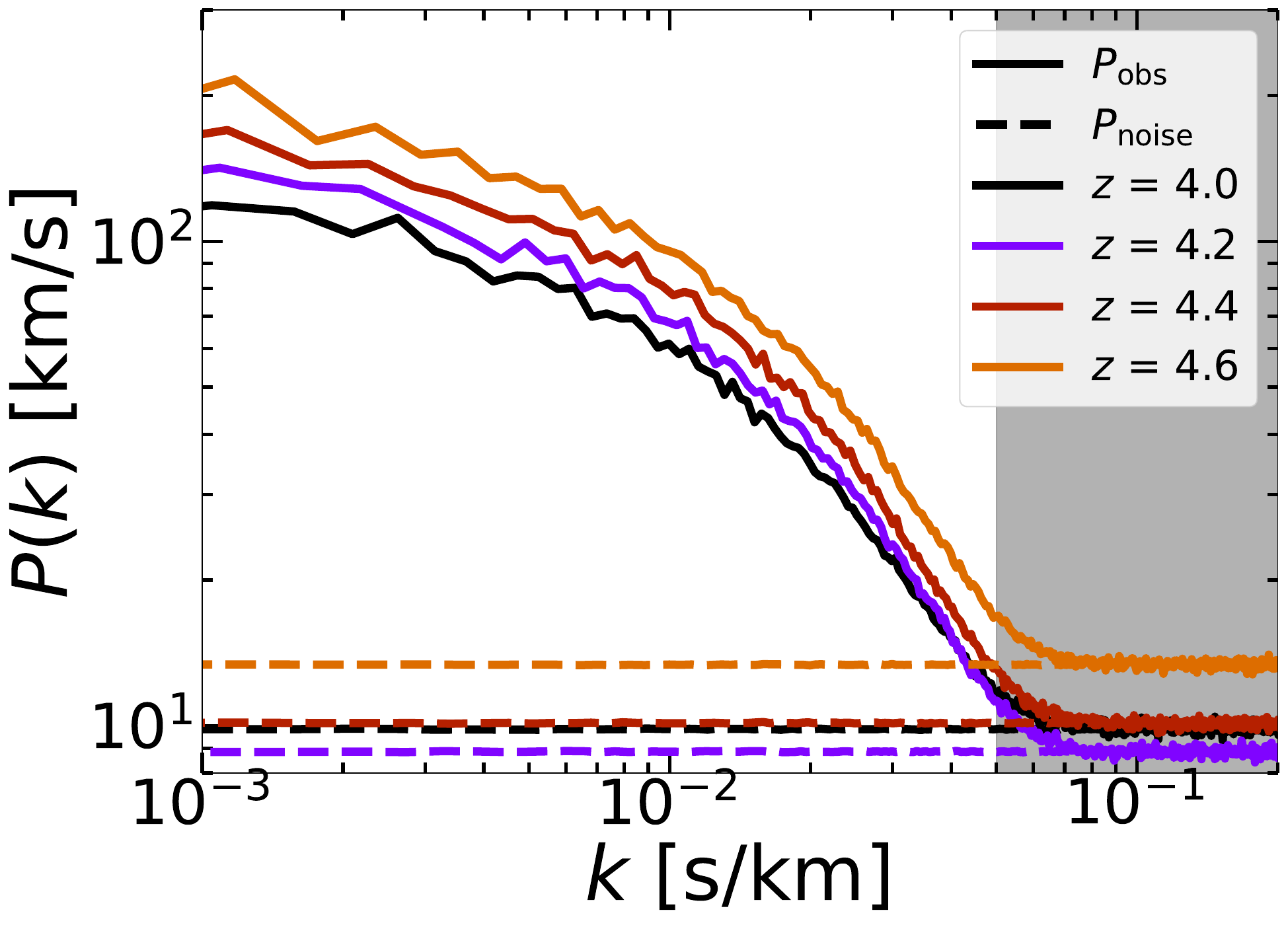}
    \vspace{-6mm}
    \caption{Comparison between the estimated noise power spectra $P_{\rm noise}$ from WEAVEify (dashed lines) and the observed power spectra $P_{\rm obs}$ (solid lines, see Equation (\ref{eqn:P1D_components}) for further details). The noise will make a significant contribution to the power spectrum uncertainty at the smallest scales (i.e., large $k$).  The grey shading shows the excluded region above the maximum wavenumber, $k_{\rm max}=0.05\rm\,s\,km^{-1}$, considered in this work.}
    \label{fig:NoiseSub}
\end{figure}

The noise level is quantified by $\sigma_{\rm f}(\lambda)$, the standard deviation of the flux. In observational studies, this quantity is typically estimated from the data reduction pipeline or using methods such as the exposure difference technique \citep[e.g.,][]{Chabanier2019, Ravoux2023}. For our mock data, however, $\sigma_{\rm f}(\lambda)$ is directly provided as an output of the WEAVEify software. The standard deviation of the noise term $\delta_{\rm noise}(\lambda)$ in Equation (\ref{eqn:delta_components}) is thus defined as
\begin{equation}
    \sigma_{\delta}(\lambda) = \frac{\sigma_{\rm f}(\lambda)}{C_{\rm QSO}(\lambda)\ \overline{F}(\lambda)}.
    \label{eqn:noisesigma}
\end{equation}
In each redshift bin, for each pixel at wavelength $\lambda$ we generate $100$ realizations of $\delta_{\rm noise}(\lambda)$ by drawing from a normal distribution with standard deviation $\sigma_{\delta}(\lambda)$. The noise power spectrum $P_{\rm noise}(k)$ in Equation (\ref{eqn:P1D_components}) is then estimated as $P_{\rm noise}(k) = \langle \ | \mathcal{F}(\delta_{\rm noise}(\lambda)) |^2 \ \rangle$, where the average is computed over all QSO spectra and $N_{\rm noise}$ realizations within the redshift bin.  Note that the use of a normal distribution for generating $\delta_{\rm noise}(\lambda)$ reflects the assumption of Gaussian noise in WEAVEify. However, the noise model can be adapted as needed to match specific instrumental characteristics or data reduction pipelines.

Figure \ref{fig:NoiseSub} displays the observed power spectra and the estimated noise power spectra for the four $P_{\rm Ly\alpha}$ redshift bins considered in this work. The noise component becomes dominant beyond $k_{\rm max} = 0.05\,\rm s\ km^{-1}$, which is also the wavenumber limit adopted for the spectral resolution correction. In a similar manner to the resolution correction uncertainty, we again estimate the uncertainty associated with noise subtraction using $1\,000$ bootstrap realizations of $N_{\rm QSO}$ sight-lines. For each realization, we compare the noise-subtracted power spectrum estimate with the corresponding noiseless mock estimate, and take the standard deviation of this correction ratio across bootstrap samples.



\subsection{Metal absorption lines}
\label{sec:P1D_met}

The metal flux contrast, $\delta_{\rm metals}$, consists of two components: one originating from intervening metal lines with rest frame wavelengths $\lambda > \lambda_{\rm Ly\alpha}$, and the other from metal lines that are correlated with the Ly$\alpha$ forest with rest frame wavelengths $\lambda \sim \lambda_{\rm Ly\alpha}$. The first category includes C\,\textsc{IV} ($\lambda \lambda$1548, 1551), Si\,\textsc{IV} ($\lambda \lambda$1394, 1403), and Mg\,\textsc{II} ($\lambda \lambda$2796, 2804) which we will refer to as uncorrelated metals. The second category is dominated by Si\,\textsc{III} ($\lambda$1207) in our modeling, which we refer to as correlated metals. We have also tested Si\,\textsc{II} $\lambda\lambda1190,1193$ in this modelling framework, but found its contribution to be sub-percent over the $k<k_{\rm max}$ range considered here; see \citet{ma2025} for further details.

\begin{figure}
    \centering
    \includegraphics[width=1\columnwidth]{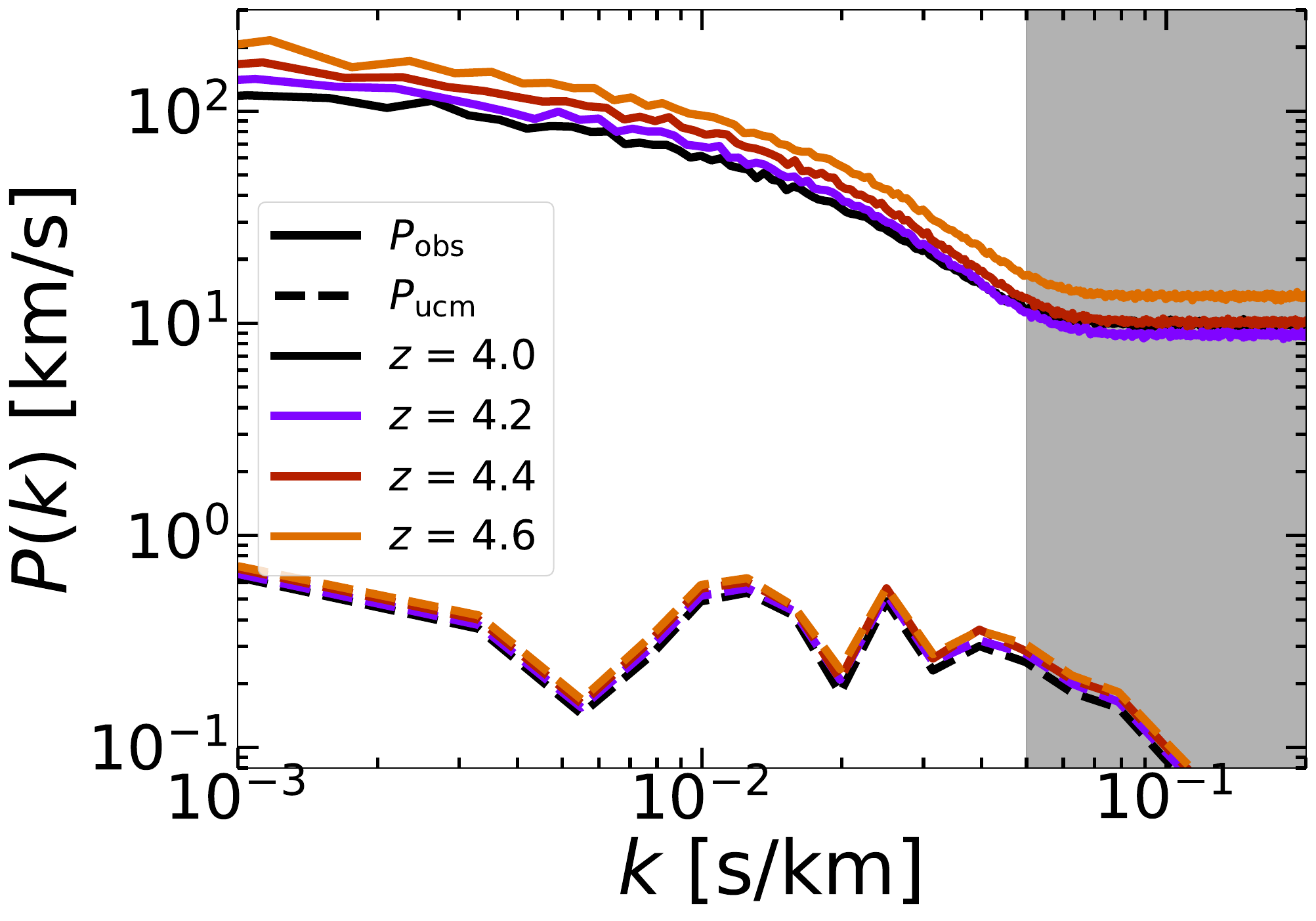}
    \vspace{-6mm}
    \caption{Comparison between the uncorrelated metal power spectrum, $P_{\rm ucm}$, predicted by Monte Carlo sampling the observed column density distribution functions for C\,\textsc{IV}, Si\,\textsc{IV}, and Mg\,\textsc{II} (dashed lines) to the total observed power spectrum, $P_{\rm obs}$. The grey shading shows the excluded region above the maximum wavenumber, $k_{\rm max}=0.05\rm\,s\,km^{-1}$, considered in this work.}
    \label{fig:MetalsUC}
\end{figure}

In practice,  the uncorrelated metal power spectrum, $P_{\rm ucm}$, is typically estimated using sidebands -- spectral regions redward of the Ly$\alpha$ rest frame wavelength that are devoid of Ly$\alpha$ absorption \citep{Palanque-Delabrouille2013, Chabanier2019, Ravoux2023}. For this forecast, we generate the baseline \lya\ mock spectra without uncorrelated metal absorption and then add synthetic uncorrelated metal features separately to estimate $P_{\rm ucm}$ and its contribution to the uncertainty budget.  We do this following the approach of \cite{Viel2013}. We first estimate the number of absorption lines for each metal line species per unit redshift path length by integrating the appropriate column density distribution function (CDDF) \citep{DOdorico2013, DOdorico2022, Mathes2017} over the column density range $10^{12} {\rm \,cm^{-2}}< N < 10^{15}$ cm$^{-2}$. Using a Monte Carlo method, we draw column densities and line widths (with $b$-parameter distributions from \cite{HuiRutledge1999} assuming $b_{\sigma} = 10\,\rm km\,s^{-1}$) until the required number of metal lines is reached. These synthetic absorption features are then randomly inserted into the Ly$\alpha$ forest region of each mock QSO spectrum. The CDDFs for C\,\textsc{IV}, Si\,\textsc{IV}, and Mg\,\textsc{II} doublets are modelled as power laws using the measurements from \cite{DOdorico2013}, \cite{DOdorico2022}, and \cite{Mathes2017}, respectively.  We use this to estimate the scale-dependent uncertainty associated with uncorrelated metal subtraction. The relative magnitude of $P_{\rm ucm}$ compared with $P_{\rm obs}$ is shown in Figure~\ref{fig:MetalsUC}. Although the contribution of $P_{\rm ucm}$ can become relevant at very small scales, $k>0.1\rm\,s,km^{-1}$ \citep{Viel2013}, at the larger scales of interest in this work it represents only a very small contribution to the total uncertainty (see Figure~\ref{fig:Errors} later).

\begin{figure}
    \centering
    \includegraphics[width=1\columnwidth]{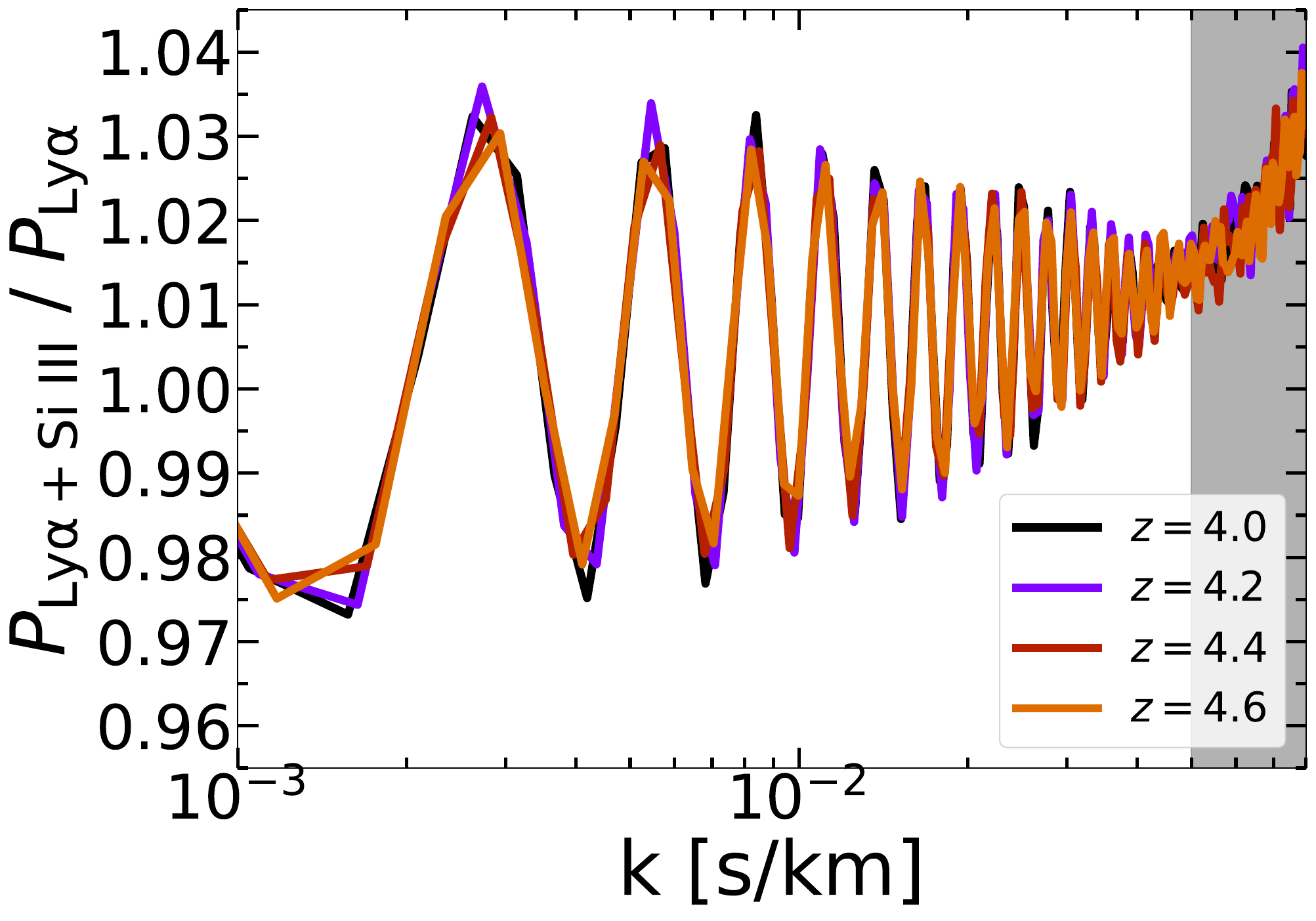}
    \vspace{-6mm}
    \caption{The ratio of the combined Ly$\alpha$ and Si\,\textsc{III} power spectrum, $P_{\rm Ly\alpha+SiIII}$, to the Ly$\alpha$ power spectrum, $P_{\rm Ly\alpha}$. The grey shading shows the excluded region above the maximum wavenumber, $k_{\rm max}=0.05\rm\,s\,km^{-1}$, considered in this work.}
    \label{fig:MetalsCL}
\end{figure}

For the correlated metals, we generate Si\,\textsc{III} absorption spectra following the method described in \citet{ma2025}. We assume the Si\,\textsc{III} optical depths scale as $10^{\rm [Si/H]}$, where the silicon abundance [Si/H] is derived from the measurements of \citet{Schaye2003, Aguirre2004} and the ionization fractions are computed using the photoionization code Cloudy \citep{Ferland1998, Chatzikos2023}.  Figure~\ref{fig:MetalsCL} illustrates the impact of Si\,\textsc{III} ($\lambda$1207) on $P_{\rm Ly\alpha}$ and the characteristic oscillations that result \citep[see also][]{McDonald2006}.   Note, however, that correlated metals like Si\,\textsc{III} are typically not removed from $P_{\rm Ly\alpha}$ analyses but are instead incorporated as additional fitting parameters. This is because their power spectrum, denoted here as $P_{\rm clm}$, cannot be directly isolated from observations as their absorption lines are blended with the Ly$\alpha$ forest. In this work, correlated metals are therefore not directly included in the systematic uncertainty estimate; instead, we assess in Section~\ref{sec:inference_framework} whether omitting an explicit Si\,\textsc{III} model affects our parameter inference.  For this purpose we use the four parameter model for Si\,\textsc{III} absorption recently introduced by \citet{ma2025}.  As we will discuss later, however, we find that the inclusion of the Si\,\textsc{III} in our mocks makes very little difference to our final results.

\subsection{High column density systems}
\label{sec:P1D_HCD}

For the purpose of estimating the systematic uncertainty associated with incomplete absorber removal, we model high column density systems (HCDs), defined here as absorbers with neutral hydrogen column densities satisfying $N_{\rm HI} > 10^{19}\rm\,cm^{-2}$.  These systems are highly relevant to this study because enhancements to the power spectrum on large scales can readily arise from column densities $N_{\rm HI}>10^{19}\rm\,cm^{-2}$ due to their extended Lorentzian damping wings.  However, the observed incidence of these absorption lines is not captured in our hydrodynamical simulations; self-shielded gas with normalised density $\rho/\langle \rho \rangle>10^{3}$ and $T<10^{5}\rm\,K$ is ignored due to the ``Quick-Ly$\alpha$'' implementation \citep{Viel2004} used in Sherwood-Relics \citep[see e.g. figure 1 in][]{Miller2019}.   We therefore incorporate these HCDs into our mock QSO spectra using a similar method to that employed for the  uncorrelated metal lines. The column density distribution function from \cite{Prochaska2014} is used to estimate the expected number of HCDs per unit redshift path length, which we then Monte Carlo sample over the range $17.2 \leq \log_{10}(N_{\rm HI}/{\rm cm}^{-2}) \leq 23.0$ defined by \cite{Prochaska2014} assuming a fixed Doppler parameter $b=20\,\rm km\,s^{-1}$. The resulting synthetic HCD absorbers are inserted as Voigt profiles at locations where the Ly$\alpha$ optical depth in our mock spectra is largest.

\begin{figure}
    \centering
    \includegraphics[width=1\columnwidth]{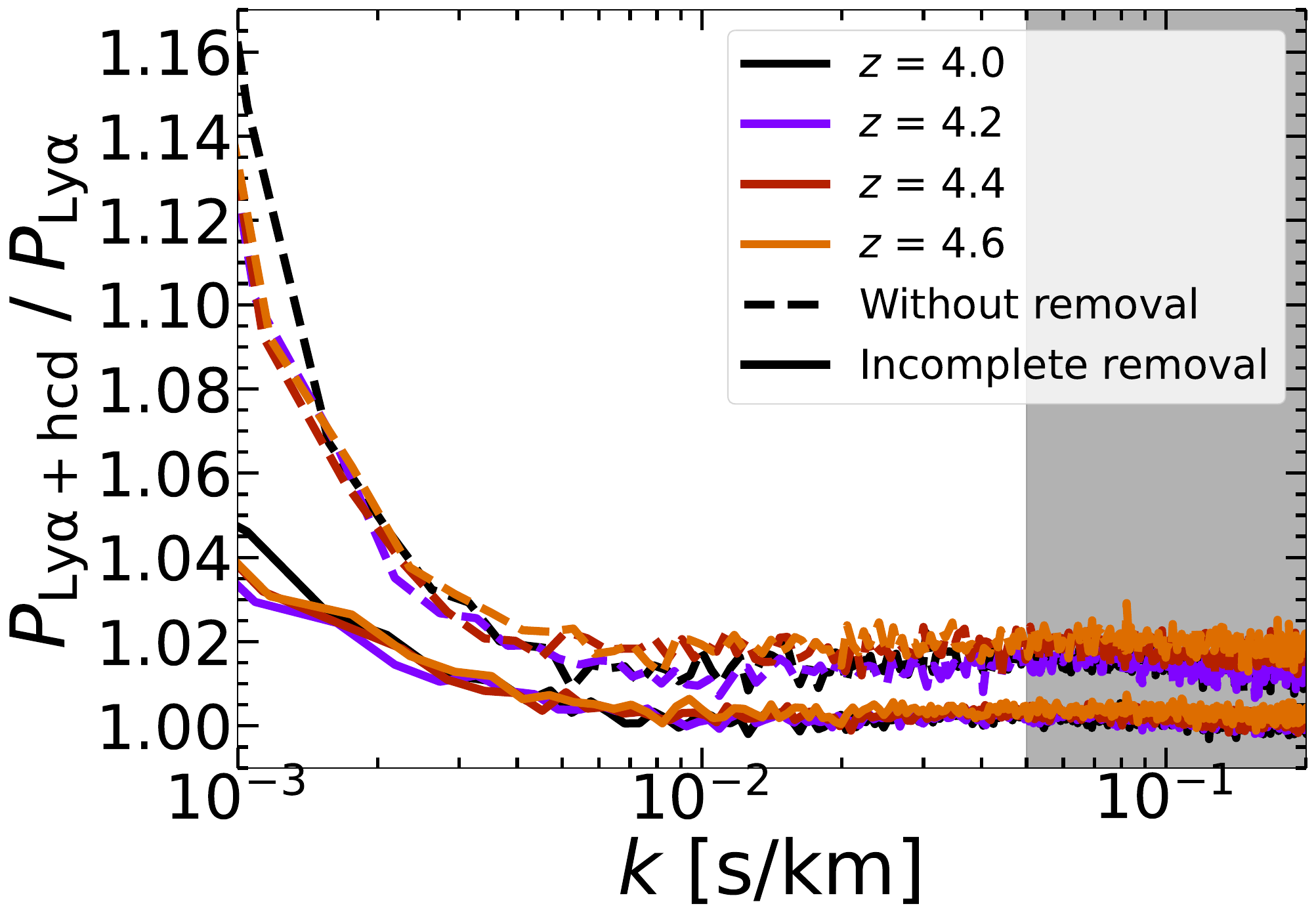}
    \vspace{-6mm}
    \caption{The ratio of the combined power spectrum for Ly$\alpha$ and HCDs, $P_{\rm Ly\alpha+hcd}$ to the Ly$\alpha$ power spectrum, $P_{\rm Ly\alpha}$.  The dashed curves show the ratio including all HCDs, whereas the solid curves give the ratio after removing a fraction of the HCDs based on the completeness function reported by \citet{Wang2022}.  The grey shading shows the excluded region above the maximum wavenumber, $k_{\rm max}=0.05\rm\,s\,km^{-1}$, considered in this work.}
    \label{fig:HCD}
\end{figure}

In practice, HCDs cannot be perfectly identified and removed from observational data. To assess the potential bias in $P_{\rm Ly\alpha}$ resulting from incomplete removal, we use the completeness function from \cite{Wang2022} for systems with $N_{\rm HI}>10^{19}\rm\,cm^{-2}$.  For each $\log_{10} N_{\rm HI}$ bin, we randomly remove HCDs according to this completeness function -- considering our mean signal-to-noise ratio -- and retain the remaining systems in the Ly$\alpha$ forest region. The impact of HCDs on $P_{\rm Ly\alpha}$, both with and without incomplete removal, is illustrated in Figure \ref{fig:HCD}. The systematic error introduced by the incomplete removal (solid curves) is added into the total uncertainty estimate in quadrature. For the incomplete-removal model shown in Figure~\ref{fig:HCD}, the residual HCD contribution to $P_{\rm Ly\alpha}$ is at most $\sim 5$ per cent at $10^{-3}\lesssim k < 0.05\,{\rm s\,km^{-1}}$, with the maximum occurring at the largest scales. Our forecast does not separately propagate uncertainty in the assumed HCD column density distribution. However, steepening and flattening its high-column-density slope by $\pm0.2$ dex, with the variation amplitude motivated by the observational uncertainties compiled by \citet{Prochaska2014}, changes the residual HCD contribution by less than 1 per cent of the underlying Ly$\alpha$ power across all four redshift bins and the fitted wavenumber range.

\subsection{Continuum shape}
\label{sec:P1D_cont}

In our mock QSO spectra a continuum template is assigned to each line of sight by WEAVEify and is subsequently divided out during the computation of $P_{\rm Ly\alpha}$ (as described in Section \ref{sec:P1D_FFT}). However, in practice the QSO continuum is not known and may be misplaced. As a conservative forecast-level test of this systematic effect, instead of dividing by the true continuum (which varies with wavelength) we divide the WEAVEified mocks by a flat, wavelength independent continuum and evaluate the resulting impact on $P_{\rm Ly\alpha}$.  The flat continuum is defined as the mean value of the true continuum over the redshift bin covered by each mock QSO spectrum.   This approach isolates shape related biases from normalization effects, which are instead marginalized over by varying the effective optical depth, $\tau_{\rm eff}$, in our parameter inference procedure.  

\begin{figure}
    \centering
    \includegraphics[width=1\columnwidth]{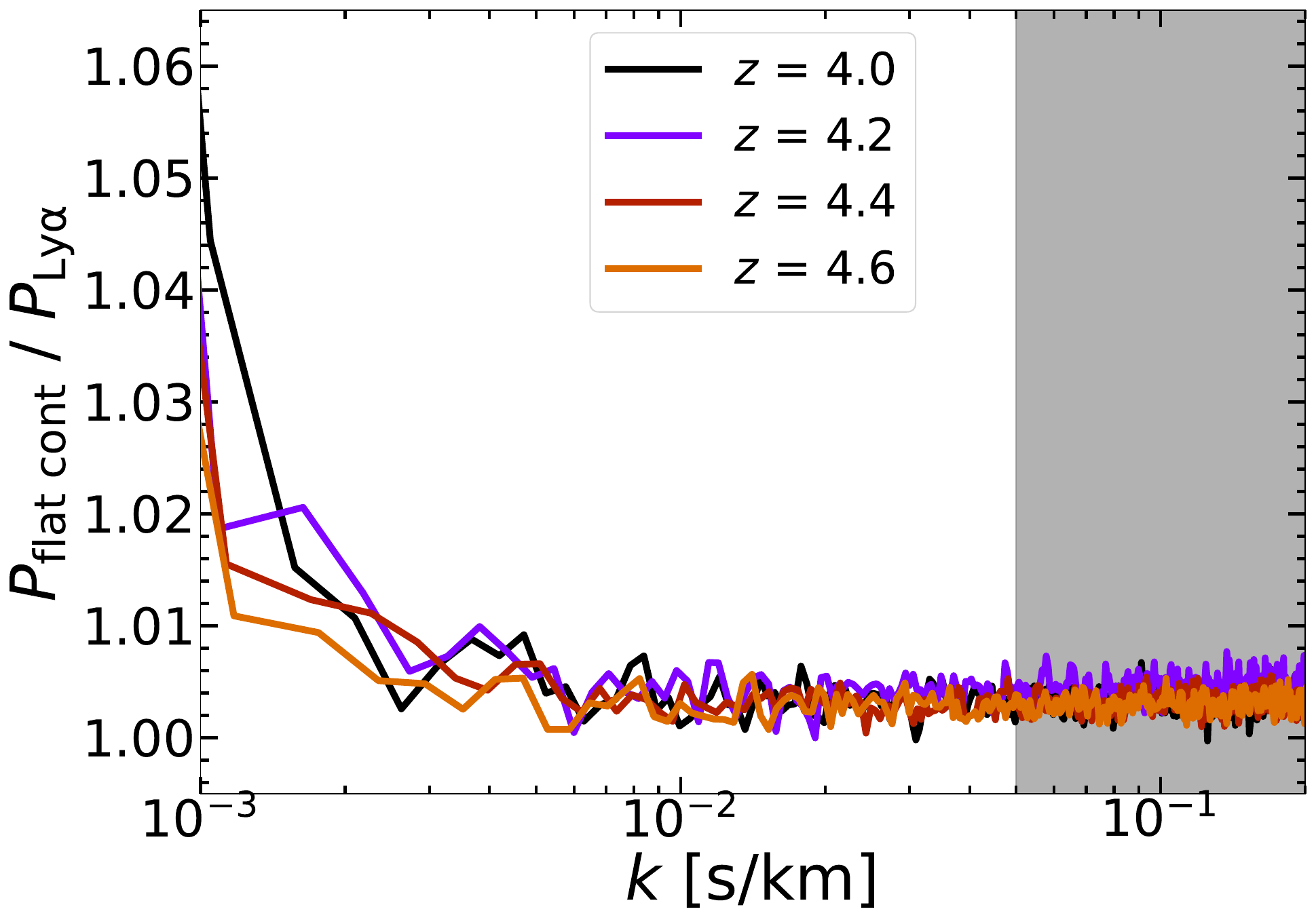}
    \vspace{-6mm}
    \caption{The ratio of the Ly$\alpha$ power spectrum measured after replacing the true continuum with a flat, wavelength-independent continuum, $P_{\rm flat\,cont}$, to the \lya\ power spectrum, $P_{\rm Ly\alpha}$, measured assuming perfect knowledge of the continuum. The grey shading shows the excluded region above the maximum wavenumber, $k_{\rm max}=0.05\rm\,s\,km^{-1}$, considered in this work.}
    \label{fig:Cont}
\end{figure}

The impact of an incorrect continuum shape on $P_{\rm Ly\alpha}$ is shown in Figure \ref{fig:Cont}. The associated systematic error is again added in quadrature to the total uncertainty estimate. Note, however, that this test does not model a specific continuum-fitting procedure. Under-fitting or over-fitting may introduce a systematic excess or suppression of large-scale power that is not explicitly accounted for here and will ideally need to be assessed for the specific continuum-fitting method adopted in the WEAVE-QSO survey. We also remark that if a smaller observational sample were obtained by selecting brighter QSOs, we might expect the associated improvements in continuum fitting accuracy (and also HCD identification) to partly offset the increase in statistical uncertainty from the reduced sample size.  However, assessing this trade-off requires modelling the performance of these procedures as a function of data quality, which is beyond the scope of the present forecast.

\subsection{Final uncertainty estimate}
\label{sec:P1D_err}

\begin{figure*}
    \centering
    \includegraphics[width=2\columnwidth]{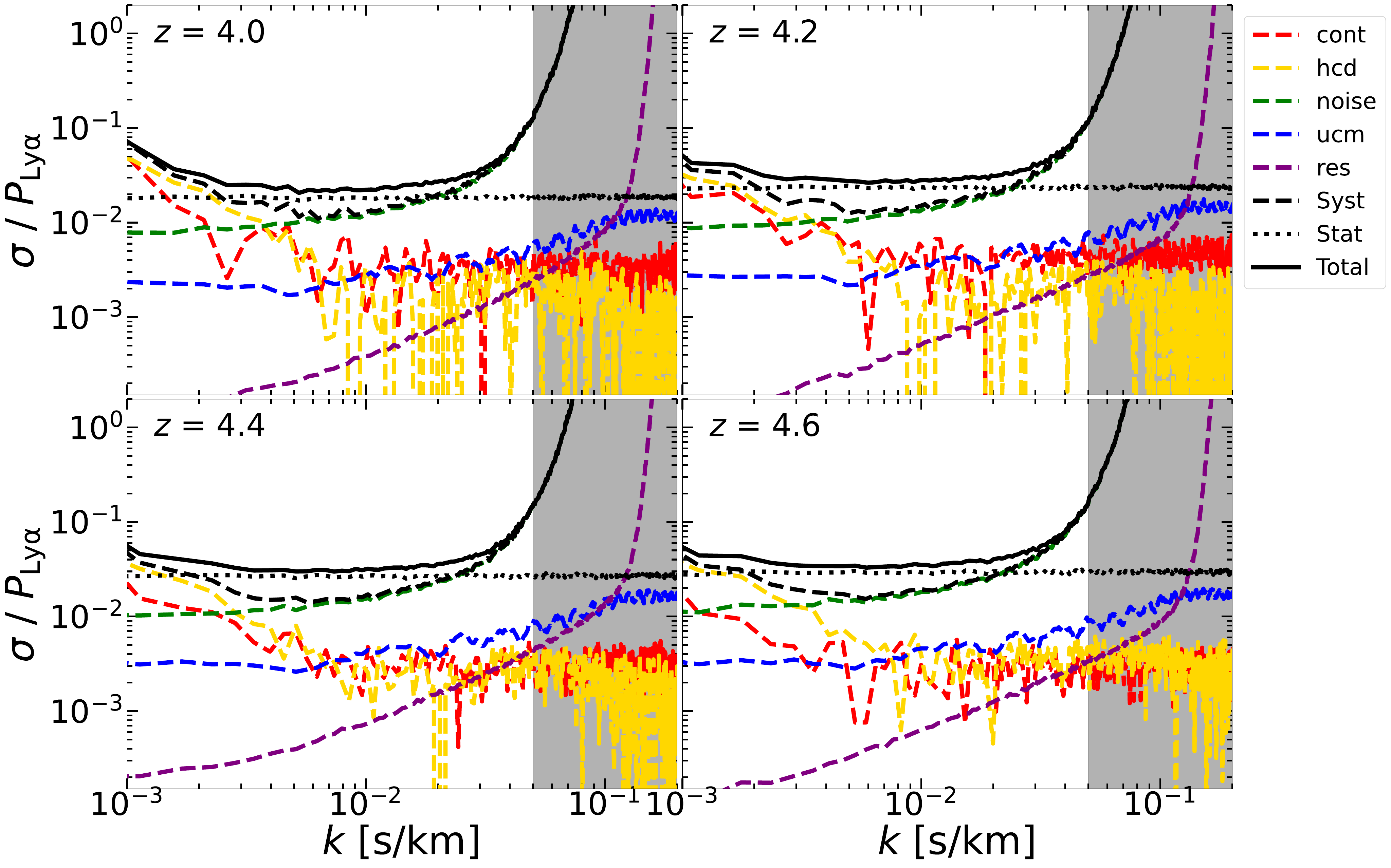}
        \vspace{-3mm}
    \caption{Forecast for the total relative uncertainty (solid black curves) on a WEAVE-QSO measurement of the 1D \lya\ forest power spectrum at $z=4.0$--$4.6$.  Each panel represents a different bin of width $\Delta z=0.2$ centred at the redshift indicated. The statistical errors (dotted black curves) are calculated by bootstrapping with replacement.  The systematic uncertainty contributions are modelled separately for each observational or astrophysical effect and  are added together in quadrature (black dashed curve). The red, yellow, green, blue, and purple dashed curves represent the contributions from continuum placement, incomplete HCD identification, noise subtraction, uncorrelated-metal contamination, and spectral resolution, respectively. The total uncertainty is obtained from the quadratic sum of the statistical and systematic uncertainty. The grey shading shows the excluded region above the maximum wavenumber, $k_{\rm max}=0.05\rm\,s\,km^{-1}$, considered in this work.  For $k<k_{\rm max}$ the relative errors at large scales are always within 10 per cent.
    }
    \label{fig:Errors}
\end{figure*}

We now combine our systematic uncertainty estimates with the expected statistical uncertainty from the WEAVE-QSO survey.  Statistical uncertainties originate from finite sample sizes and are estimated relative to the mean $P_{\rm Ly\alpha}$.  We use the predicted number of QSOs (denoted $N_{\rm QSO}$) expected in a $10\,000\,\rm deg^2$ survey after applying an r-band cut of $m_{\rm r} < 21.9$, as described in Section~\ref{sec:QSM_mock}. For each of the four redshift bins considered here, we then generate $1\,000$ bootstrap realizations of the mean $P_{\rm Ly\alpha}$ by randomly sampling $N_{\rm QSO}$ sightlines with replacement. The statistical errors are derived as the square root of the diagonal elements of the resulting covariance matrix.

Next, for systematics associated with spectral resolution, noise subtraction, and uncorrelated metals, we use the mock spectra to construct bootstrap estimates of the residual uncertainty associated with each correction. The corresponding quantities shown in Figure~\ref{fig:Errors} are the standard deviations of the corresponding correction ratios across $1\,000$ bootstrap realizations of $N_{\rm QSO}$ randomly sampled sightlines. For HCDs and continuum placement, we estimate the uncertainty from the fractional change in $P_{\rm Ly\alpha}$. The HCD contribution is modelled using the ratio between power spectra with incomplete HCD removal and the pure Ly$\alpha$ spectrum (solid curves in Figure~\ref{fig:HCD}). The continuum placement contribution is evaluated analogously (solid curves in Figure~\ref{fig:Cont}). The total systematic uncertainty is computed as the quadratic sum of all individual contributions. We note that the forecast uncertainty estimate assumes perfect knowledge of the spectral-resolution window and of the noise model. In real WEAVE-QSO data, uncertainties in both quantities will need to be tested carefully. We have tested the effect of including a 10 per cent uncertainty in $R$, finding that the resulting fractional systematic uncertainty is roughly 25--50 per cent near $k_{\rm max}$, much larger than the current bootstrap resolution term and also larger than the noise term near $k_{\rm max}$. However, this contribution remains below 1 per cent at $\log_{10}(k/{\rm s\,km^{-1}})\lesssim -2.0$ (and is $\sim 0.01$ per cent at $\log_{10}(k/{\rm s\,km^{-1}})\sim -3.0$) and therefore does not affect the large-scale modes that dominate the patchy reionization detection forecast presented in this paper.


Figure \ref{fig:Errors} shows the resulting estimate for the 1$\sigma$ relative uncertainties. The statistical and systematic contributions are indicated by dotted and dashed lines, respectively, while the solid lines represent their quadratic sum.  As before, the grey shaded region shows the maximum wavenumber for the $P_{\rm Ly\alpha}$ measurements that we consider in this work, $k_{\rm max} = 0.05\ \rm s\ km^{-1}$. The total systematic uncertainty at large $k$ reflects the dominance of the noise subtraction term on small scales, whereas the increase in the statistical error with redshift is due to the smaller number of QSOs in the higher redshift bins. At large scales ($k \lesssim 0.01\,\rm km\,s^{-1}$), the total uncertainties remain below 10 per cent for all redshift bins. The average uncertainty at $k < k_{\rm max}$ over all four redshift bins is 5.2 per cent. Over the range $-2.6 \lesssim \log_{10}(k/{\rm s\ km^{-1}}) \lesssim -1.7$, the relative uncertainties on $P_{\rm Ly\alpha}$ are dominated by the statistical contribution (i.e. where the dotted line is above the dashed line in Figure~\ref{fig:Errors}). Within this range, we forecast an average relative uncertainty of 3.1 per cent. These estimates are used to construct the covariance matrix for a WEAVE-QSO $P_{\rm Ly\alpha}$ measurement, which we now use as an input within our Markov Chain Monte Carlo (MCMC) parameter inference framework.


\section{The imprint of patchy reionization at z>4}
\label{sec:inference}

\subsection{Parameter inference framework}
\label{sec:inference_framework}

\begin{figure*}
    \centering
    \includegraphics[width=2\columnwidth]{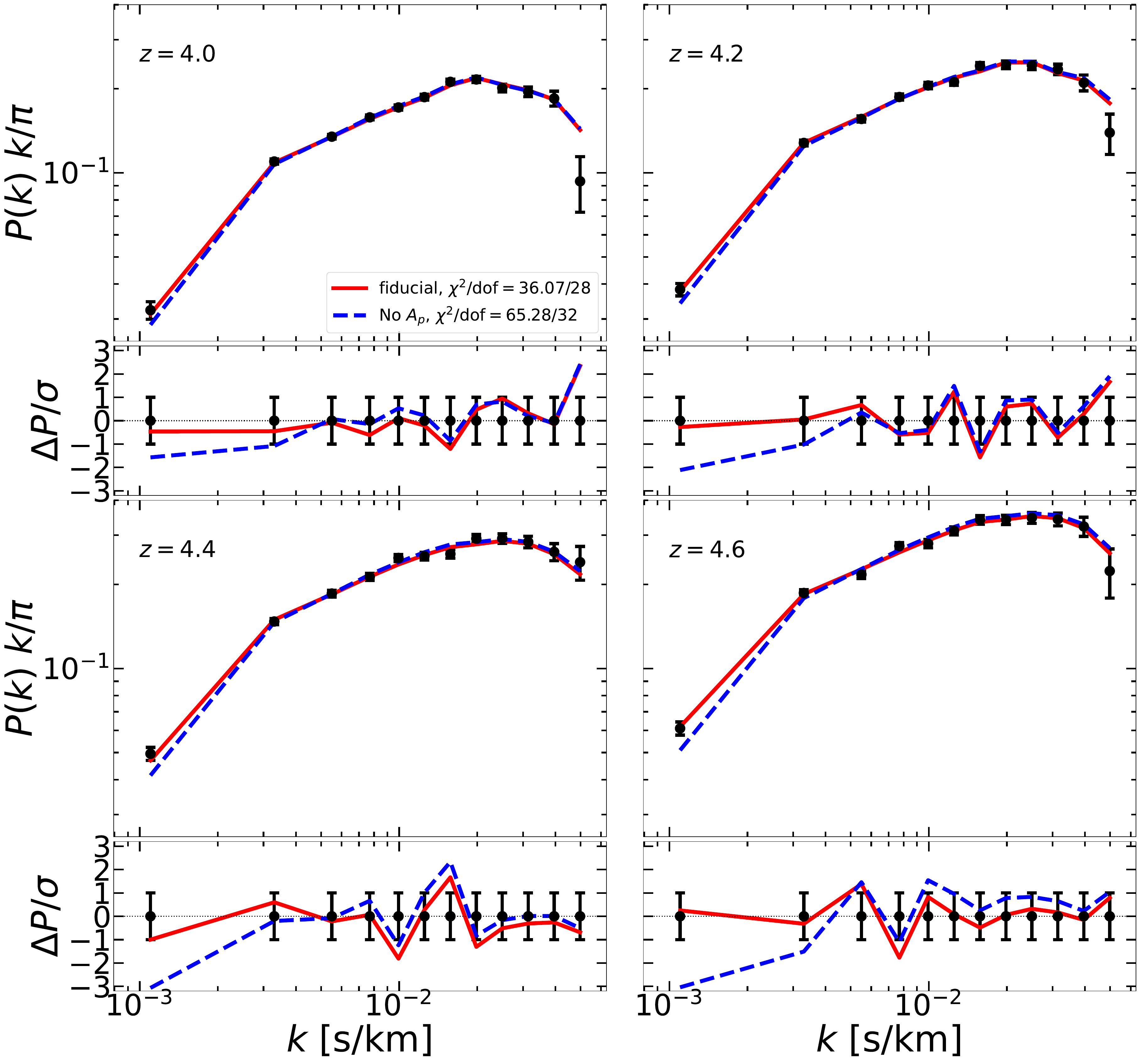}
        \vspace{-3mm}
    \caption{Comparison between the WEAVE-QSO mock $P_{\rm Ly\alpha}$ (black data points) and the best-fit models from the MCMC parameter inference. The redshift bin is indicated in the upper left of each panel.  The red solid curves show the fiducial fit, in which the patchy reionization parameter $A_{\rm p}$ is allowed to vary independently in each redshift bin.  The blue dashed curves show the no-$A_{\rm p}$ comparison fit, where the patchy correction is disabled within the inference framework. The lower panels show the residuals relative to the mock data uncertainties, $(P_{\rm fit}-P_{\rm data})/\sigma_{\rm data}$. For the combined fit across all four redshift bins, allowing $A_{\rm p}$ to vary improves the fit from $\chi^2/{\rm dof}=65.28/32$ to $36.07/28$, corresponding to $\Delta\chi^2=29.21$ for four additional parameters.  This corresponds to a two-sided Gaussian-equivalent significance of $4.5\sigma$ for a large scale power enhancement due to patchy reionization.}
    \label{fig:mcmc_p1d_comparison}
\end{figure*}


We now use our mock 1D \lya\ power spectrum covariance matrix to test whether a WEAVE-QSO measurement can recover the large-scale imprint of patchy reionization.  Our MCMC parameter inference framework follows the approach used in earlier work, where the measured $P_{\rm Ly\alpha}$ is compared against a grid of simulated power spectra while marginalizing over astrophysical parameters \citep[e.g.][]{Viel2013,Irsic2017b,Molaro2023}.  For each redshift bin, our Ly$\alpha$ forest model is described by the effective optical depth $\tau_{\rm eff}$, the temperature at mean density $T_0$, the slope of the temperature-density relation $\gamma$, and the cumulative thermal energy per unit mass $u_0$ \citep{Nasir2016}. These parameters control the mean transmission and thermal state of the IGM, and hence the amplitude and shape of the intrinsic $P_{\rm Ly\alpha}$.  We do not vary any cosmological parameters in this analysis \citep[see][for further discussion of this point]{Molaro2023}.

In addition, we include an extra parameter that describes the effect of patchy reionization on the \lya\ forest at $z>4$.  For this purpose we adopt the patchy reionization correction used by \citet{Molaro2023}, based on the tabulated data presented in table 2 of \citet{Molaro2022}.  These corrections were calculated using the Sherwood-Relics simulation suite \citep{Puchwein2023}, by taking the ratio of $P_{\rm Ly\alpha}$ obtained from hybrid radiative transfer simulations of patchy reionization to models using a spatially uniform UV background with \emph{the same volume averaged reionisation history} as the patchy model.  On the scales relevant for this work the correction enhances the large-scale power at $\log_{10}(k/{\rm s\,km^{-1}})\lesssim -1.5$, primarily due to spatial variations in the IGM ionization and thermal state that persist following the completion of reionization \citep{Cen2009,DAloisio2015,Molaro2022}.

Following \citet{Molaro2023}, we parameterize this patchy reionization correction using the dimensionless parameter, $A_{\rm p}$, where
\begin{equation}
    A_{\rm p} = \frac{z-3.45}{2.55},
    \label{eq:Ap_definition}
\end{equation}
and $z$ is the redshift at which the correction is linearly interpolated from table 2 of \citet{Molaro2022}.  We allow $A_{\rm p}$ to vary over the range $0\leq A_{\rm p}\leq 1$ by linearly extrapolating to a lower and upper redshift limit of $z = 3.45$ (i.e., for no correction) and $z = 6$, respectively. Note that $A_{\rm p}$ is used only as a numerical parameter controlling the amplitude of the patchy correction and should not be interpreted as a measurement of the redshift of reionization. In our MCMC analysis we allow one independent $A_{\rm p}$ parameter in each redshift bin.

We compare two classes of fit to the same mock data set, which we construct by applying our covariance matrix estimate to $P_{\rm Ly\alpha}$ from the Sherwood-Relics model used in this work. To facilitate a direct comparison to the earlier work of \citet{Molaro2023}, throughout we adopt the wavenumber binning introduced by \citet{Karacayli2022} and add a Gaussian noise realization to $P_{\rm Ly\alpha}$ drawn from the forecast covariance matrix.  The first fit includes $A_{\rm p}$ and therefore allows the large-scale patchy reionization correction to vary in each $P_{\rm Ly\alpha}$ redshift bin.  We add this to our simulated mocks such that $A_{\rm p}= 0.216,\,0.294,\,0.373,$ and $0.451$ in the redshift bins at $z=4.0,\,4.2,\,4.4,\,$ and $4.6$, respectively.  The second fit removes this degree of freedom by disabling the patchy reionization correction.  In both cases, we apply a broad top-hat prior on $\tau_{\rm eff}$ spanning $0.4$--$1.6$ times the reference effective optical depth, following \citet{Boera2019}.  The thermal parameters $T_0$, $\gamma$, and $u_0$ are assigned Gaussian priors centered on the reference thermal history of the Sherwood-Relics simulation used here, with a 10 per cent width in each redshift bin. We additionally impose a hard $T_0$--$u_0$ consistency prior, analogous to the simulation-grid prior used by \citet{Molaro2023}, to exclude thermal histories outside the modeled region. As both fits are applied to the same mock data  the difference in the goodness-of-fit provides a test of the significance that a WEAVE-QSO $P_{\rm Ly\alpha}$ measurement would favour an enhancement of large scale power from reionization.

Finally, we also include the impact of correlated Si\,\textsc{III} absorption following \citet{ma2025}. As shown in Figure~\ref{fig:MetalsCL}, the correlated Si\,\textsc{III} contribution has an oscillatory dependence on wavenumber. For the relatively sparse large-scale $k$-binning that we adopt here, however, this oscillatory structure is always weakly sampled.  Consequently we find negligible bias in the recovered $A_{\rm p}$ posteriors when the Si\,\textsc{III} absorption model is included in our inference framework.  We therefore do not consider this issue further.

\begin{figure*}
    \centering
    \includegraphics[width=2\columnwidth]{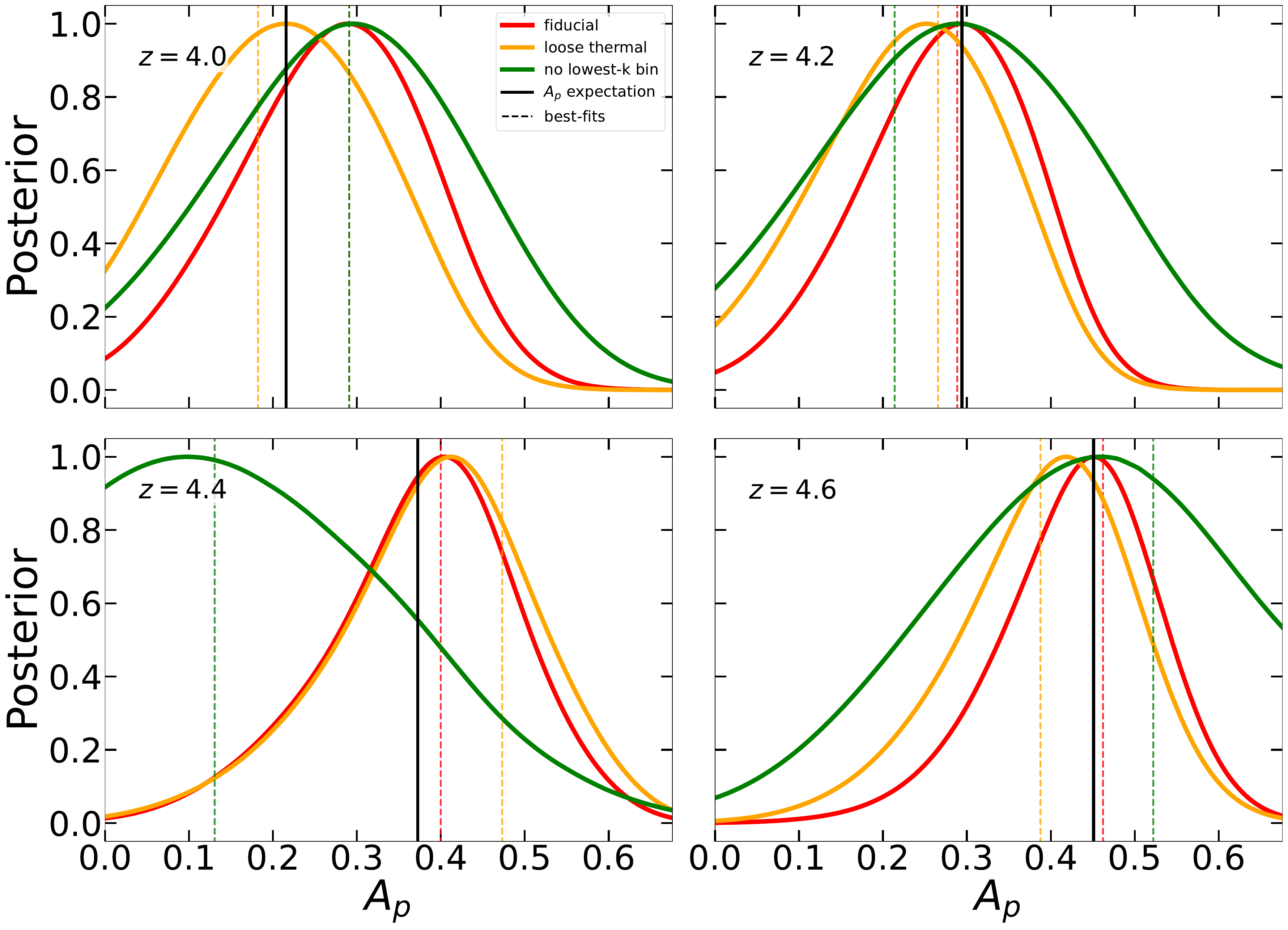}
        \vspace{-3mm}
    \caption{One dimensional posterior distributions for the patchy reionization parameter $A_{\rm p}$ in the four $P_{\rm Ly\alpha}$ redshift bins considered in this work.  The vertical black lines show the expected values used in the mocks (see Eq.~(\ref{eq:Ap_definition})).  The dashed vertical lines show the best-fit values for each run.  The fiducial analysis (red curves) recovers the amplitudes in all redshift bins.  Relaxing the thermal-history priors (yellow curves) produces slightly broader posteriors but does not break the recovery.  However, removing the lowest-$k$ bin from each redshift bin further broadens the constraints and leads to only an upper limit at $z=4.4$.  This confirms the importance of the largest-scale modes for detecting the enhanced large-scale power from patchy reionization.}
    \label{fig:ap_recovery}
\end{figure*}

\begin{table}
\centering
\renewcommand{\arraystretch}{1.25}
\begin{tabular}{llccc}
    \hline
    \hline
    Run & $z$ & Expected $A_p$ & Posterior peak & Best-fit \\
    \hline
    Fiducial & 4.0 & 0.216 & $0.289_{-0.121}^{+0.105}$ & 0.291 \\
             & 4.2 & 0.294 & $0.294_{-0.108}^{+0.094}$ & 0.289 \\
             & 4.4 & 0.373 & $0.404_{-0.121}^{+0.100}$ & 0.400 \\
             & 4.6 & 0.451 & $0.451_{-0.099}^{+0.084}$ & 0.462 \\
    \hline
    Loose thermal & 4.0 & 0.216 & $0.216_{-0.124}^{+0.116}$ & 0.182 \\
                  & 4.2 & 0.294 & $0.252_{-0.118}^{+0.104}$ & 0.266 \\
                  & 4.4 & 0.373 & $0.412_{-0.123}^{+0.111}$ & 0.473 \\
                  & 4.6 & 0.451 & $0.418_{-0.115}^{+0.093}$ & 0.388 \\
    \hline
    No lowest-$k$ bin & 4.0 & 0.216 & $0.295_{-0.151}^{+0.142}$ & 0.291 \\
                      & 4.2 & 0.294 & $0.289_{-0.158}^{+0.164}$ & 0.214 \\
                      & 4.4 & 0.373 & $< 0.387$ & 0.131 \\
                      & 4.6 & 0.451 & $0.460_{-0.223}^{+0.212}$ & 0.522 \\
    \hline
  \end{tabular}
  \caption{Recovered constraints on the patchy reionization amplitude parameter, $A_{\rm p}$, for the three cases displayed in Figure~\ref{fig:ap_recovery}. The expected values are the amplitudes used in the mock data vector. The posterior peak column reports the marginalized one-dimensional posterior peak with the 68 per cent credible interval or one-sided upper limit, while the final column reports the best-fit value.}
  \label{tab:ap_recovery}
\end{table}

\subsection{Recovery of the large scale power enhancement}
\label{sec:inference_p1d_comparison}

Figure~\ref{fig:mcmc_p1d_comparison} shows the mock $P_{\rm Ly\alpha}$ data in the four redshift bins considered in this work, together with the best-fit model predictions obtained from the MCMC analysis.  Note again the uncertainties are derived from the uncertainty estimate summarised in Section~\ref{sec:P1D_err}. This comparison is performed using the same mock data set and covariance matrix for all fits, such that any difference in the likelihood is driven only by the model assumptions. The red solid curves show the fiducial fit in which one independent $A_{\rm p}$ parameter is varied in each redshift bin.  The blue dashed curves show the best-fit excluding the $A_{\rm p}$ parameter while all other IGM parameters are allowed to vary as before (the posterior distributions for all parameters are provided in Appendix~\ref{app:posteriors}).

For the combined fit across all four redshift bins, the fiducial model provides an acceptable fit to the mock, with $\chi^2/{\rm dof} = 36.07/28 $. By contrast, disabling the patchy correction degrades the fit to $\chi^2/{\rm dof} = 65.28/32$. The improvement is $ \Delta \chi^2 = 29.21 $ for four additional patchy reionization amplitude parameters. Interpreting this likelihood ratio statistic using Wilks' theorem gives $p\simeq 7.1\times10^{-6}$, corresponding to a two-sided Gaussian-equivalent significance of $\simeq 4.5\sigma$ for the fiducial patchy reionization model. Thus, even after allowing the no-$A_p$ fit to adjust the thermal and ionization parameters the mock data strongly prefer the additional large-scale correction associated with patchy reionization.  This demonstrates the patchy reionization imprint inserted into the mock is not fully absorbed by variations in $\tau_{\rm eff}$, $T_0$, $\gamma$, or $u_0$, and remains significant even after a detailed accounting of WEAVE-QSO survey uncertainties.

The posterior distributions for $A_{\rm p}$ are displayed in Figure~\ref{fig:ap_recovery} and Table~\ref{tab:ap_recovery}. In addition to the fiducial fit (red curves) here we consider two further cases. First, we repeat the analysis with broader thermal priors to test whether the inferred patchy reionization amplitudes are driven by restrictive assumptions about the IGM thermal state (orange curves).  In this relaxed-prior run, we replace the fiducial Gaussian priors on $T_0$, $\gamma$, and $u_0$ with broad top-hat priors over the simulation grid ranges, while keeping the same $\tau_{\rm eff}$ prior and the same $T_0$--$u_0$ consistency cut. Specifically, we use $T_0 \in [0.5,1.5]\times10^4\,{\rm K}$ and $\gamma \in [1.0,1.7]$, with $u_0$ allowed to vary over the corresponding simulation-grid range in each redshift bin.  Second, we remove the lowest-$k$ bin at $k\sim 10^{-3}\rm\,s\,km^{-1}$ from each redshift bin to test the sensitivity of the $A_{\rm p}$ recovery to the largest scale mode included in the fit (green curves).

The fiducial analysis recovers the expected patchy amplitudes in all bins. In each case, the expected value lies within the 68 per cent credible interval of the posterior distribution.   Relaxing the thermal parameter priors leads to minor shifts and slightly broader posteriors, but the expected $A_{\rm p}$ values remain consistent with the inferred distributions. This indicates that the recovery is not simply imposed by restrictive priors on $T_0$, $\gamma$, or the integrated thermal history. However, removing the lowest-$k$ bin from each redshift bin weakens the constraints. This is expected because the patchy reionization correction is largest on this scale. The posteriors broaden by factors of approximately 1.3--2.4 relative to the fiducial case.  In the corresponding no-lowest-$k$ bin model comparison, the improvement from including $A_{\rm p}$ is reduced to $\Delta\chi^2=5.40$ for four additional parameters, corresponding to a two-sided Gaussian-equivalent preference of only $\simeq 1.2\sigma$.  In this case there is no longer a preference for large scale power enhancement from patchy reionization. Nevertheless, the expected $A_{\rm p}$ values remain well recovered with the exception of the $z=4.4$, where only an upper limit is obtained. This demonstrates that most of the constraining power on $A_{\rm p}$ comes from retaining modes with $k<0.0033\rm\,s\,km^{-1}$.  We also emphasise  that the forecast uncertainties in the lowest-$k$ bins are dominated by the systematic uncertainty associated with damping wings from high column density absorbers and to a lesser extent the continuum placement.  We have verified this remains the case even if lowering $N_{\rm QSO}$ in each redshift bin by a factor of $\sim 2$ from our fiducial values.  This implies that the WEAVE-QSO survey configuration of approximately $10\,000\rm\,deg^{2}$ with $m_{\rm r}<22$ should comfortably exceed the minimum requirement on QSO sample size.

In summary, our results indicate a WEAVE-QSO $P_{\rm Ly\alpha}$ measurement at $z\geq 4$ with the forecast covariance adopted here can recover the redshift dependent patchy reionization amplitude at a two-sided Gaussian-equivalent significance of $4.5\sigma$. This forecast significance is higher than the $2.7\sigma$ preference for additional large scale power reported by \citet{Molaro2023} from an analysis of the observed 1D power spectrum measurements presented by \citet{Karacayli2022}. While our analysis is based on mock WEAVE-QSO data with a forecast covariance, this comparison suggests that WEAVE-QSO should have sufficient statistical precision to turn the tentative evidence seen in current data into a significant detection of the large scale patchy reionization signature.  However, the largest scale mode provides substantial constraining power: when the lowest-$k$ bin is removed, the model-comparison preference for including $A_{\rm p}$ is weakened to $\simeq1.2\sigma$ and is no longer significant.


\section{Summary and conclusions}

In this work, we have developed a framework to forecast the sensitivity of a massive spectroscopic survey to the large-scale imprint of patchy hydrogen reionization in the one-dimensional Ly$\alpha$ forest flux power spectrum, $P_{\rm Ly\alpha}$.  We have opted to focus on a WEAVE-QSO survey configuration \citep{Pieri2016,Jin2024}, targeting four high-redshift bins of width $\Delta z=0.2$ centred at $z=4.0,4.2,4.4,$ and $4.6$, where relic temperature fluctuations from inhomogeneous reionization are expected to remain observable in $P_{\rm Ly\alpha}$ \citep{Wu2019,Molaro2022,Molaro2023}.  However, we anticipate that other current and forthcoming massive spectroscopic surveys that aim to constrain large-scale power in the \lya forest at wavenumbers $k\sim 10^{-3}\rm\,s\,km^{-1}$ at $z>4$ should also be capable of this novel measurement.

Our mock spectra are based on the Sherwood-Relics simulation suite \citep{Puchwein2023} and are processed with the WEAVEify pipeline to include the main observational ingredients relevant for WEAVE-QSO spectra. We construct a realistic high-redshift QSO population using luminosity-function-based Monte Carlo sampling \citep{Palanque-Delabrouille2016} and adopt an $r$-band magnitude cut of $m_{\rm r}<21.9$. For a $10\,000\,{\rm deg}^2$ survey area, this selection yields expected QSO numbers of $2942$, $1822$, $1385$, and $1155$ in the four redshift bins considered here.

We use our pipeline to quantify the overall uncertainty for a WEAVE-QSO $P_{\rm Ly\alpha}$ measurement. The covariance matrix estimate includes statistical errors and systematic contributions from spectral resolution, noise subtraction, continuum misplacement, uncorrelated metal contamination, and the incomplete identification of high column density absorbers with $N_{\rm HI}>10^{19}\rm\,cm^{-2}$.  Within our adopted maximum wavenumber cut, $k<0.05\,{\rm s\,km^{-1}}$, the estimated relative uncertainties remain below 10 per cent on the largest scales in all redshift bins. Averaged over all bins and scales below this cut, the total uncertainty is 5.2 per cent. Over the range $-2.6 \lesssim \log_{10}(k/{\rm s\,km^{-1}}) \lesssim -1.7$, the uncertainties are dominated by the statistical contribution and the average uncertainty is 3.1 per cent.

Using this forecast covariance, we construct a mock $P_{\rm Ly\alpha}$ data vector containing the fiducial patchy reionization signal and test whether it can be distinguished from an IGM that has been heated and ionised by a spatially uniform UV background.  This inference test follows the patchy reionization parameterization and wavenumber binning used in earlier work \citep{Molaro2023}. Allowing one independent patchy reionization amplitude parameter, $A_{\rm p}$, in each redshift bin improves the fit from $\chi^2/{\rm dof}=65.28/32$ to $\chi^2/{\rm dof}=36.07/28$, corresponding to $\Delta\chi^2=29.21$ for four additional patchy amplitude parameters and a $\simeq4.5\sigma$ preference for additional power at large scales from patchy reionization.  This result remains robust to changes in the prior adopted for IGM thermal parameters and is driven almost entirely by the lowest-$k$ bin from each redshift bin.

Our results indicate that a WEAVE-QSO measurement of the high-redshift Ly$\alpha$ forest power spectrum should be capable of detecting and characterizing the large-scale relic imprint of patchy reionization, provided that the observational covariance is controlled at the level forecast here. More broadly, the framework developed in this work provides a route for assessing the sensitivity of current and future massive spectroscopic surveys to the timing of reionization through large-scale $P_{\rm Ly\alpha}$ measurements.

\section*{Acknowledgements}

KM gratefully acknowledges the financial support provided by the China Scholarship Council programme (Project ID: CSC NO. 202309110040).  JSB is supported by STFC consolidated grant ST/X000982/1.  The modelling in this work was performed using the Joliot Curie supercomputer at the Tr{\'e}s Grand Centre de Calcul (TGCC) and the DiRAC Data Intensive service (CSD3) at the University of Cambridge, managed by the University of Cambridge University Information Services on behalf of the STFC DiRAC HPC facility (www.dirac.ac.uk).  We acknowledge the Partnership for Advanced Computing in Europe (PRACE) for awarding us time on Joliot Curie in the 16th call. The DiRAC component of CSD3 at Cambridge was funded by BEIS, UKRI and STFC capital funding and STFC operations grants.  DiRAC is part of the National e-Infrastructure.   We thank Volker Springel for making P-Gadget-3 available and Philip Parry for technical support.


\section*{Data Availability}
All data and analysis code used in this work are available from the
first author on reasonable request.  Further guidance on accessing the publicly available Sherwood-Relics simulation data may also be found at
\url{https://www.nottingham.ac.uk/astronomy/sherwood-relics/}.



\bibliographystyle{mnras}
\bibliography{bibliography}

@ARTICLE{Aguirre2004,
       author = {{Aguirre}, Anthony and {Schaye}, Joop and {Kim}, Tae-Sun and {Theuns}, Tom and {Rauch}, Michael and {Sargent}, Wallace L.~W.},
        title = "{Metallicity of the Intergalactic Medium Using Pixel Statistics. III. Silicon}",
      journal = {\apj},
         year = 2004,
        month = feb,
       volume = {602},
       number = {1},
        pages = {38-50},
          doi = {10.1086/380961},
archivePrefix = {arXiv},
       eprint = {astro-ph/0310664},
 primaryClass = {astro-ph},
       adsurl = {https://ui.adsabs.harvard.edu/abs/2004ApJ...602...38A}
}

@ARTICLE{Becker2013,
       author = {{Becker}, George D. and {Hewett}, Paul C. and {Worseck}, G{\'a}bor and {Prochaska}, J. Xavier},
        title = "{A refined measurement of the mean transmitted flux in the Ly{\ensuremath{\alpha}} forest over 2 < z < 5 using composite quasar spectra}",
      journal = {\mnras},
         year = 2013,
        month = apr,
       volume = {430},
       number = {3},
        pages = {2067-2081},
          doi = {10.1093/mnras/stt031},
archivePrefix = {arXiv},
       eprint = {1208.2584},
 primaryClass = {astro-ph.CO},
       adsurl = {https://ui.adsabs.harvard.edu/abs/2013MNRAS.430.2067B}
}

@ARTICLE{Boera2019,
       author = {{Boera}, Elisa and {Becker}, George D. and {Bolton}, James S. and {Nasir}, Fahad},
        title = "{Revealing Reionization with the Thermal History of the Intergalactic Medium: New Constraints from the Ly{\ensuremath{\alpha}} Flux Power Spectrum}",
      journal = {\apj},
         year = 2019,
        month = feb,
       volume = {872},
       number = {1},
          eid = {101},
        pages = {101},
          doi = {10.3847/1538-4357/aafee4},
archivePrefix = {arXiv},
       eprint = {1809.06980},
 primaryClass = {astro-ph.CO},
       adsurl = {https://ui.adsabs.harvard.edu/abs/2019ApJ...872..101B}
}

@ARTICLE{Bolton2005,
       author = {{Bolton}, James S. and {Haehnelt}, Martin G. and {Viel}, Matteo and {Springel}, Volker},
        title = "{The Lyman {\ensuremath{\alpha}} forest opacity and the metagalactic hydrogen ionization rate at z\raisebox{-0.5ex}\textasciitilde 2-4}",
      journal = {\mnras},
         year = 2005,
        month = mar,
       volume = {357},
       number = {4},
        pages = {1178-1188},
          doi = {10.1111/j.1365-2966.2005.08704.x},
archivePrefix = {arXiv},
       eprint = {astro-ph/0411072},
 primaryClass = {astro-ph},
       adsurl = {https://ui.adsabs.harvard.edu/abs/2005MNRAS.357.1178B}
}

@ARTICLE{Bolton2017,
       author = {{Bolton}, James S. and {Puchwein}, Ewald and {Sijacki}, Debora and {Haehnelt}, Martin G. and {Kim}, Tae-Sun and {Meiksin}, Avery and {Regan}, John A. and {Viel}, Matteo},
        title = "{The Sherwood simulation suite: overview and data comparisons with the Lyman {\ensuremath{\alpha}} forest at redshifts 2 {\ensuremath{\leq}} z {\ensuremath{\leq}} 5}",
      journal = {\mnras},
         year = 2017,
        month = jan,
       volume = {464},
       number = {1},
        pages = {897-914},
          doi = {10.1093/mnras/stw2397},
archivePrefix = {arXiv},
       eprint = {1605.03462},
 primaryClass = {astro-ph.CO},
       adsurl = {https://ui.adsabs.harvard.edu/abs/2017MNRAS.464..897B}
}

@ARTICLE{BoltonBecker2009,
       author = {{Bolton}, James S. and {Becker}, George D.},
        title = "{Resolving the high redshift Ly{\ensuremath{\alpha}} forest in smoothed particle hydrodynamics simulations}",
      journal = {\mnras},
         year = 2009,
        month = sep,
       volume = {398},
       number = {1},
        pages = {L26-L30},
          doi = {10.1111/j.1745-3933.2009.00700.x},
archivePrefix = {arXiv},
       eprint = {0906.2861},
 primaryClass = {astro-ph.CO},
       adsurl = {https://ui.adsabs.harvard.edu/abs/2009MNRAS.398L..26B}
}

@ARTICLE{Cain2024,
       author = {{Cain}, Christopher and {Scannapieco}, Evan and {McQuinn}, Matthew and {D'Aloisio}, Anson and {Trac}, Hy},
        title = "{The hydrodynamic response of small-scale structure to reionization drives large IGM temperature fluctuations that persist to z = 4}",
      journal = {\mnras},
         year = 2024,
        month = sep,
       volume = {533},
       number = {1},
        pages = {L100-L106},
          doi = {10.1093/mnrasl/slae067},
archivePrefix = {arXiv},
       eprint = {2405.02397},
 primaryClass = {astro-ph.CO},
       adsurl = {https://ui.adsabs.harvard.edu/abs/2024MNRAS.533L.100C}
}

@ARTICLE{Cen2009,
       author = {{Cen}, Renyue and {McDonald}, Patrick and {Trac}, Hy and {Loeb}, Abraham},
        title = "{Probing the Epoch of Reionization with the Ly{\ensuremath{\alpha}} Forest at z \raisebox{-0.5ex}\textasciitilde 4-5}",
      journal = {\apjl},
         year = 2009,
        month = nov,
       volume = {706},
       number = {1},
        pages = {L164-L167},
          doi = {10.1088/0004-637X/706/1/L164},
archivePrefix = {arXiv},
       eprint = {0907.0735},
 primaryClass = {astro-ph.CO},
       adsurl = {https://ui.adsabs.harvard.edu/abs/2009ApJ...706L.164C}
}

@ARTICLE{ChavesMontero2026,
       author = {{Chaves-Montero}, J. and {Font-Ribera}, A. and {McDonald}, P. and {Armengaud}, E. and {Chebat}, D. and {Garcia-Quintero}, C. and {Kara{\c{c}}ayl{\i}}, N.~G. and {Ravoux}, C. and {Satyavolu}, S. and {Sch{\"o}neberg}, N. and {Walther}, M. and {Aguilar}, J. and {Ahlen}, S. and {Bailey}, S. and {Bianchi}, D. and {Brooks}, D. and {Claybaugh}, T. and {Cuceu}, A. and {de la Macorra}, A. and {Doel}, P. and {Ferraro}, S. and {Forero-Romero}, J.~E. and {Gazta{\~n}aga}, E. and {Gontcho}, S. Gontcho A. and {Gonzalez-Morales}, A.~X. and {Gutierrez}, G. and {Guy}, J. and {Hahn}, C. and {Herrera-Alcantar}, H.~K. and {Honscheid}, K. and {Ishak}, M. and {Joyce}, R. and {Juneau}, S. and {Kirkby}, D. and {Kremin}, A. and {Lahav}, O. and {Lamman}, C. and {Landriau}, M. and {Le Goff}, J.~M. and {Le Guillou}, L. and {Leauthaud}, A. and {Levi}, M.~E. and {Manera}, M. and {Martini}, P. and {Meisner}, A. and {Miquel}, R. and {Moustakas}, J. and {Nadathur}, S. and {Niz}, G. and {Palanque-Delabrouille}, N. and {Percival}, W.~J. and {Prada}, F. and {P{\'e}rez-R{\`a}fols}, I. and {Rossi}, G. and {Sanchez}, E. and {Schlegel}, D. and {Schubnell}, M. and {Seo}, H. and {Silber}, J. and {Sprayberry}, D. and {Tan}, T. and {Tarl{\'e}}, G. and {Weaver}, B.~A. and {Y{\`e}che}, C. and {Zhou}, R. and {Zou}, H.},
        title = "{Cosmological analysis of the DESI DR1 Ly{\ensuremath{\alpha}} 1D power spectrum}",
      journal = {\jcap},
         year = 2026,
        month = jun,
       volume = {2026},
       number = {6},
          eid = {040},
        pages = {040},
          doi = {10.1088/1475-7516/2026/06/040},
archivePrefix = {arXiv},
       eprint = {2601.21432},
 primaryClass = {astro-ph.CO},
       adsurl = {https://ui.adsabs.harvard.edu/abs/2026JCAP...06..040C}
}

@ARTICLE{Chabanier2019,
       author = {{Chabanier}, Sol{\`e}ne and {Palanque-Delabrouille}, Nathalie and {Y{\`e}che}, Christophe and {Le Goff}, Jean-Marc and {Armengaud}, Eric and {Bautista}, Julian and {Blomqvist}, Michael and {Busca}, Nicolas and {Dawson}, Kyle and {Etourneau}, Thomas and {Font-Ribera}, Andreu and {Lee}, Youngbae and {du Mas des Bourboux}, H{\'e}lion and {Pieri}, Matthew and {Rich}, James and {Rossi}, Graziano and {Schneider}, Donald and {Slosar}, An{\v{z}}e},
        title = "{The one-dimensional power spectrum from the SDSS DR14 Ly{\ensuremath{\alpha}} forests}",
      journal = {\jcap},
         year = 2019,
        month = jul,
       volume = {2019},
       number = {7},
          eid = {017},
        pages = {017},
          doi = {10.1088/1475-7516/2019/07/017},
archivePrefix = {arXiv},
       eprint = {1812.03554},
 primaryClass = {astro-ph.CO},
       adsurl = {https://ui.adsabs.harvard.edu/abs/2019JCAP...07..017C}
}

@ARTICLE{Chatzikos2023,
       author = {{Chatzikos}, M. and {Bianchi}, S. and {Camilloni}, F. and {Chakraborty}, P. and {Gunasekera}, C.~M. and {Guzm{\'a}n}, F. and {Milby}, J.~S. and {Sarkar}, A. and {Shaw}, G. and {van Hoof}, P.~A.~M. and {Ferland}, G.~J.},
        title = "{The 2023 Release of Cloudy}",
      journal = {\rmxaa},
         year = 2023,
        month = oct,
       volume = {59},
        pages = {327-343},
          doi = {10.22201/ia.01851101p.2023.59.02.12},
archivePrefix = {arXiv},
       eprint = {2308.06396},
 primaryClass = {astro-ph.GA},
       adsurl = {https://ui.adsabs.harvard.edu/abs/2023RMxAA..59..327C}
}

@ARTICLE{Croft2002,
       author = {{Croft}, Rupert A.~C. and {Weinberg}, David H. and {Bolte}, Mike and {Burles}, Scott and {Hernquist}, Lars and {Katz}, Neal and {Kirkman}, David and {Tytler}, David},
        title = "{Toward a Precise Measurement of Matter Clustering: Ly{\ensuremath{\alpha}} Forest Data at Redshifts 2-4}",
      journal = {\apj},
         year = 2002,
        month = dec,
       volume = {581},
       number = {1},
        pages = {20-52},
          doi = {10.1086/344099},
archivePrefix = {arXiv},
       eprint = {astro-ph/0012324},
 primaryClass = {astro-ph},
       adsurl = {https://ui.adsabs.harvard.edu/abs/2002ApJ...581...20C}
}

@ARTICLE{Carswell1982,
       author = {{Carswell}, R.~F. and {Whelan}, J.~A.~J. and {Smith}, M.~G. and {Boksenberg}, A. and {Tytler}, D.},
        title = "{Observations of the spectra of Q0122-380 and Q1101-264.}",
      journal = {\mnras},
         year = 1982,
        month = jan,
       volume = {198},
        pages = {91-110},
          doi = {10.1093/mnras/198.1.91},
       adsurl = {https://ui.adsabs.harvard.edu/abs/1982MNRAS.198...91C}
}

@ARTICLE{davies2018,
       author = {{Davies}, Frederick B. and {Hennawi}, Joseph F. and {Ba{\~n}ados}, Eduardo and {Simcoe}, Robert A. and {Decarli}, Roberto and {Fan}, Xiaohui and {Farina}, Emanuele P. and {Mazzucchelli}, Chiara and {Rix}, Hans-Walter and {Venemans}, Bram P. and {Walter}, Fabian and {Wang}, Feige and {Yang}, Jinyi},
        title = "{Predicting Quasar Continua Near Lyman-$\alpha$ with Principal Component Analysis}",
      journal = {arXiv e-prints},
         year = 2018,
        month = jan,
          eid = {arXiv:1801.07679},
        pages = {arXiv:1801.07679},
          doi = {10.48550/arXiv.1801.07679},
archivePrefix = {arXiv},
       eprint = {1801.07679},
 primaryClass = {astro-ph.GA},
       adsurl = {https://ui.adsabs.harvard.edu/abs/2018arXiv180107679D}
}

@ARTICLE{DeJong2019,
       author = {{de Jong}, R.~S. and {Agertz}, O. and {Berbel}, A.~A. and {Aird}, J. and {Alexander}, D.~A. and {Amarsi}, A. and {Anders}, F. and {Andrae}, R. and {Ansarinejad}, B. and {Ansorge}, W. and {Antilogus}, P. and {Anwand-Heerwart}, H. and {Arentsen}, A. and {Arnadottir}, A. and {Asplund}, M. and {Auger}, M. and {Azais}, N. and {Baade}, D. and {Baker}, G. and {Baker}, S. and {Balbinot}, E. and {Baldry}, I.~K. and {Banerji}, M. and {Barden}, S. and {Barklem}, P. and {Barth{\'e}l{\'e}my-Mazot}, E. and {Battistini}, C. and {Bauer}, S. and {Bell}, C.~P.~M. and {Bellido-Tirado}, O. and {Bellstedt}, S. and {Belokurov}, V. and {Bensby}, T. and {Bergemann}, M. and {Bestenlehner}, J.~M. and {Bielby}, R. and {Bilicki}, M. and {Blake}, C. and {Bland-Hawthorn}, J. and {Boeche}, C. and {Boland}, W. and {Boller}, T. and {Bongard}, S. and {Bongiorno}, A. and {Bonifacio}, P. and {Boudon}, D. and {Brooks}, D. and {Brown}, M.~J.~I. and {Brown}, R. and {Br{\"u}ggen}, M. and {Brynnel}, J. and {Brzeski}, J. and {Buchert}, T. and {Buschkamp}, P. and {Caffau}, E. and {Caillier}, P. and {Carrick}, J. and {Casagrande}, L. and {Case}, S. and {Casey}, A. and {Cesarini}, I. and {Cescutti}, G. and {Chapuis}, D. and {Chiappini}, C. and {Childress}, M. and {Christlieb}, N. and {Church}, R. and {Cioni}, M. -R.~L. and {Cluver}, M. and {Colless}, M. and {Collett}, T. and {Comparat}, J. and {Cooper}, A. and {Couch}, W. and {Courbin}, F. and {Croom}, S. and {Croton}, D. and {Daguis{\'e}}, E. and {Dalton}, G. and {Davies}, L.~J.~M. and {Davis}, T. and {de Laverny}, P. and {Deason}, A. and {Dionies}, F. and {Disseau}, K. and {Doel}, P. and {D{\"o}scher}, D. and {Driver}, S.~P. and {Dwelly}, T. and {Eckert}, D. and {Edge}, A. and {Edvardsson}, B. and {Youssoufi}, D.~E. and {Elhaddad}, A. and {Enke}, H. and {Erfanianfar}, G. and {Farrell}, T. and {Fechner}, T. and {Feiz}, C. and {Feltzing}, S. and {Ferreras}, I. and {Feuerstein}, D. and {Feuillet}, D. and {Finoguenov}, A. and {Ford}, D. and {Fotopoulou}, S. and {Fouesneau}, M. and {Frenk}, C. and {Frey}, S. and {Gaessler}, W. and {Geier}, S. and {Gentile Fusillo}, N. and {Gerhard}, O. and {Giannantonio}, T. and {Giannone}, D. and {Gibson}, B. and {Gillingham}, P. and {Gonz{\'a}lez-Fern{\'a}ndez}, C. and {Gonzalez-Solares}, E. and {Gottloeber}, S. and {Gould}, A. and {Grebel}, E.~K. and {Gueguen}, A. and {Guiglion}, G. and {Haehnelt}, M. and {Hahn}, T. and {Hansen}, C.~J. and {Hartman}, H. and {Hauptner}, K. and {Hawkins}, K. and {Haynes}, D. and {Haynes}, R. and {Heiter}, U. and {Helmi}, A. and {Aguayo}, C.~H. and {Hewett}, P. and {Hinton}, S. and {Hobbs}, D. and {Hoenig}, S. and {Hofman}, D. and {Hook}, I. and {Hopgood}, J. and {Hopkins}, A. and {Hourihane}, A. and {Howes}, L. and {Howlett}, C. and {Huet}, T. and {Irwin}, M. and {Iwert}, O. and {Jablonka}, P. and {Jahn}, T. and {Jahnke}, K. and {Jarno}, A. and {Jin}, S. and {Jofre}, P. and {Johl}, D. and {Jones}, D. and {J{\"o}nsson}, H. and {Jordan}, C. and {Karovicova}, I. and {Khalatyan}, A. and {Kelz}, A. and {Kennicutt}, R. and {King}, D. and {Kitaura}, F. and {Klar}, J. and {Klauser}, U. and {Kneib}, J. -P. and {Koch}, A. and {Koposov}, S. and {Kordopatis}, G. and {Korn}, A. and {Kosmalski}, J. and {Kotak}, R. and {Kovalev}, M. and {Kreckel}, K. and {Kripak}, Y. and {Krumpe}, M. and {Kuijken}, K. and {Kunder}, A. and {Kushniruk}, I. and {Lam}, M.~I. and {Lamer}, G. and {Laurent}, F. and {Lawrence}, J. and {Lehmitz}, M. and {Lemasle}, B. and {Lewis}, J. and {Li}, B. and {Lidman}, C. and {Lind}, K. and {Liske}, J. and {Lizon}, J. -L. and {Loveday}, J. and {Ludwig}, H. -G. and {McDermid}, R.~M. and {Maguire}, K. and {Mainieri}, V. and {Mali}, S. and {Mandel}, H.},
        title = "{4MOST: Project overview and information for the First Call for Proposals}",
      journal = {The Messenger},
         year = 2019,
        month = mar,
       volume = {175},
        pages = {3-11},
          doi = {10.18727/0722-6691/5117},
archivePrefix = {arXiv},
       eprint = {1903.02464},
 primaryClass = {astro-ph.IM},
       adsurl = {https://ui.adsabs.harvard.edu/abs/2019Msngr.175....3D}
}

@ARTICLE{DESI2016,
       author = {{DESI Collaboration} and {Aghamousa}, Amir and {Aguilar}, Jessica and {Ahlen}, Steve and {Alam}, Shadab and {Allen}, Lori E. and {Allende Prieto}, Carlos and {Annis}, James and {Bailey}, Stephen and {Balland}, Christophe and {Ballester}, Otger and {Baltay}, Charles and {Beaufore}, Lucas and {Bebek}, Chris and {Beers}, Timothy C. and {Bell}, Eric F. and {Bernal}, Jos{\'e} Luis and {Besuner}, Robert and {Beutler}, Florian and {Blake}, Chris and {Bleuler}, Hannes and {Blomqvist}, Michael and {Blum}, Robert and {Bolton}, Adam S. and {Briceno}, Cesar and {Brooks}, David and {Brownstein}, Joel R. and {Buckley-Geer}, Elizabeth and {Burden}, Angela and {Burtin}, Etienne and {Busca}, Nicolas G. and {Cahn}, Robert N. and {Cai}, Yan-Chuan and {Cardiel-Sas}, Laia and {Carlberg}, Raymond G. and {Carton}, Pierre-Henri and {Casas}, Ricard and {Castander}, Francisco J. and {Cervantes-Cota}, Jorge L. and {Claybaugh}, Todd M. and {Close}, Madeline and {Coker}, Carl T. and {Cole}, Shaun and {Comparat}, Johan and {Cooper}, Andrew P. and {Cousinou}, M. -C. and {Crocce}, Martin and {Cuby}, Jean-Gabriel and {Cunningham}, Daniel P. and {Davis}, Tamara M. and {Dawson}, Kyle S. and {de la Macorra}, Axel and {De Vicente}, Juan and {Delubac}, Timoth{\'e}e and {Derwent}, Mark and {Dey}, Arjun and {Dhungana}, Govinda and {Ding}, Zhejie and {Doel}, Peter and {Duan}, Yutong T. and {Ealet}, Anne and {Edelstein}, Jerry and {Eftekharzadeh}, Sarah and {Eisenstein}, Daniel J. and {Elliott}, Ann and {Escoffier}, St{\'e}phanie and {Evatt}, Matthew and {Fagrelius}, Parker and {Fan}, Xiaohui and {Fanning}, Kevin and {Farahi}, Arya and {Farihi}, Jay and {Favole}, Ginevra and {Feng}, Yu and {Fernandez}, Enrique and {Findlay}, Joseph R. and {Finkbeiner}, Douglas P. and {Fitzpatrick}, Michael J. and {Flaugher}, Brenna and {Flender}, Samuel and {Font-Ribera}, Andreu and {Forero-Romero}, Jaime E. and {Fosalba}, Pablo and {Frenk}, Carlos S. and {Fumagalli}, Michele and {Gaensicke}, Boris T. and {Gallo}, Giuseppe and {Garcia-Bellido}, Juan and {Gaztanaga}, Enrique and {Pietro Gentile Fusillo}, Nicola and {Gerard}, Terry and {Gershkovich}, Irena and {Giannantonio}, Tommaso and {Gillet}, Denis and {Gonzalez-de-Rivera}, Guillermo and {Gonzalez-Perez}, Violeta and {Gott}, Shelby and {Graur}, Or and {Gutierrez}, Gaston and {Guy}, Julien and {Habib}, Salman and {Heetderks}, Henry and {Heetderks}, Ian and {Heitmann}, Katrin and {Hellwing}, Wojciech A. and {Herrera}, David A. and {Ho}, Shirley and {Holland}, Stephen and {Honscheid}, Klaus and {Huff}, Eric and {Hutchinson}, Timothy A. and {Huterer}, Dragan and {Hwang}, Ho Seong and {Illa Laguna}, Joseph Maria and {Ishikawa}, Yuzo and {Jacobs}, Dianna and {Jeffrey}, Niall and {Jelinsky}, Patrick and {Jennings}, Elise and {Jiang}, Linhua and {Jimenez}, Jorge and {Johnson}, Jennifer and {Joyce}, Richard and {Jullo}, Eric and {Juneau}, St{\'e}phanie and {Kama}, Sami and {Karcher}, Armin and {Karkar}, Sonia and {Kehoe}, Robert and {Kennamer}, Noble and {Kent}, Stephen and {Kilbinger}, Martin and {Kim}, Alex G. and {Kirkby}, David and {Kisner}, Theodore and {Kitanidis}, Ellie and {Kneib}, Jean-Paul and {Koposov}, Sergey and {Kovacs}, Eve and {Koyama}, Kazuya and {Kremin}, Anthony and {Kron}, Richard and {Kronig}, Luzius and {Kueter-Young}, Andrea and {Lacey}, Cedric G. and {Lafever}, Robin and {Lahav}, Ofer and {Lambert}, Andrew and {Lampton}, Michael and {Landriau}, Martin and {Lang}, Dustin and {Lauer}, Tod R. and {Le Goff}, Jean-Marc and {Le Guillou}, Laurent and {Le Van Suu}, Auguste and {Lee}, Jae Hyeon and {Lee}, Su-Jeong and {Leitner}, Daniela and {Lesser}, Michael and {Levi}, Michael E. and {L'Huillier}, Benjamin and {Li}, Baojiu and {Liang}, Ming and {Lin}, Huan and {Linder}, Eric and {Loebman}, Sarah R. and {Luki{\'c}}, Zarija and {Ma}, Jun and {MacCrann}, Niall and {Magneville}, Christophe and {Makarem}, Laleh and {Manera}, Marc and {Manser}, Christopher J. and {Marshall}, Robert and {Martini}, Paul and {Massey}, Richard and {Matheson}, Thomas and {McCauley}, Jeremy and {McDonald}, Patrick and {McGreer}, Ian D. and {Meisner}, Aaron and {Metcalfe}, Nigel and {Miller}, Timothy N. and {Miquel}, Ramon and {Moustakas}, John and {Myers}, Adam and {Naik}, Milind and {Newman}, Jeffrey A. and {Nichol}, Robert C. and {Nicola}, Andrina and {Nicolati da Costa}, Luiz and {Nie}, Jundan and {Niz}, Gustavo and {Norberg}, Peder and {Nord}, Brian and {Norman}, Dara and {Nugent}, Peter and {O'Brien}, Thomas and {Oh}, Minji and {Olsen}, Knut A.~G.},
        title = "{The DESI Experiment Part I: Science,Targeting, and Survey Design}",
      journal = {arXiv e-prints},
         year = 2016,
        month = oct,
          eid = {arXiv:1611.00036},
        pages = {arXiv:1611.00036},
          doi = {10.48550/arXiv.1611.00036},
archivePrefix = {arXiv},
       eprint = {1611.00036},
 primaryClass = {astro-ph.IM},
       adsurl = {https://ui.adsabs.harvard.edu/abs/2016arXiv161100036D}
}

@ARTICLE{DOdorico2013,
       author = {{D'Odorico}, V. and {Cupani}, G. and {Cristiani}, S. and {Maiolino}, R. and {Molaro}, P. and {Nonino}, M. and {Centuri{\'o}n}, M. and {Cimatti}, A. and {di Serego Alighieri}, S. and {Fiore}, F. and {Fontana}, A. and {Gallerani}, S. and {Giallongo}, E. and {Mannucci}, F. and {Marconi}, A. and {Pentericci}, L. and {Viel}, M. and {Vladilo}, G.},
        title = "{Metals in the IGM approaching the re-ionization epoch: results from X-shooter at the VLT}",
      journal = {\mnras},
         year = 2013,
        month = oct,
       volume = {435},
       number = {2},
        pages = {1198-1232},
          doi = {10.1093/mnras/stt1365},
archivePrefix = {arXiv},
       eprint = {1306.4604},
 primaryClass = {astro-ph.CO},
       adsurl = {https://ui.adsabs.harvard.edu/abs/2013MNRAS.435.1198D}
}

@ARTICLE{DOdorico2022,
       author = {{D'Odorico}, V. and {Finlator}, K. and {Cristiani}, S. and {Cupani}, G. and {Perrotta}, S. and {Calura}, F. and {C{\`e}nturion}, M. and {Becker}, G. and {Berg}, T.~A.~M. and {Lopez}, S. and {Ellison}, S. and {Pomante}, E.},
        title = "{The evolution of the Si IV content in the Universe from the epoch of reionization to cosmic noon}",
      journal = {\mnras},
         year = 2022,
        month = may,
       volume = {512},
       number = {2},
        pages = {2389-2401},
          doi = {10.1093/mnras/stac545},
archivePrefix = {arXiv},
       eprint = {2202.12206},
 primaryClass = {astro-ph.GA},
       adsurl = {https://ui.adsabs.harvard.edu/abs/2022MNRAS.512.2389D}
}

@ARTICLE{DAloisio2015,
       author = {{D'Aloisio}, Anson and {McQuinn}, Matthew and {Trac}, Hy},
        title = "{Large Opacity Variations in the High-redshift Ly{\ensuremath{\alpha}} Forest: The Signature of Relic Temperature Fluctuations from Patchy Reionization}",
      journal = {\apjl},
         year = 2015,
        month = nov,
       volume = {813},
       number = {2},
          eid = {L38},
        pages = {L38},
          doi = {10.1088/2041-8205/813/2/L38},
archivePrefix = {arXiv},
       eprint = {1509.02523},
 primaryClass = {astro-ph.CO},
       adsurl = {https://ui.adsabs.harvard.edu/abs/2015ApJ...813L..38D}
}

@ARTICLE{Ferland1998,
       author = {{Ferland}, G.~J. and {Korista}, K.~T. and {Verner}, D.~A. and {Ferguson}, J.~W. and {Kingdon}, J.~B. and {Verner}, E.~M.},
        title = "{CLOUDY 90: Numerical Simulation of Plasmas and Their Spectra}",
      journal = {\pasp},
         year = 1998,
        month = jul,
       volume = {110},
       number = {749},
        pages = {761-778},
          doi = {10.1086/316190},
       adsurl = {https://ui.adsabs.harvard.edu/abs/1998PASP..110..761F}
}

@ARTICLE{Gaikwad2021,
       author = {{Gaikwad}, Prakash and {Srianand}, Raghunathan and {Haehnelt}, Martin G. and {Choudhury}, Tirthankar Roy},
        title = "{A consistent and robust measurement of the thermal state of the IGM at 2 {\ensuremath{\leq}} z {\ensuremath{\leq}} 4 from a large sample of Ly {\ensuremath{\alpha}} forest spectra: evidence for late and rapid He II reionization}",
      journal = {\mnras},
         year = 2021,
        month = sep,
       volume = {506},
       number = {3},
        pages = {4389-4412},
          doi = {10.1093/mnras/stab2017},
archivePrefix = {arXiv},
       eprint = {2009.00016},
 primaryClass = {astro-ph.CO},
       adsurl = {https://ui.adsabs.harvard.edu/abs/2021MNRAS.506.4389G}
}

@ARTICLE{Greene2022,
       author = {{Greene}, Jenny and {Bezanson}, Rachel and {Ouchi}, Masami and {Silverman}, John and {the PFS Galaxy Evolution Working Group}},
        title = "{The Prime Focus Spectrograph Galaxy Evolution Survey}",
      journal = {arXiv e-prints},
         year = 2022,
        month = jun,
          eid = {arXiv:2206.14908},
        pages = {arXiv:2206.14908},
          doi = {10.48550/arXiv.2206.14908},
archivePrefix = {arXiv},
       eprint = {2206.14908},
 primaryClass = {astro-ph.GA},
       adsurl = {https://ui.adsabs.harvard.edu/abs/2022arXiv220614908G}
}

@ARTICLE{GarciaGallego2026,
       author = {{Garcia-Gallego}, Olga and {Ir{\v{s}}i{\v{c}}}, Vid and {Viel}, Matteo and {Haehnelt}, Martin G. and {Bolton}, James S.},
        title = "{Post-inflationary axion constraints from the Lyman-$α$ forest}",
      journal = {arXiv e-prints},
         year = 2026,
        month = mar,
          eid = {arXiv:2603.04401},
        pages = {arXiv:2603.04401},
          doi = {10.48550/arXiv.2603.04401},
archivePrefix = {arXiv},
       eprint = {2603.04401},
 primaryClass = {astro-ph.CO},
       adsurl = {https://ui.adsabs.harvard.edu/abs/2026arXiv260304401G}
}

@ARTICLE{Garcia2025,
       author = {{Garcia-Gallego}, Olga and {Ir{\v{s}}i{\v{c}}}, Vid and {Haehnelt}, Martin G. and {Viel}, Matteo and {Bolton}, James S.},
        title = "{Constraining mixed dark matter models with high-redshift Lyman-alpha forest data}",
      journal = {\prd},
         year = 2025,
        month = aug,
       volume = {112},
       number = {4},
          eid = {043502},
        pages = {043502},
          doi = {10.1103/4k29-h99l},
archivePrefix = {arXiv},
       eprint = {2504.06367},
 primaryClass = {astro-ph.CO},
       adsurl = {https://ui.adsabs.harvard.edu/abs/2025PhRvD.112d3502G}
}

@ARTICLE{Hirata2018,
       author = {{Hirata}, Christopher M.},
        title = "{Small-scale structure and the Lyman-{\ensuremath{\alpha}} forest baryon acoustic oscillation feature}",
      journal = {\mnras},
         year = 2018,
        month = feb,
       volume = {474},
       number = {2},
        pages = {2173-2193},
          doi = {10.1093/mnras/stx2854},
archivePrefix = {arXiv},
       eprint = {1707.03358},
 primaryClass = {astro-ph.CO},
       adsurl = {https://ui.adsabs.harvard.edu/abs/2018MNRAS.474.2173H}
}

@ARTICLE{Hooper2022,
       author = {{Hooper}, Deanna C. and {Sch{\"o}neberg}, Nils and {Murgia}, Riccardo and {Archidiacono}, Maria and {Lesgourgues}, Julien and {Viel}, Matteo},
        title = "{One likelihood to bind them all: Lyman-{\ensuremath{\alpha}} constraints on non-standard dark matter}",
      journal = {\jcap},
         year = 2022,
        month = oct,
       volume = {2022},
       number = {10},
          eid = {032},
        pages = {032},
          doi = {10.1088/1475-7516/2022/10/032},
archivePrefix = {arXiv},
       eprint = {2206.08188},
 primaryClass = {astro-ph.CO},
       adsurl = {https://ui.adsabs.harvard.edu/abs/2022JCAP...10..032H}
}

@ARTICLE{HuiRutledge1999,
       author = {{Hui}, Lam and {Rutledge}, Robert E.},
        title = "{The b Distribution and the Velocity Structure of Absorption Peaks in the Ly{\ensuremath{\alpha}} Forest}",
      journal = {\apj},
         year = 1999,
        month = jun,
       volume = {517},
       number = {2},
        pages = {541-548},
          doi = {10.1086/307202},
archivePrefix = {arXiv},
       eprint = {astro-ph/9709100},
 primaryClass = {astro-ph},
       adsurl = {https://ui.adsabs.harvard.edu/abs/1999ApJ...517..541H}
}

@article{HuiHaiman2003,
doi = {10.1086/377229},
url = {https://doi.org/10.1086/377229},
year = {2003},
month = {oct},
publisher = {},
volume = {596},
number = {1},
pages = {9},
author = {Hui, Lam and Haiman, Zoltán},
title = {The Thermal Memory of Reionization History},
journal = {The Astrophysical Journal}
}

@ARTICLE{Irsic2017a,
       author = {{Ir{\v{s}}i{\v{c}}}, Vid and {Viel}, Matteo and {Haehnelt}, Martin G. and {Bolton}, James S. and {Cristiani}, Stefano and {Becker}, George D. and {D'Odorico}, Valentina and {Cupani}, Guido and {Kim}, Tae-Sun and {Berg}, Trystyn A.~M. and {L{\'o}pez}, Sebastian and {Ellison}, Sara and {Christensen}, Lise and {Denney}, Kelly D. and {Worseck}, G{\'a}bor},
        title = "{New constraints on the free-streaming of warm dark matter from intermediate and small scale Lyman-{\ensuremath{\alpha}} forest data}",
      journal = {\prd},
         year = 2017,
        month = jul,
       volume = {96},
       number = {2},
          eid = {023522},
        pages = {023522},
          doi = {10.1103/PhysRevD.96.023522},
archivePrefix = {arXiv},
       eprint = {1702.01764},
 primaryClass = {astro-ph.CO},
       adsurl = {https://ui.adsabs.harvard.edu/abs/2017PhRvD..96b3522I}
}

@ARTICLE{Irsic2017b,
       author = {{Ir{\v{s}}i{\v{c}}}, Vid and {Viel}, Matteo and {Berg}, Trystyn A.~M. and {D'Odorico}, Valentina and {Haehnelt}, Martin G. and {Cristiani}, Stefano and {Cupani}, Guido and {Kim}, Tae-Sun and {L{\'o}pez}, Sebastian and {Ellison}, Sara and {Becker}, George D. and {Christensen}, Lise and {Denney}, Kelly D. and {Worseck}, G{\'a}bor and {Bolton}, James S.},
        title = "{The Lyman {\ensuremath{\alpha}} forest power spectrum from the XQ-100 Legacy Survey}",
      journal = {\mnras},
         year = 2017,
        month = apr,
       volume = {466},
       number = {4},
        pages = {4332-4345},
          doi = {10.1093/mnras/stw3372},
archivePrefix = {arXiv},
       eprint = {1702.01761},
 primaryClass = {astro-ph.CO},
       adsurl = {https://ui.adsabs.harvard.edu/abs/2017MNRAS.466.4332I}
}

@ARTICLE{Irsic2024,
       author = {{Ir{\v{s}}i{\v{c}}}, Vid and {Viel}, Matteo and {Haehnelt}, Martin G. and {Bolton}, James S. and {Molaro}, Margherita and {Puchwein}, Ewald and {Boera}, Elisa and {Becker}, George D. and {Gaikwad}, Prakash and {Keating}, Laura C. and {Kulkarni}, Girish},
        title = "{Unveiling dark matter free streaming at the smallest scales with the high redshift Lyman-alpha forest}",
      journal = {\prd},
         year = 2024,
        month = feb,
       volume = {109},
       number = {4},
          eid = {043511},
        pages = {043511},
          doi = {10.1103/PhysRevD.109.043511},
archivePrefix = {arXiv},
       eprint = {2309.04533},
 primaryClass = {astro-ph.CO},
       adsurl = {https://ui.adsabs.harvard.edu/abs/2024PhRvD.109d3511I}
}

@ARTICLE{Jin2024,
       author = {{Jin}, Shoko and {Trager}, Scott C. and {Dalton}, Gavin B. and {Aguerri}, J. Alfonso L. and {Drew}, J.~E. and {Falc{\'o}n-Barroso}, Jes{\'u}s and {G{\"a}nsicke}, Boris T. and {Hill}, Vanessa and {Iovino}, Angela and {Pieri}, Matthew M. and {Poggianti}, Bianca M. and {Smith}, D.~J.~B. and {Vallenari}, Antonella and {Abrams}, Don Carlos and {Aguado}, David S. and {Antoja}, Teresa and {Arag{\'o}n-Salamanca}, Alfonso and {Ascasibar}, Yago and {Babusiaux}, Carine and {Balcells}, Marc and {Barrena}, R. and {Battaglia}, Giuseppina and {Belokurov}, Vasily and {Bensby}, Thomas and {Bonifacio}, Piercarlo and {Bragaglia}, Angela and {Carrasco}, Esperanza and {Carrera}, Ricardo and {Cornwell}, Daniel J. and {Dom{\'\i}nguez-Palmero}, Lilian and {Duncan}, Kenneth J. and {Famaey}, Benoit and {Fari{\~n}a}, Cecilia and {Gonzalez}, Oscar A. and {Guest}, Steve and {Hatch}, Nina A. and {Hess}, Kelley M. and {Hoskin}, Matthew J. and {Irwin}, Mike and {Knapen}, Johan H. and {Koposov}, Sergey E. and {Kuchner}, Ulrike and {Laigle}, Clotilde and {Lewis}, Jim and {Longhetti}, Marcella and {Lucatello}, Sara and {M{\'e}ndez-Abreu}, Jairo and {Mercurio}, Amata and {Molaeinezhad}, Alireza and {Mongui{\'o}}, Maria and {Morrison}, Sean and {Murphy}, David N.~A. and {Peralta de Arriba}, Luis and {P{\'e}rez}, Isabel and {P{\'e}rez-R{\`a}fols}, Ignasi and {Pic{\'o}}, Sergio and {Raddi}, Roberto and {Romero-G{\'o}mez}, Merc{\`e} and {Royer}, Fr{\'e}d{\'e}ric and {Siebert}, Arnaud and {Seabroke}, George M. and {Som}, Debopam and {Terrett}, David and {Thomas}, Guillaume and {Wesson}, Roger and {Worley}, C. Clare and {Alfaro}, Emilio J. and {Allende Prieto}, Carlos and {Alonso-Santiago}, Javier and {Amos}, Nicholas J. and {Ashley}, Richard P. and {Balaguer-N{\'u}{\~n}ez}, Lola and {Balbinot}, Eduardo and {Bellazzini}, Michele and {Benn}, Chris R. and {Berlanas}, Sara R. and {Bernard}, Edouard J. and {Best}, Philip and {Bettoni}, Daniela and {Bianco}, Andrea and {Bishop}, Georgia and {Blomqvist}, Michael and {Boeche}, Corrado and {Bolzonella}, Micol and {Bonoli}, Silvia and {Bosma}, Albert and {Britavskiy}, Nikolay and {Busarello}, Gianni and {Caffau}, Elisabetta and {Cantat-Gaudin}, Tristan and {Castro-Ginard}, Alfred and {Couto}, Guilherme and {Carbajo-Hijarrubia}, Juan and {Carter}, David and {Casamiquela}, Laia and {Conrado}, Ana M. and {Corcho-Caballero}, Pablo and {Costantin}, Luca and {Deason}, Alis and {de Burgos}, Abel and {De Grandi}, Sabrina and {Di Matteo}, Paola and {Dom{\'\i}nguez-G{\'o}mez}, Jes{\'u}s and {Dorda}, Ricardo and {Drake}, Alyssa and {Dutta}, Rajeshwari and {Erkal}, Denis and {Feltzing}, Sofia and {Ferr{\'e}-Mateu}, Anna and {Feuillet}, Diane and {Figueras}, Francesca and {Fossati}, Matteo and {Franciosini}, Elena and {Frasca}, Antonio and {Fumagalli}, Michele and {Gallazzi}, Anna and {Garc{\'\i}a-Benito}, Rub{\'e}n and {Gentile Fusillo}, Nicola and {Gebran}, Marwan and {Gilbert}, James and {Gledhill}, T.~M. and {Gonz{\'a}lez Delgado}, Rosa M. and {Greimel}, Robert and {Guarcello}, Mario Giuseppe and {Guerra}, Jose and {Gullieuszik}, Marco and {Haines}, Christopher P. and {Hardcastle}, Martin J. and {Harris}, Amy and {Haywood}, Misha and {Helmi}, Amina and {Hernandez}, Nauzet and {Herrero}, Artemio and {Hughes}, Sarah and {Ir{\v{s}}i{\v{c}}}, Vid and {Jablonka}, Pascale and {Jarvis}, Matt J. and {Jordi}, Carme and {Kondapally}, Rohit and {Kordopatis}, Georges and {Krogager}, Jens-Kristian and {La Barbera}, Francesco and {Lam}, Man I. and {Larsen}, S{\o}ren S. and {Lemasle}, Bertrand and {Lewis}, Ian J. and {Lhom{\'e}}, Emilie and {Lind}, Karin and {Lodi}, Marcello and {Longobardi}, Alessia and {Lonoce}, Ilaria and {Magrini}, Laura and {Ma{\'\i}z Apell{\'a}niz}, Jes{\'u}s and {Marchal}, Olivier and {Marco}, Amparo and {Martin}, Nicolas F. and {Matsuno}, Tadafumi and {Maurogordato}, Sophie and {Merluzzi}, Paola and {Miralda-Escud{\'e}}, Jordi and {Molinari}, Emilio and {Monari}, Giacomo and {Morelli}, Lorenzo and {Mottram}, Christopher J. and {Naylor}, Tim and {Negueruela}, Ignacio and {O{\~n}orbe}, Jose and {Pancino}, Elena and {Peirani}, S{\'e}bastien and {Peletier}, Reynier F. and {Pozzetti}, Lucia and {Rainer}, Monica and {Ramos}, Pau and {Read}, Shaun C. and {Rossi}, Elena Maria and {R{\"o}ttgering}, Huub J.~A. and {Rubi{\~n}o-Mart{\'\i}n}, Jose Alberto and {Sabater}, Jose and {San Juan}, Jos{\'e} and {Sanna}, Nicoletta and {Schallig}, Ellen and {Schiavon}, Ricardo P. and {Schultheis}, Mathias and {Serra}, Paolo and {Shimwell}, Timothy W. and {Sim{\'o}n-D{\'\i}az}, Sergio and {Smith}, Russell J. and {Sordo}, Rosanna and {Sorini}, Daniele and {Soubiran}, Caroline and {Starkenburg}, Else and {Steele}, Iain A. and {Stott}, John and {Stuik}, Remko and {Tolstoy}, Eline and {Tortora}, Crescenzo and {Tsantaki}, Maria and {Van der Swaelmen}, Mathieu and {van Weeren}, Reinout J. and {Vergani}, Daniela},
        title = "{The wide-field, multiplexed, spectroscopic facility WEAVE: Survey design, overview, and simulated implementation}",
      journal = {\mnras},
         year = 2024,
        month = may,
       volume = {530},
       number = {3},
        pages = {2688-2730},
          doi = {10.1093/mnras/stad557},
archivePrefix = {arXiv},
       eprint = {2212.03981},
 primaryClass = {astro-ph.IM},
       adsurl = {https://ui.adsabs.harvard.edu/abs/2024MNRAS.530.2688J}
}

@ARTICLE{Karacayli2020,
       author = {{Kara{\c{c}}ayl{\i}}, Naim G{\"o}ksel and {Font-Ribera}, Andreu and {Padmanabhan}, Nikhil},
        title = "{Optimal 1D Ly {\ensuremath{\alpha}} forest power spectrum estimation - I. DESI-lite spectra}",
      journal = {\mnras},
         year = 2020,
        month = oct,
       volume = {497},
       number = {4},
        pages = {4742-4752},
          doi = {10.1093/mnras/staa2331},
archivePrefix = {arXiv},
       eprint = {2008.06421},
 primaryClass = {astro-ph.CO},
       adsurl = {https://ui.adsabs.harvard.edu/abs/2020MNRAS.497.4742K}
}

@ARTICLE{Karacayli2022,
       author = {{Kara{\c{c}}ayl{\i}}, Naim G{\"o}ksel and {Padmanabhan}, Nikhil and {Font-Ribera}, Andreu and {Ir{\v{s}}i{\v{c}}}, Vid and {Walther}, Michael and {Brooks}, David and {Gazta{\~n}aga}, Enrique and {Kehoe}, Robert and {Levi}, Michael and {Ntelis}, Pierros and {Palanque-Delabrouille}, Nathalie and {Tarl{\'e}}, Gregory},
        title = "{Optimal 1D Ly {\ensuremath{\alpha}} forest power spectrum estimation - II. KODIAQ, SQUAD, and XQ-100}",
      journal = {\mnras},
         year = 2022,
        month = jan,
       volume = {509},
       number = {2},
        pages = {2842-2855},
          doi = {10.1093/mnras/stab3201},
archivePrefix = {arXiv},
       eprint = {2108.10870},
 primaryClass = {astro-ph.CO},
       adsurl = {https://ui.adsabs.harvard.edu/abs/2022MNRAS.509.2842K}
}

@ARTICLE{Karacayli2024,
       author = {{Kara{\c{c}}ayl{\i}}, Naim G{\"o}ksel and {Martini}, Paul and {Guy}, Julien and {Ravoux}, Corentin and {Abdul Karim}, Marie Lynn and {Armengaud}, Eric and {Walther}, Michael and {Aguilar}, J. and {Ahlen}, S. and {Bailey}, S. and {Bautista}, J. and {Beltran}, S.~F. and {Brooks}, D. and {Cabayol-Garcia}, L. and {Chabanier}, S. and {Chaussidon}, E. and {Chaves-Montero}, J. and {Dawson}, K. and {de la Cruz}, R. and {de la Macorra}, A. and {Doel}, P. and {Font-Ribera}, A. and {Forero-Romero}, J.~E. and {Gontcho}, S. Gontcho A. and {Gonzalez-Morales}, A.~X. and {Gordon}, C. and {Herrera-Alcantar}, H.~K. and {Honscheid}, K. and {Ir{\v{s}}i{\v{c}}}, V. and {Ishak}, M. and {Kehoe}, R. and {Kisner}, T. and {Kremin}, A. and {Landriau}, M. and {Le Guillou}, L. and {Levi}, M.~E. and {Luki{\'c}}, Z. and {Meisner}, A. and {Miquel}, R. and {Moustakas}, J. and {Mueller}, E. and {Mu{\~n}oz-Guti{\'e}rrez}, A. and {Napolitano}, L. and {Nie}, J. and {Niz}, G. and {Palanque-Delabrouille}, N. and {Percival}, W.~J. and {Pieri}, M. and {Poppett}, C. and {Prada}, F. and {P{\'e}rez-R{\`a}fols}, I. and {Ram{\'\i}rez-P{\'e}rez}, C. and {Rossi}, G. and {Sanchez}, E. and {Seo}, H. and {Sinigaglia}, F. and {Tan}, T. and {Tarl{\'e}}, G. and {Wang}, B. and {Weaver}, B.~A. and {Y{\'e}che}, C. and {Zhou}, Z.},
        title = "{Optimal 1D Ly {\ensuremath{\alpha}} forest power spectrum estimation - III. DESI early data}",
      journal = {\mnras},
         year = 2024,
        month = mar,
       volume = {528},
       number = {3},
        pages = {3941-3963},
          doi = {10.1093/mnras/stae171},
archivePrefix = {arXiv},
       eprint = {2306.06316},
 primaryClass = {astro-ph.CO},
       adsurl = {https://ui.adsabs.harvard.edu/abs/2024MNRAS.528.3941K}
}

@ARTICLE{Karacayli2025,
       author = {{Kara{\c{c}}ayl{\i}}, N.~G. and {Martini}, P. and {Aguilar}, J. and {Ahlen}, S. and {Armengaud}, E. and {Bailey}, S. and {Bault}, A. and {Bianchi}, D. and {Brodzeller}, A. and {Brooks}, D. and {Chaves-Montero}, J. and {Claybaugh}, T. and {Cuceu}, A. and {de la Macorra}, A. and {Dey}, A. and {Dey}, B. and {Doel}, P. and {Ferraro}, S. and {Font-Ribera}, A. and {Forero-Romero}, J.~E. and {Gazta{\~n}aga}, E. and {Gontcho}, S. Gontcho A and {Gutierrez}, G. and {Guy}, J. and {Hahn}, C. and {Herrera-Alcantar}, H.~K. and {Honscheid}, K. and {Ishak}, M. and {Kehoe}, R. and {Kirkby}, D. and {Kremin}, A. and {Landriau}, M. and {Le Goff}, J.~M. and {Le Guillou}, L. and {Levi}, M.~E. and {Manera}, M. and {Meisner}, A. and {Miquel}, R. and {Montero-Camacho}, P. and {Nadathur}, S. and {Niz}, G. and {Palanque-Delabrouille}, N. and {Pan}, Z. and {Percival}, W.~J. and {Pieri}, Matthew M. and {Prada}, F. and {P{\'e}rez-R{\`a}fols}, I. and {Ravoux}, C. and {Rossi}, G. and {Sanchez}, E. and {Saulder}, C. and {Schlegel}, D. and {Schubnell}, M. and {Seo}, H. and {Siudek}, M. and {Sprayberry}, D. and {Tan}, T. and {Tang}, Ji-Jia and {Tarl{\'e}}, G. and {Walther}, M. and {Weaver}, B.~A. and {Yu}, J. and {Zhou}, R. and {Zou}, H.},
        title = "{DESI DR1 Ly$α$ 1D power spectrum: The optimal estimator measurement}",
      journal = {arXiv e-prints},
         year = 2025,
        month = may,
          eid = {arXiv:2505.07974},
        pages = {arXiv:2505.07974},
          doi = {10.48550/arXiv.2505.07974},
archivePrefix = {arXiv},
       eprint = {2505.07974},
 primaryClass = {astro-ph.CO},
       adsurl = {https://ui.adsabs.harvard.edu/abs/2025arXiv250507974K}
}

@ARTICLE{Keating2018,
       author = {{Keating}, Laura C. and {Puchwein}, Ewald and {Haehnelt}, Martin G.},
        title = "{Spatial fluctuations of the intergalactic temperature-density relation after hydrogen reionization}",
      journal = {\mnras},
         year = 2018,
        month = jul,
       volume = {477},
       number = {4},
        pages = {5501-5516},
          doi = {10.1093/mnras/sty968},
archivePrefix = {arXiv},
       eprint = {1709.05351},
 primaryClass = {astro-ph.CO},
       adsurl = {https://ui.adsabs.harvard.edu/abs/2018MNRAS.477.5501K}
}

@ARTICLE{Kim2007,
       author = {{Kim}, T. -S. and {Bolton}, J.~S. and {Viel}, M. and {Haehnelt}, M.~G. and {Carswell}, R.~F.},
        title = "{An improved measurement of the flux distribution of the Ly{\ensuremath{\alpha}} forest in QSO absorption spectra: the effect of continuum fitting, metal contamination and noise properties}",
      journal = {\mnras},
         year = 2007,
        month = dec,
       volume = {382},
       number = {4},
        pages = {1657-1674},
          doi = {10.1111/j.1365-2966.2007.12406.x},
archivePrefix = {arXiv},
       eprint = {0711.1862},
 primaryClass = {astro-ph},
       adsurl = {https://ui.adsabs.harvard.edu/abs/2007MNRAS.382.1657K}
}

@ARTICLE{Lopez2016,
       author = {{L{\'o}pez}, S. and {D'Odorico}, V. and {Ellison}, S.~L. and {Becker}, G.~D. and {Christensen}, L. and {Cupani}, G. and {Denney}, K.~D. and {P{\^a}ris}, I. and {Worseck}, G. and {Berg}, T.~A.~M. and {Cristiani}, S. and {Dessauges-Zavadsky}, M. and {Haehnelt}, M. and {Hamann}, F. and {Hennawi}, J. and {Ir{\v{s}}i{\v{c}}}, V. and {Kim}, T. -S. and {L{\'o}pez}, P. and {Lund Saust}, R. and {M{\'e}nard}, B. and {Perrotta}, S. and {Prochaska}, J.~X. and {S{\'a}nchez-Ram{\'\i}rez}, R. and {Vestergaard}, M. and {Viel}, M. and {Wisotzki}, L.},
        title = "{XQ-100: A legacy survey of one hundred 3.5 {\ensuremath{\lesssim}} z {\ensuremath{\lesssim}} 4.5 quasars observed with VLT/X-shooter}",
      journal = {\aap},
         year = 2016,
        month = oct,
       volume = {594},
          eid = {A91},
        pages = {A91},
          doi = {10.1051/0004-6361/201628161},
archivePrefix = {arXiv},
       eprint = {1607.08776},
 primaryClass = {astro-ph.GA},
       adsurl = {https://ui.adsabs.harvard.edu/abs/2016A&A...594A..91L}
}

@ARTICLE{Mathes2017,
       author = {{Mathes}, Nigel L. and {Churchill}, Christopher W. and {Murphy}, Michael T.},
        title = "{The Vulture Survey I: Analyzing the Evolution of ${\MgII}$ Absorbers}",
      journal = {arXiv e-prints},
         year = 2017,
        month = jan,
          eid = {arXiv:1701.05624},
        pages = {arXiv:1701.05624},
          doi = {10.48550/arXiv.1701.05624},
archivePrefix = {arXiv},
       eprint = {1701.05624},
 primaryClass = {astro-ph.GA},
       adsurl = {https://ui.adsabs.harvard.edu/abs/2017arXiv170105624M}
}

@ARTICLE{McDonald2005,
       author = {{McDonald}, Patrick and {Seljak}, Uro{\v{s}} and {Cen}, Renyue and {Shih}, David and {Weinberg}, David H. and {Burles}, Scott and {Schneider}, Donald P. and {Schlegel}, David J. and {Bahcall}, Neta A. and {Briggs}, John W. and {Brinkmann}, J. and {Fukugita}, Masataka and {Ivezi{\'c}}, {\v{Z}}eljko and {Kent}, Stephen and {Vanden Berk}, Daniel E.},
        title = "{The Linear Theory Power Spectrum from the Ly{\ensuremath{\alpha}} Forest in the Sloan Digital Sky Survey}",
      journal = {\apj},
         year = 2005,
        month = dec,
       volume = {635},
       number = {2},
        pages = {761-783},
          doi = {10.1086/497563},
archivePrefix = {arXiv},
       eprint = {astro-ph/0407377},
 primaryClass = {astro-ph},
       adsurl = {https://ui.adsabs.harvard.edu/abs/2005ApJ...635..761M}
}

@ARTICLE{McDonald2006,
       author = {{McDonald}, Patrick and {Seljak}, Uro{\v{s}} and {Burles}, Scott and {Schlegel}, David J. and {Weinberg}, David H. and {Cen}, Renyue and {Shih}, David and {Schaye}, Joop and {Schneider}, Donald P. and {Bahcall}, Neta A. and {Briggs}, John W. and {Brinkmann}, J. and {Brunner}, Robert J. and {Fukugita}, Masataka and {Gunn}, James E. and {Ivezi{\'c}}, {\v{Z}}eljko and {Kent}, Stephen and {Lupton}, Robert H. and {Vanden Berk}, Daniel E.},
        title = "{The Ly{\ensuremath{\alpha}} Forest Power Spectrum from the Sloan Digital Sky Survey}",
      journal = {\apjs},
         year = 2006,
        month = mar,
       volume = {163},
       number = {1},
        pages = {80-109},
          doi = {10.1086/444361},
archivePrefix = {arXiv},
       eprint = {astro-ph/0405013},
 primaryClass = {astro-ph},
       adsurl = {https://ui.adsabs.harvard.edu/abs/2006ApJS..163...80M}
}

@ARTICLE{McQuinn2016,
       author = {{McQuinn}, Matthew},
        title = "{The Evolution of the Intergalactic Medium}",
      journal = {\araa},
         year = 2016,
        month = sep,
       volume = {54},
        pages = {313-362},
          doi = {10.1146/annurev-astro-082214-122355},
archivePrefix = {arXiv},
       eprint = {1512.00086},
 primaryClass = {astro-ph.CO},
       adsurl = {https://ui.adsabs.harvard.edu/abs/2016ARA&A..54..313M}
}

@ARTICLE{Meiksin2009,
       author = {{Meiksin}, Avery A.},
        title = "{The physics of the intergalactic medium}",
      journal = {Reviews of Modern Physics},
         year = 2009,
        month = oct,
       volume = {81},
       number = {4},
        pages = {1405-1469},
          doi = {10.1103/RevModPhys.81.1405},
archivePrefix = {arXiv},
       eprint = {0711.3358},
 primaryClass = {astro-ph},
       adsurl = {https://ui.adsabs.harvard.edu/abs/2009RvMP...81.1405M}
}

@ARTICLE{Molaro2022,
       author = {{Molaro}, Margherita and {Ir{\v{s}}i{\v{c}}}, Vid and {Bolton}, James S. and {Keating}, Laura C. and {Puchwein}, Ewald and {Gaikwad}, Prakash and {Haehnelt}, Martin G. and {Kulkarni}, Girish and {Viel}, Matteo},
        title = "{The effect of inhomogeneous reionization on the Lyman {\ensuremath{\alpha}} forest power spectrum at redshift z > 4: implications for thermal parameter recovery}",
      journal = {\mnras},
         year = 2022,
        month = feb,
       volume = {509},
       number = {4},
        pages = {6119-6137},
          doi = {10.1093/mnras/stab3416},
archivePrefix = {arXiv},
       eprint = {2109.06897},
 primaryClass = {astro-ph.CO},
       adsurl = {https://ui.adsabs.harvard.edu/abs/2022MNRAS.509.6119M}
}

@ARTICLE{Molaro2023,
       author = {{Molaro}, Margherita and {Ir{\v{s}}i{\v{c}}}, Vid and {Bolton}, James S. and {Lieu}, Maggie and {Keating}, Laura C. and {Puchwein}, Ewald and {Haehnelt}, Martin G. and {Viel}, Matteo},
        title = "{Possible evidence for a large-scale enhancement in the Lyman-{\ensuremath{\alpha}} forest power spectrum at redshift z {\ensuremath{\geq}} 4}",
      journal = {\mnras},
         year = 2023,
        month = may,
       volume = {521},
       number = {1},
        pages = {1489-1501},
          doi = {10.1093/mnras/stad598},
archivePrefix = {arXiv},
       eprint = {2303.05167},
 primaryClass = {astro-ph.CO},
       adsurl = {https://ui.adsabs.harvard.edu/abs/2023MNRAS.521.1489M}
}

@ARTICLE{Montero2021,
 author = {{Montero-Camacho}, Paulo and {Mao}, Yi},
 title = "{Extracting the astrophysics of reionization from the Ly{\ensuremath{\alpha}} forest power spectrum: a first forecast}",
 journal = {\mnras},
 year = 2021,
 month = nov,
 volume = {508},
 number = {1},
 pages = {1262-1279},
 doi = {10.1093/mnras/stab2569},
archivePrefix = {arXiv},
 eprint = {2106.14492},
 primaryClass = {astro-ph.CO},
 adsurl = {https://ui.adsabs.harvard.edu/abs/2021MNRAS.508.1262M}
}

@ARTICLE{Murphy2019,
       author = {{Murphy}, Michael T. and {Kacprzak}, Glenn G. and {Savorgnan}, Giulia A.~D. and {Carswell}, Robert F.},
        title = "{The UVES Spectral Quasar Absorption Database (SQUAD) data release 1: the first 10 million seconds}",
      journal = {\mnras},
         year = 2019,
        month = jan,
       volume = {482},
       number = {3},
        pages = {3458-3479},
          doi = {10.1093/mnras/sty2834},
archivePrefix = {arXiv},
       eprint = {1810.06136},
 primaryClass = {astro-ph.GA},
       adsurl = {https://ui.adsabs.harvard.edu/abs/2019MNRAS.482.3458M}
}

@ARTICLE{Miller2019,
       author = {{Miller}, Joel S.~A. and {Bolton}, James S. and {Hatch}, Nina},
        title = "{Searching for the shadows of giants: characterizing protoclusters with line of sight Lyman-{\ensuremath{\alpha}} absorption}",
      journal = {\mnras},
         year = 2019,
        month = nov,
       volume = {489},
       number = {4},
        pages = {5381-5397},
          doi = {10.1093/mnras/stz2504},
archivePrefix = {arXiv},
       eprint = {1909.02513},
 primaryClass = {astro-ph.CO},
       adsurl = {https://ui.adsabs.harvard.edu/abs/2019MNRAS.489.5381M}
}

@ARTICLE{Murdoch1986,
       author = {{Murdoch}, H.~S. and {Hunstead}, R.~W. and {Pettini}, M. and {Blades}, J.~C.},
        title = "{Absorption Spectrum of the Z = 3.78 QSO 2000-330. II. The Redshift and Equivalent Width Distributions of Primordial Hydrogen Clouds}",
      journal = {\apj},
         year = 1986,
        month = oct,
       volume = {309},
        pages = {19},
          doi = {10.1086/164573},
       adsurl = {https://ui.adsabs.harvard.edu/abs/1986ApJ...309...19M}
}

@ARTICLE{Mainieri2024,
       author = {{Mainieri}, Vincenzo and {Anderson}, Richard I. and {Brinchmann}, Jarle and {Cimatti}, Andrea and {Ellis}, Richard S. and {Hill}, Vanessa and {Kneib}, Jean-Paul and {McLeod}, Anna F. and {Opitom}, Cyrielle and {Roth}, Martin M. and {Sanchez-Saez}, Paula and {Smiljanic}, Rodolfo and {Tolstoy}, Eline and {Bacon}, Roland and {Randich}, Sofia and {Adamo}, Angela and {Annibali}, Francesca and {Arevalo}, Patricia and {Audard}, Marc and {Barsanti}, Stefania and {Battaglia}, Giuseppina and {Bayo Aran}, Amelia M. and {Belfiore}, Francesco and {Bellazzini}, Michele and {Bellini}, Emilio and {Beltran}, Maria Teresa and {Berni}, Leda and {Bianchi}, Simone and {Biazzo}, Katia and {Bisero}, Sofia and {Bisogni}, Susanna and {Bland-Hawthorn}, Joss and {Blondin}, Stephane and {Bodensteiner}, Julia and {Boffin}, Henri M.~J. and {Bonito}, Rosaria and {Bono}, Giuseppe and {Bouche}, Nicolas F. and {Bowman}, Dominic and {Braga}, Vittorio F. and {Bragaglia}, Angela and {Branchesi}, Marica and {Brucalassi}, Anna and {Bryant}, Julia J. and {Bryson}, Ian and {Busa}, Innocenza and {Camera}, Stefano and {Carbone}, Carmelita and {Casali}, Giada and {Casali}, Mark and {Casasola}, Viviana and {Castro}, Norberto and {Catelan}, Marcio and {Cavallo}, Lorenzo and {Chiappini}, Cristina and {Cioni}, Maria-Rosa and {Colless}, Matthew and {Colzi}, Laura and {Contarini}, Sofia and {Couch}, Warrick and {D'Ammando}, Filippo and {d'Assignies D.}, William and {D'Orazi}, Valentina and {da Silva}, Ronaldo and {Dainotti}, Maria Giovanna and {Damiani}, Francesco and {Danielski}, Camilla and {De Cia}, Annalisa and {de Jong}, Roelof S. and {Dhawan}, Suhail and {Dierickx}, Philippe and {Driver}, Simon P. and {Dupletsa}, Ulyana and {Escoffier}, Stephanie and {Escorza}, Ana and {Fabrizio}, Michele and {Fiorentino}, Giuliana and {Fontana}, Adriano and {Fontani}, Francesco and {Forero Sanchez}, Daniel and {Franois}, Patrick and {Galindo-Guil}, Francisco Jose and {Gallazzi}, Anna Rita and {Galli}, Daniele and {Garcia}, Miriam and {Garcia-Rojas}, Jorge and {Garilli}, Bianca and {Grand}, Robert and {Guarcello}, Mario Giuseppe and {Hazra}, Nandini and {Helmi}, Amina and {Herrero}, Artemio and {Iglesias}, Daniela and {Ilic}, Dragana and {Irsic}, Vid and {Ivanov}, Valentin D. and {Izzo}, Luca and {Jablonka}, Pascale and {Joachimi}, Benjamin and {Kakkad}, Darshan and {Kamann}, Sebastian and {Koposov}, Sergey and {Kordopatis}, Georges and {Kovacevic}, Andjelka B. and {Kraljic}, Katarina and {Kuncarayakti}, Hanindyo and {Kwon}, Yuna and {La Forgia}, Fiorangela and {Lahav}, Ofer and {Laigle}, Clotilde and {Lazzarin}, Monica and {Leaman}, Ryan and {Leclercq}, Floriane and {Lee}, Khee-Gan and {Lee}, David and {Lehnert}, Matt D. and {Lira}, Paulina and {Loffredo}, Eleonora and {Lucatello}, Sara and {Magrini}, Laura and {Maguire}, Kate and {Mahler}, Guillaume and {Zahra Majidi}, Fatemeh and {Malavasi}, Nicola and {Mannucci}, Filippo and {Marconi}, Marcella and {Martin}, Nicolas and {Marulli}, Federico and {Massari}, Davide and {Matsuno}, Tadafumi and {Mattheee}, Jorryt and {McGee}, Sean and {Merc}, Jaroslav and {Merle}, Thibault and {Miglio}, Andrea and {Migliorini}, Alessandra and {Minchev}, Ivan and {Minniti}, Dante and {Miret-Roig}, Nuria and {Monreal Ibero}, Ana and {Montano}, Federico and {Montet}, Ben T. and {Moresco}, Michele and {Moretti}, Chiara and {Moscardini}, Lauro and {Moya}, Andres and {Mueller}, Oliver and {Nanayakkara}, Themiya and {Nicholl}, Matt and {Nordlander}, Thomas and {Onori}, Francesca and {Padovani}, Marco and {Pala}, Anna Francesca and {Panda}, Swayamtrupta and {Pandey-Pommier}, Mamta and {Pasquini}, Luca and {Pawlak}, Michal and {Pessi}, Priscila J. and {Pisani}, Alice and {Popovic}, Lukav C. and {Prisinzano}, Loredana and {Raddi}, Roberto and {Rainer}, Monica and {Rebassa-Mansergas}, Alberto and {Richard}, Johan and {Rigault}, Mickael and {Rocher}, Antoine and {Romano}, Donatella and {Rosati}, Piero and {Sacco}, Germano and {Sanchez-Janssen}, Ruben and {Sander}, Andreas A.~C. and {Sanders}, Jason L. and {Sargent}, Mark and {Sarpa}, Elena and {Schimd}, Carlo and {Schipani}, Pietro and {Sefusatti}, Emiliano and {Smith}, Graham P. and {Spina}, Lorenzo and {Steinmetz}, Matthias and {Tacchella}, Sandro and {Tautvaisiene}, Grazina and {Theissen}, Christopher and {Thomas}, Guillaume and {Ting}, Yuan-Sen and {Travouillon}, Tony and {Tresse}, Laurence and {Trivedi}, Oem and {Tsantaki}, Maria and {Tsedrik}, Maria and {Urrutia}, Tanya and {Valenti}, Elena and {Van der Swaelmen}, Mathieu and {Van Eck}, Sophie and {Verdiani}, Francesco and {Verdier}, Aurelien and {Vergani}, Susanna Diana and {Verhamme}, Anne and {Vernet}, Joel},
        title = "{The Wide-field Spectroscopic Telescope (WST) Science White Paper}",
      journal = {arXiv e-prints},
         year = 2024,
        month = mar,
          eid = {arXiv:2403.05398},
        pages = {arXiv:2403.05398},
          doi = {10.48550/arXiv.2403.05398},
archivePrefix = {arXiv},
       eprint = {2403.05398},
 primaryClass = {astro-ph.IM},
       adsurl = {https://ui.adsabs.harvard.edu/abs/2024arXiv240305398M}
}

@ARTICLE{ma2025,
       author = {{Ma}, Ke and {Bolton}, James S. and {Ir{\v{s}}i{\v{c}}}, Vid and {Gaikwad}, Prakash and {Puchwein}, Ewald},
        title = "{An improved model for the effect of correlated Si III absorption on the one-dimensional Lyman-{\ensuremath{\alpha}} forest power spectrum}",
      journal = {\mnras},
         year = 2026,
        month = feb,
       volume = {546},
       number = {1},
          eid = {staf2262},
        pages = {staf2262},
          doi = {10.1093/mnras/staf2262},
archivePrefix = {arXiv},
       eprint = {2509.08613},
 primaryClass = {astro-ph.CO},
       adsurl = {https://ui.adsabs.harvard.edu/abs/2026MNRAS.546f2262M}
}

@ARTICLE{Nasir2016,
       author = {{Nasir}, Fahad and {Bolton}, James S. and {Becker}, George D.},
        title = "{Inferring the IGM thermal history during reionization with the Lyman {\ensuremath{\alpha}} forest power spectrum at redshift z ≃ 5}",
      journal = {\mnras},
         year = 2016,
        month = dec,
       volume = {463},
       number = {3},
        pages = {2335-2347},
          doi = {10.1093/mnras/stw2147},
archivePrefix = {arXiv},
       eprint = {1605.04155},
 primaryClass = {astro-ph.CO},
       adsurl = {https://ui.adsabs.harvard.edu/abs/2016MNRAS.463.2335N}
}

@ARTICLE{OMeara2015,
       author = {{O'Meara}, J.~M. and {Lehner}, N. and {Howk}, J.~C. and {Prochaska}, J.~X. and {Fox}, A.~J. and {Swain}, M.~A. and {Gelino}, C.~R. and {Berriman}, G.~B. and {Tran}, H.},
        title = "{The First Data Release of the KODIAQ Survey}",
      journal = {\aj},
         year = 2015,
        month = oct,
       volume = {150},
       number = {4},
          eid = {111},
        pages = {111},
          doi = {10.1088/0004-6256/150/4/111},
archivePrefix = {arXiv},
       eprint = {1505.03529},
 primaryClass = {astro-ph.CO},
       adsurl = {https://ui.adsabs.harvard.edu/abs/2015AJ....150..111O}
}

@ARTICLE{Palanque-Delabrouille2013,
       author = {{Palanque-Delabrouille}, Nathalie and {Y{\`e}che}, Christophe and {Borde}, Arnaud and {Le Goff}, Jean-Marc and {Rossi}, Graziano and {Viel}, Matteo and {Aubourg}, {\'E}ric and {Bailey}, Stephen and {Bautista}, Julian and {Blomqvist}, Michael and {Bolton}, Adam and {Bolton}, James S. and {Busca}, Nicol{\'a}s G. and {Carithers}, Bill and {Croft}, Rupert A.~C. and {Dawson}, Kyle S. and {Delubac}, Timoth{\'e}e and {Font-Ribera}, Andreu and {Ho}, Shirley and {Kirkby}, David and {Lee}, Khee-Gan and {Margala}, Daniel and {Miralda-Escud{\'e}}, Jordi and {Muna}, Demitri and {Myers}, Adam D. and {Noterdaeme}, Pasquier and {P{\^a}ris}, Isabelle and {Petitjean}, Patrick and {Pieri}, Matthew M. and {Rich}, James and {Rollinde}, Emmanuel and {Ross}, Nicholas P. and {Schlegel}, David J. and {Schneider}, Donald P. and {Slosar}, An{\v{z}}e and {Weinberg}, David H.},
        title = "{The one-dimensional Ly{\ensuremath{\alpha}} forest power spectrum from BOSS}",
      journal = {\aap},
         year = 2013,
        month = nov,
       volume = {559},
          eid = {A85},
        pages = {A85},
          doi = {10.1051/0004-6361/201322130},
archivePrefix = {arXiv},
       eprint = {1306.5896},
 primaryClass = {astro-ph.CO},
       adsurl = {https://ui.adsabs.harvard.edu/abs/2013A&A...559A..85P}
}

@ARTICLE{Palanque-Delabrouille2016,
       author = {{Palanque-Delabrouille}, N. and {Magneville}, Ch. and {Y{\`e}che}, Ch. and {P{\^a}ris}, I. and {Petitjean}, P. and {Burtin}, E. and {Dawson}, K. and {McGreer}, I. and {Myers}, A.~D. and {Rossi}, G. and {Schlegel}, D. and {Schneider}, D. and {Streblyanska}, A. and {Tinker}, J.},
        title = "{The extended Baryon Oscillation Spectroscopic Survey: Variability selection and quasar luminosity function}",
      journal = {\aap},
         year = 2016,
        month = mar,
       volume = {587},
          eid = {A41},
        pages = {A41},
          doi = {10.1051/0004-6361/201527392},
archivePrefix = {arXiv},
       eprint = {1509.05607},
 primaryClass = {astro-ph.CO},
       adsurl = {https://ui.adsabs.harvard.edu/abs/2016A&A...587A..41P}
}

@ARTICLE{paris2011,
       author = {{P{\^a}ris}, I. and {Petitjean}, P. and {Rollinde}, E. and {Aubourg}, E. and {Busca}, N. and {Charlassier}, R. and {Delubac}, T. and {Hamilton}, J. -Ch. and {Le Goff}, J. -M. and {Palanque-Delabrouille}, N. and {Peirani}, S. and {Pichon}, Ch. and {Rich}, J. and {Vargas-Maga{\~n}a}, M. and {Y{\`e}che}, Ch.},
        title = "{A principal component analysis of quasar UV spectra at z \raisebox{-0.5ex}\textasciitilde 3}",
      journal = {\aap},
         year = 2011,
        month = jun,
       volume = {530},
          eid = {A50},
        pages = {A50},
          doi = {10.1051/0004-6361/201016233},
archivePrefix = {arXiv},
       eprint = {1104.2024},
 primaryClass = {astro-ph.CO},
       adsurl = {https://ui.adsabs.harvard.edu/abs/2011A&A...530A..50P}
}

@INPROCEEDINGS{Pieri2016,
       author = {{Pieri}, M.~M. and {Bonoli}, S. and {Chaves-Montero}, J. and {P{\^a}ris}, I. and {Fumagalli}, M. and {Bolton}, J.~S. and {Viel}, M. and {Noterdaeme}, P. and {Miralda-Escud{\'e}}, J. and {Busca}, N.~G. and {Rahmani}, H. and {Peroux}, C. and {Font-Ribera}, A. and {Trager}, S.~C.},
        title = "{WEAVE-QSO: A Massive Intergalactic Medium Survey for the William Herschel Telescope}",
    booktitle = {SF2A-2016: Proceedings of the Annual meeting of the French Society of Astronomy and Astrophysics},
         year = 2016,
       editor = {{Reyl{\'e}}, C. and {Richard}, J. and {Cambr{\'e}sy}, L. and {Deleuil}, M. and {P{\'e}contal}, E. and {Tresse}, L. and {Vauglin}, I.},
        month = dec,
        pages = {259-266},
          doi = {10.48550/arXiv.1611.09388},
archivePrefix = {arXiv},
       eprint = {1611.09388},
 primaryClass = {astro-ph.CO},
       adsurl = {https://ui.adsabs.harvard.edu/abs/2016sf2a.conf..259P}
}

@ARTICLE{Pistis2025,
       author = {{Pistis}, Francesco and {Fumagalli}, Michele and {Fossati}, Matteo and {Berg}, Trystyn and {Mangola}, Elena S. and {Dutta}, Rajeshwari and {Grespan}, Margherita and {Iovino}, Angela and {Ma{\l}ek}, Katarzyna and {Morrison}, Sean and {Murphy}, David N.~A. and {Pearson}, William J. and {P{\'e}rez-R{\'a}fols}, Ignasi and {Pieri}, Matthew M. and {Pollo}, Agnieszka and {Vergani}, Daniela},
        title = "{Automated quasar continuum estimation using neural networks: A comparative study of deep-learning architectures}",
      journal = {\aap},
         year = 2025,
        month = jun,
       volume = {698},
          eid = {A292},
        pages = {A292},
          doi = {10.1051/0004-6361/202453377},
archivePrefix = {arXiv},
       eprint = {2505.10976},
 primaryClass = {astro-ph.GA},
       adsurl = {https://ui.adsabs.harvard.edu/abs/2025A&A...698A.292P}
}

@ARTICLE{Prochaska2014,
       author = {{Prochaska}, J. Xavier and {Madau}, Piero and {O'Meara}, John M. and {Fumagalli}, Michele},
        title = "{Towards a unified description of the intergalactic medium at redshift z {\ensuremath{\approx}} 2.5}",
      journal = {\mnras},
         year = 2014,
        month = feb,
       volume = {438},
       number = {1},
        pages = {476-486},
          doi = {10.1093/mnras/stt2218},
archivePrefix = {arXiv},
       eprint = {1310.0052},
 primaryClass = {astro-ph.CO},
       adsurl = {https://ui.adsabs.harvard.edu/abs/2014MNRAS.438..476P}
}

@ARTICLE{Puchwein2023,
       author = {{Puchwein}, Ewald and {Bolton}, James S. and {Keating}, Laura C. and {Molaro}, Margherita and {Gaikwad}, Prakash and {Kulkarni}, Girish and {Haehnelt}, Martin G. and {Ir{\v{s}}i{\v{c}}}, Vid and {{\v{S}}oltinsk{\'y}}, Tom{\'a}{\v{s}} and {Viel}, Matteo and {Aubert}, Dominique and {Becker}, George D. and {Meiksin}, Avery},
        title = "{The Sherwood-Relics simulations: overview and impact of patchy reionization and pressure smoothing on the intergalactic medium}",
      journal = {\mnras},
         year = 2023,
        month = mar,
       volume = {519},
       number = {4},
        pages = {6162-6183},
          doi = {10.1093/mnras/stac3761},
archivePrefix = {arXiv},
       eprint = {2207.13098},
 primaryClass = {astro-ph.CO},
       adsurl = {https://ui.adsabs.harvard.edu/abs/2023MNRAS.519.6162P}
}

@article{Pavicevic2025,
  title = {Constraints on Primordial Magnetic Fields from the Lyman-$\ensuremath{\alpha}$ Forest},
  author = {Pavi\ifmmode \check{c}\else \v{c}\fi{}evi\ifmmode \acute{c}\else \'{c}\fi{}, Mak and Ir\ifmmode \check{s}\else \v{s}\fi{}i\ifmmode \check{c}\else \v{c}\fi{}, Vid and Viel, Matteo and Bolton, James S. and Haehnelt, Martin G. and Martin-Alvarez, Sergio and Puchwein, Ewald and Ralegankar, Pranjal},
  journal = {Phys. Rev. Lett.},
  volume = {135},
  issue = {7},
  pages = {071001},
  numpages = {7},
  year = {2025},
  month = {Aug},
  publisher = {American Physical Society},
  doi = {10.1103/77rd-vkpz},
  url = {https://link.aps.org/doi/10.1103/77rd-vkpz}
}

@ARTICLE{Ravoux2023,
       author = {{Ravoux}, Corentin and {Abdul Karim}, Marie Lynn and {Armengaud}, Eric and {Walther}, Michael and {Kara{\c{c}}ayl{\i}}, Naim G{\"o}ksel and {Martini}, Paul and {Guy}, Julien and {Aguilar}, Jessica Nicole and {Ahlen}, Steven and {Bailey}, Stephen and {Bautista}, Julian and {Beltran}, Sergio Felipe and {Brooks}, David and {Cabayol-Garcia}, Laura and {Chabanier}, Sol{\`e}ne and {Chaussidon}, Edmond and {Chaves-Montero}, Jon{\'a}s and {Dawson}, Kyle and {de la Cruz}, Rodrigo and {de la Macorra}, Axel and {Doel}, Peter and {Fanning}, Kevin and {Font-Ribera}, Andreu and {Forero-Romero}, Jaime and {Gontcho A Gontcho}, Satya and {Gonzalez-Morales}, Alma X. and {Gordon}, Calum and {Herrera-Alcantar}, Hiram K. and {Honscheid}, Klaus and {Ir{\v{s}}i{\v{c}}}, Vid and {Ishak}, Mustapha and {Kehoe}, Robert and {Kisner}, Theodore and {Kremin}, Anthony and {Landriau}, Martin and {Le Guillou}, Laurent and {Levi}, Michael and {Luki{\'c}}, Zarija and {Magneville}, Christophe and {Meisner}, Aaron and {Miquel}, Ramon and {Moustakas}, John and {Mueller}, Eva-Maria and {Mu{\~n}oz-Guti{\'e}rrez}, Andrea and {Napolitano}, Lucas and {Nie}, Jundan and {Niz}, Gustavo and {Palanque-Delabrouille}, Nathalie and {Percival}, Will and {P{\'e}rez-R{\`a}fols}, Ignasi and {Pieri}, Matthew and {Poppett}, Claire and {Prada}, Francisco and {Ram{\'\i}rez P{\'e}rez}, C{\'e}sar and {Rossi}, Graziano and {Sanchez}, Eusebio and {Schlegel}, David and {Schubnell}, Michael and {Seo}, Hee-Jong and {Sinigaglia}, Francesco and {Tan}, Ting and {Tarl{\'e}}, Gregory and {Wang}, Ben and {Weaver}, Benjamin and {Y{\`e}che}, Christophe and {Zhou}, Zhimin},
        title = "{The Dark Energy Spectroscopic Instrument: one-dimensional power spectrum from first Ly {\ensuremath{\alpha}} forest samples with Fast Fourier Transform}",
      journal = {\mnras},
         year = 2023,
        month = dec,
       volume = {526},
       number = {4},
        pages = {5118-5140},
          doi = {10.1093/mnras/stad3008},
archivePrefix = {arXiv},
       eprint = {2306.06311},
 primaryClass = {astro-ph.CO},
       adsurl = {https://ui.adsabs.harvard.edu/abs/2023MNRAS.526.5118R}
}

@ARTICLE{Rauch1997,
       author = {{Rauch}, Michael and {Miralda-Escud{\'e}}, Jordi and {Sargent}, Wallace L.~W. and {Barlow}, Tom A. and {Weinberg}, David H. and {Hernquist}, Lars and {Katz}, Neal and {Cen}, Renyue and {Ostriker}, Jeremiah P.},
        title = "{The Opacity of the Ly{\ensuremath{\alpha}} Forest and Implications for {\ensuremath{\Omega}}$_{b}$ and the Ionizing Background}",
      journal = {\apj},
         year = 1997,
        month = nov,
       volume = {489},
       number = {1},
        pages = {7-20},
          doi = {10.1086/304765},
archivePrefix = {arXiv},
       eprint = {astro-ph/9612245},
 primaryClass = {astro-ph},
       adsurl = {https://ui.adsabs.harvard.edu/abs/1997ApJ...489....7R}
}

@ARTICLE{Rogers2025,
       author = {{Rogers}, Keir K. and {Poulin}, Vivian},
        title = "{5{\ensuremath{\sigma}} tension between Planck cosmic microwave background and eBOSS Lyman-alpha forest and constraints on physics beyond {\ensuremath{\Lambda}}CDM}",
      journal = {Physical Review Research},
         year = 2025,
        month = jan,
       volume = {7},
       number = {1},
          eid = {L012018},
        pages = {L012018},
          doi = {10.1103/PhysRevResearch.7.L012018},
archivePrefix = {arXiv},
       eprint = {2311.16377},
 primaryClass = {astro-ph.CO},
       adsurl = {https://ui.adsabs.harvard.edu/abs/2025PhRvR...7a2018R}
}

@ARTICLE{Schaye2003,
       author = {{Schaye}, Joop and {Aguirre}, Anthony and {Kim}, Tae-Sun and {Theuns}, Tom and {Rauch}, Michael and {Sargent}, Wallace L.~W.},
        title = "{Metallicity of the Intergalactic Medium Using Pixel Statistics. II. The Distribution of Metals as Traced by C IV}",
      journal = {\apj},
         year = 2003,
        month = oct,
       volume = {596},
       number = {2},
        pages = {768-796},
          doi = {10.1086/378044},
archivePrefix = {arXiv},
       eprint = {astro-ph/0306469},
 primaryClass = {astro-ph},
       adsurl = {https://ui.adsabs.harvard.edu/abs/2003ApJ...596..768S}
}

@ARTICLE{Springel2005,
       author = {{Springel}, Volker},
        title = "{The cosmological simulation code GADGET-2}",
      journal = {\mnras},
         year = 2005,
        month = dec,
       volume = {364},
       number = {4},
        pages = {1105-1134},
          doi = {10.1111/j.1365-2966.2005.09655.x},
archivePrefix = {arXiv},
       eprint = {astro-ph/0505010},
 primaryClass = {astro-ph},
       adsurl = {https://ui.adsabs.harvard.edu/abs/2005MNRAS.364.1105S}
}

@ARTICLE{Szakacs2023,
       author = {{Szakacs}, Roland and {P{\'e}roux}, C{\'e}line and {Nelson}, Dylan and {Zwaan}, Martin A. and {Gr{\"u}n}, Daniel and {Weng}, Simon and {Fresco}, Alejandra Y. and {Bollo}, Victoria and {Casavecchia}, Benedetta},
        title = "{The BarYon Cycle project (ByCycle): identifying and localizing Mg II metal absorbers with machine learning}",
      journal = {\mnras},
         year = 2023,
        month = dec,
       volume = {526},
       number = {3},
        pages = {3744-3756},
          doi = {10.1093/mnras/stad2431},
archivePrefix = {arXiv},
       eprint = {2305.17970},
 primaryClass = {astro-ph.GA},
       adsurl = {https://ui.adsabs.harvard.edu/abs/2023MNRAS.526.3744S}
}

@ARTICLE{Schaye2000,
       author = {{Schaye}, Joop and {Theuns}, Tom and {Rauch}, Michael and {Efstathiou}, George and {Sargent}, Wallace L.~W.},
        title = "{The thermal history of the intergalactic medium$^{*}$}",
      journal = {\mnras},
         year = 2000,
        month = nov,
       volume = {318},
       number = {3},
        pages = {817-826},
          doi = {10.1046/j.1365-8711.2000.03815.x},
archivePrefix = {arXiv},
       eprint = {astro-ph/9912432},
 primaryClass = {astro-ph},
       adsurl = {https://ui.adsabs.harvard.edu/abs/2000MNRAS.318..817S}
}

@article{Theuns2002,
doi = {10.1086/339998},
url = {https://doi.org/10.1086/339998},
year = {2002},
month = {feb},
publisher = {},
volume = {567},
number = {2},
pages = {L103},
author = {Theuns, Tom and Schaye, Joop and Zaroubi, Saleem and Kim, Tae-Sun and Tzanavaris, Panayiotis and Carswell, Bob},
title = {Constraints on Reionization from the Thermal History of the Intergalactic Medium* **},
journal = {The Astrophysical Journal}
}

@ARTICLE{Viel2004,
       author = {{Viel}, Matteo and {Haehnelt}, Martin G. and {Springel}, Volker},
        title = "{Inferring the dark matter power spectrum from the Lyman {\ensuremath{\alpha}} forest in high-resolution QSO absorption spectra}",
      journal = {\mnras},
         year = 2004,
        month = nov,
       volume = {354},
       number = {3},
        pages = {684-694},
          doi = {10.1111/j.1365-2966.2004.08224.x},
archivePrefix = {arXiv},
       eprint = {astro-ph/0404600},
 primaryClass = {astro-ph},
       adsurl = {https://ui.adsabs.harvard.edu/abs/2004MNRAS.354..684V}
}

@ARTICLE{Viel2013,
       author = {{Viel}, Matteo and {Becker}, George D. and {Bolton}, James S. and {Haehnelt}, Martin G.},
        title = "{Warm dark matter as a solution to the small scale crisis: New constraints from high redshift Lyman-{\ensuremath{\alpha}} forest data}",
      journal = {\prd},
         year = 2013,
        month = aug,
       volume = {88},
       number = {4},
          eid = {043502},
        pages = {043502},
          doi = {10.1103/PhysRevD.88.043502},
archivePrefix = {arXiv},
       eprint = {1306.2314},
 primaryClass = {astro-ph.CO},
       adsurl = {https://ui.adsabs.harvard.edu/abs/2013PhRvD..88d3502V}
}

@ARTICLE{Villasenor2023,
       author = {{Villasenor}, Bruno and {Robertson}, Brant and {Madau}, Piero and {Schneider}, Evan},
        title = "{New constraints on warm dark matter from the Lyman-{\ensuremath{\alpha}} forest power spectrum}",
      journal = {\prd},
         year = 2023,
        month = jul,
       volume = {108},
       number = {2},
          eid = {023502},
        pages = {023502},
          doi = {10.1103/PhysRevD.108.023502},
archivePrefix = {arXiv},
       eprint = {2209.14220},
 primaryClass = {astro-ph.CO},
       adsurl = {https://ui.adsabs.harvard.edu/abs/2023PhRvD.108b3502V}
}

@ARTICLE{Wang2022,
       author = {{Wang}, Ben and {Zou}, Jiaqi and {Cai}, Zheng and {Prochaska}, J. Xavier and {Sun}, Zechang and {Ding}, Jiani and {Font-Ribera}, Andreu and {Gonzalez}, Alma and {Herrera-Alcantar}, Hiram K. and {Irsic}, Vid and {Lin}, Xiaojing and {Brooks}, David and {Chabanier}, Sol{\'e}ne and {de Belsunce}, Roger and {Palanque-Delabrouille}, Nathalie and {Tarle}, Gregory and {Zhou}, Zhimin},
        title = "{Deep Learning of Dark Energy Spectroscopic Instrument Mock Spectra to Find Damped Ly{\ensuremath{\alpha}} Systems}",
      journal = {\apjs},
         year = 2022,
        month = mar,
       volume = {259},
       number = {1},
          eid = {28},
        pages = {28},
          doi = {10.3847/1538-4365/ac4504},
archivePrefix = {arXiv},
       eprint = {2201.00827},
 primaryClass = {astro-ph.GA},
       adsurl = {https://ui.adsabs.harvard.edu/abs/2022ApJS..259...28W}
}

@ARTICLE{Wu2019,
       author = {{Wu}, Xiaohan and {McQuinn}, Matthew and {Kannan}, Rahul and {D'Aloisio}, Anson and {Bird}, Simeon and {Marinacci}, Federico and {Dav{\'e}}, Romeel and {Hernquist}, Lars},
        title = "{Imprints of temperature fluctuations on the z {\ensuremath{\sim}} 5 Lyman-{\ensuremath{\alpha}} forest: a view from radiation-hydrodynamic simulations of reionization}",
      journal = {\mnras},
         year = 2019,
        month = dec,
       volume = {490},
       number = {3},
        pages = {3177-3195},
          doi = {10.1093/mnras/stz2807},
archivePrefix = {arXiv},
       eprint = {1907.04860},
 primaryClass = {astro-ph.CO},
       adsurl = {https://ui.adsabs.harvard.edu/abs/2019MNRAS.490.3177W}
}

@ARTICLE{Zheng2026,
       author = {{Zheng}, Yifan and {Montero-Camacho}, Paulo and {Cai}, Zheng and {Mao}, Yi},
        title = "{Tantalizing evidence of reionization relics in the eBOSS DR16 Ly {\ensuremath{\alpha}} forest correlations: a preference for early reionization}",
      journal = {\mnras},
         year = 2026,
        month = feb,
       volume = {545},
       number = {4},
          eid = {staf2014},
        pages = {staf2014},
          doi = {10.1093/mnras/staf2014},
archivePrefix = {arXiv},
       eprint = {2503.09971},
 primaryClass = {astro-ph.CO},
       adsurl = {https://ui.adsabs.harvard.edu/abs/2026MNRAS.545f2014Z}
}

\appendix

\section{Posterior distributions}
\label{app:posteriors}

\begin{figure*}
    \centering
    \includegraphics[width=2\columnwidth]{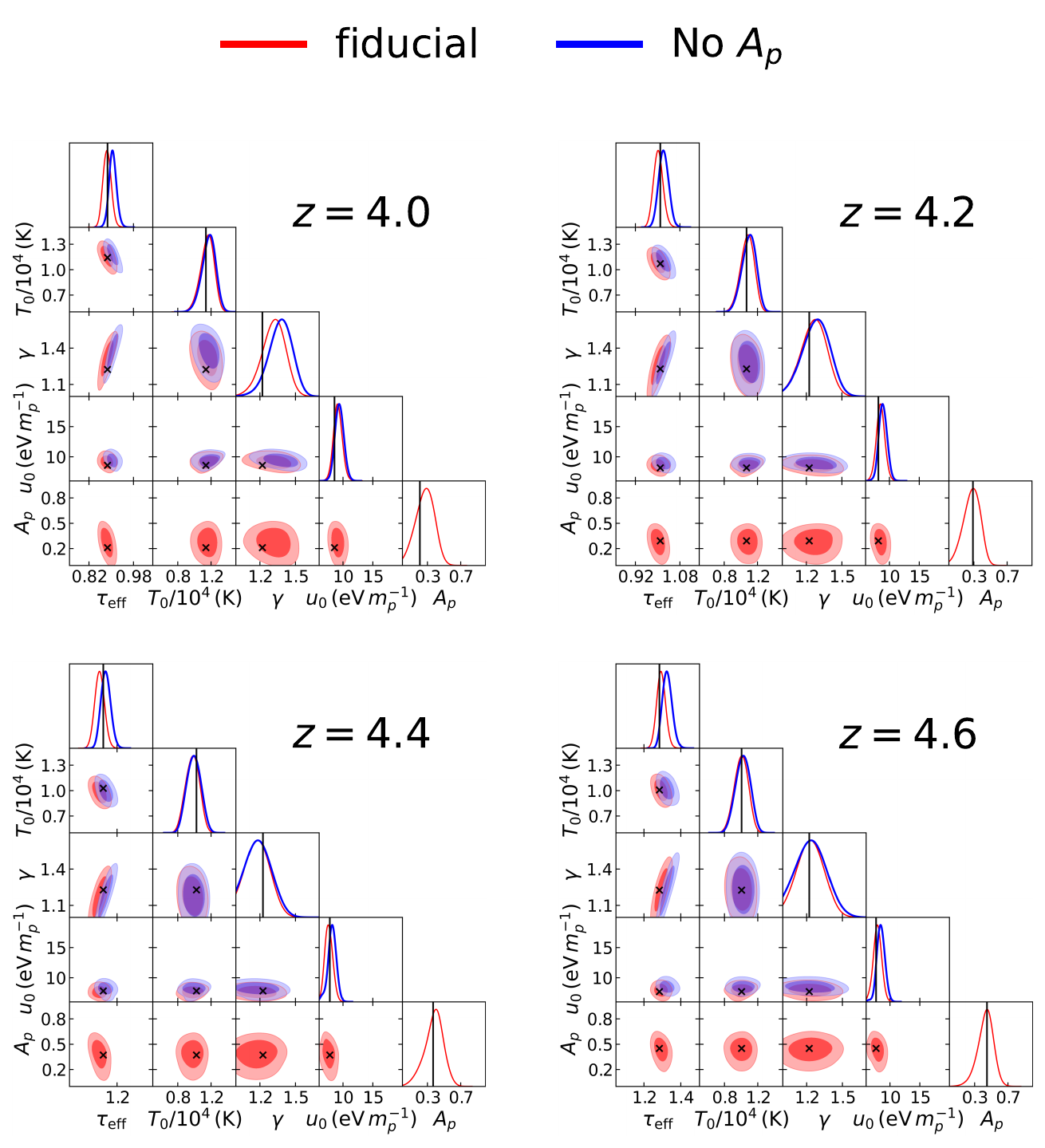}
    \caption{The 1D and 2D posterior distributions for $\tau_{\rm eff}$, $T_0$, $\gamma$, $u_0$, and $A_p$ for our fiducial model (red contours) and no-$A_{\rm p}$ model (blue contours). The shaded regions show the 68 and 95 per cent confidence intervals for the joint distributions. The four panels are for the redshift bins used in this study, namely $z = 4.0$, $4.2$, $4.4$ and $4.6$. Expected parameter values are shown as black crosses in the contour panels and vertical black lines in the one-dimensional posterior panels. Note that only the fiducial posterior is shown in panels involving $A_{\rm p}$. The similarity of the posteriors in the two cases indicates the preference for patchy reionization is not driven by significant shifts in the other IGM parameters.}
    \label{fig:full_posteriors_fid_noap}
\end{figure*}

\begin{figure*}
    \centering
    \includegraphics[width=2\columnwidth]{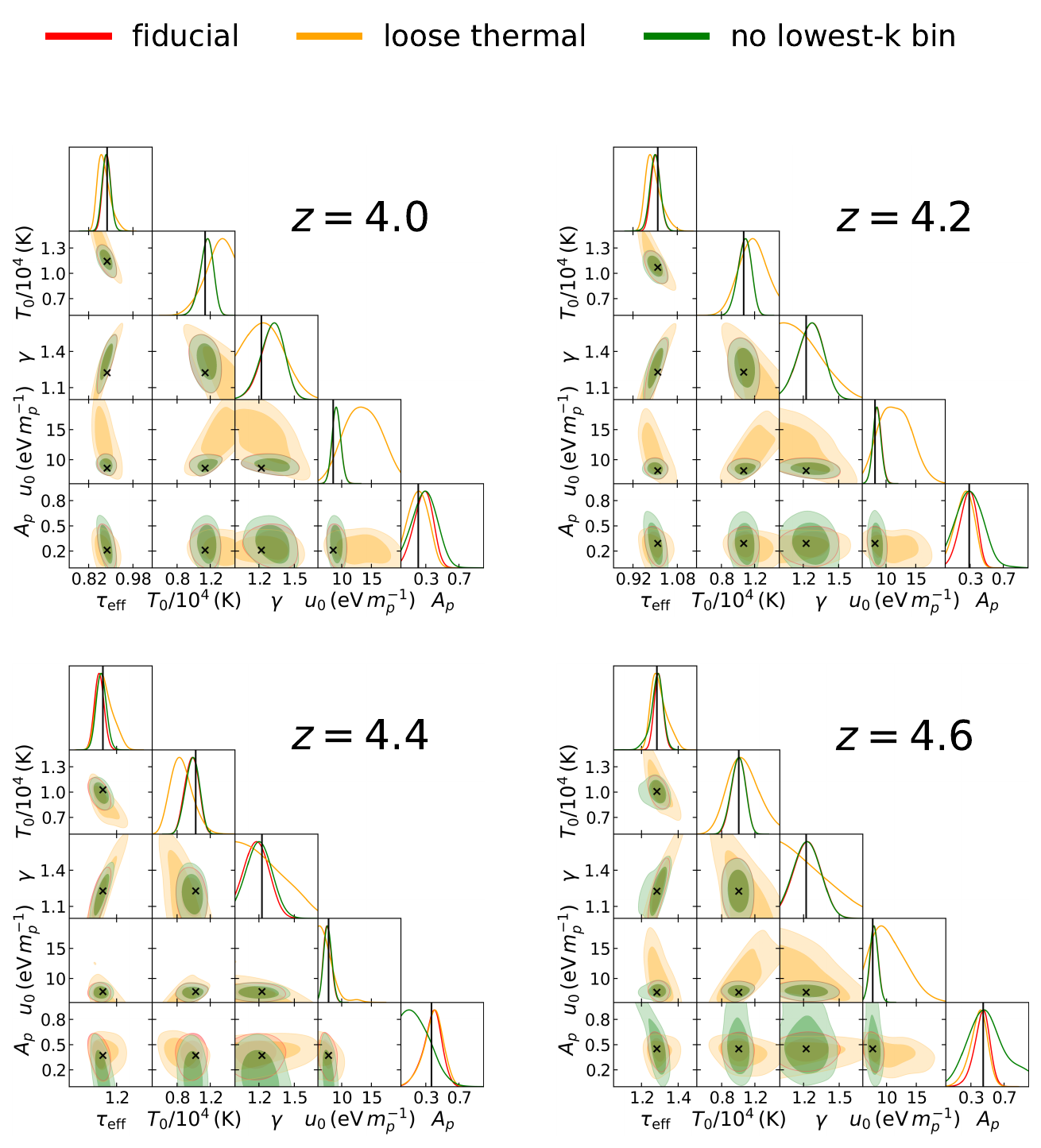}
    \caption{As for Figure~\ref{fig:full_posteriors_fid_noap}, but now showing our fiducial model (red contours), loose thermal prior model (orange contours), and no-lowest-$k$ model (green contours).  The posterior for $A_{\rm p}$ does not change significantly with the thermal prior, although the constraint on $A_{\rm p}$ is significantly weakened when ignoring the largest-scale $k$ bin.}
    \label{fig:full_posteriors_robustness}
\end{figure*}

Figures~\ref{fig:full_posteriors_fid_noap} and~\ref{fig:full_posteriors_robustness} show the marginalized posterior distributions for all five parameters ($\tau_{\rm eff}$, $T_0$, $\gamma$, $u_0$, $A_{\rm p}$) varied in each $P_{\rm Ly\alpha}$ redshift bin. Figure~\ref{fig:full_posteriors_fid_noap} corresponds to the best fit $P_{\rm Ly\alpha}$ displayed in Figure~\ref{fig:mcmc_p1d_comparison}, while Figure \ref{fig:full_posteriors_robustness} corresponds to the 1D posteriors displayed in Figure~\ref{fig:ap_recovery}.

\end{document}